\documentstyle{article}
\begin{document}
\makeatletter
\@addtoreset{equation}{section}
\makeatother
\tolerance=10000
\renewcommand{\theequation}{\thesection.\arabic{equation}}      
\def\nn{\nonumber}
\def\NSNS{${NS}\otimes {NS}$}
\def\RR{${\cal R}\otimes {\cal R}$}
\def\calF{{\cal F}}
\def\calS{{\cal S}}
\def\calG{{\cal G}}
\def\calL{{\cal L}}
\def\calt{{\cal T}}
\def\calV{{\cal V}}
\def\calN{{\cal N}}
\def\calP{{\cal P}}
\def\calD{{\cal D}}
\def\calO{{\cal O}}
\def\mh{{\hat \mu}}
\def\nh{{\hat \nu}}
\def\rh{{\hat \rho}}
\def\tr{{\rm tr}}
\def\half{{\textstyle {1 \over 2}}}
\def\spin{{\rm spin}}
\gdef\angbr#1{<#1>}
\def\Re{\rm Re}
\def\ie{{\it i.e.}}
\def\half{{\textstyle {1 \over 2}}}
\def\quart{{\textstyle {1 \over 4}}}
\def\thh{{\textstyle {3 \over 2}}}
\def\gtilde{\tilde \sqrt{-g}g}
\def\AdS5s5{$AdS_5\times S^5$}
\def\gy{g_{_{YM}}}
\def\al {\alpha'}

\def\thbar{{\bar \theta}}
\def\Sl2{SL(2,\IZ)}
\def\np{Nucl. Phys.}
\def\pl{Phys. Lett.}
\def\pr{Phys. Rev.}
\def\prl{Phys. Rev. Lett.}
\def\cmp{Comm. Math. Phys.}
\def\gtilde{\tilde \sqrt{-g}g}
\def\AdS5s5{$AdS_5\times S^5$}

\def\Re{\rm Re}
\def\Im{\rm Im}
\def\titlestyle#1{\par\begingroup \interlinepenalty=9999
\fourteenpoint
\noindent #1\par\endgroup }
\def\tr{{\rm tr}}
\def\Tr{{\rm Tr}}
\def\half{{\textstyle {1 \over 2}}}
\def\calt{{\cal T}}
\def\Rfour{\calR^4}
\def\lam16{\lambda^{16}}
\def\psih{\hat \psi}
\def\hI{\hat I}
\def\ie{{\it i.e.}}
\def\quart{{\textstyle {1 \over 4}}}
\def\gtilde{\tilde \sqrt{-g}g}
\def\lr { \lref}

\def\pano{\par\noindent}
\def\smno{\smallskip\noindent}
\def\meno{\medskip\noindent}
\def\bigno{\bigskip\noindent}
\font\cmss=cmss10
\font\cmsss=cmss10 at 7pt
\def\rlx{\relax\leavevmode}
\def\inbar{\vrule height1.5ex width.4pt depth0pt}
\def\IC{\relax\,\hbox{$\inbar\kern-.3em{\rm C}$}}
\def\IN{\relax{\rm I\kern-.18em N}}
\def\IP{\relax{\rm I\kern-.18em P}}
\def\ZZ{\rlx\leavevmode\ifmmode\mathchoice{\hbox{\cmss Z\kern-.4em Z}}
 {\hbox{\cmss Z\kern-.4em Z}}{\lower.9pt\hbox{\cmsss Z\kern-.36em Z}}
 {\lower1.2pt\hbox{\cmsss Z\kern-.36em Z}}\else{\cmss Z\kern-.4em
 Z}\fi}
\def\IZ{\relax\ifmmode\mathchoice
{\hbox{\cmss Z\kern-.4em Z}}{\hbox{\cmss Z\kern-.4em Z}}
{\lower.9pt\hbox{\cmsss Z\kern-.4em Z}}
{\lower1.2pt\hbox{\cmsss Z\kern-.4em Z}}\else{\cmss Z\kern-.4em
Z}\fi}

\def\narrowplus{\kern -.04truein + \kern -.03truein}
\def\narrowminus{- \kern -.04truein}
\def\narrowminussub{\kern -.02truein - \kern -.01truein}
\font\ninerm=cmr9
\def\bps{Bogomol'nyi}
\def\YM{Yang-Mills}
\def\kh{K\"{a}hler}
\def\half{{1\over 2}}
\def\sub{\scriptscriptstyle}
\def\calR{{\cal R}}
\def\sF{{\cal F}}
\def\sG{{\cal G}}
\def\sP{{\cal P}}
\def\ap{{\alpha^{\prime}}}  
\def\ker{{\rm Ker}}
\def\cok{{\rm Cok}}
\def\cl{\centerline}
\def\ext{{\rm Ext}}
\def\hom{{\rm Hom}}
\def\zbar{\overline{z}}
\def\vac{\vert 0\rangle}
\def\lvac{\langle 0 \vert}
\def\oo{\overline}
\def\cA{{\cal A}}
\def\cB{{\cal B}}
\def\Mt{\widetilde{M}}
\def\b{{\beta}}
\def\a{{\alpha}}
\def\g{{\gamma}}
\def\e{{\epsilon}}
\def\vol{{\rm vol}}
\def\Pf{{\rm Pf}}
\def\l{{\lambda}}
\def\w{{\omega}}
\def\ph{{\phi}}
\def\r{{\rightarrow}}
\def\cp{{\overline \psi}}
\def\frac#1#2{{#1\over #2}}
\def\x{{\tilde x}}
\def\com#1#2{{ \left[ #1, #2 \right] }}
\def\h#1#2{{ h_{#1}^{ \{ #2 \} }  }}
\def\acom#1#2{{ \left\{ #1, #2 \right\} }}
\def\xp{{ x\cdot p}}
\def\pp{{ p\cdot p}}
\def\xx{{ x\cdot x}}

\def\CA{{\cal A}}
\def\CC{{\cal C}}
\def\CD{{\cal D}}
\def\CE{{\cal E}}
\def\CF{{\cal F}}
\def\CT{{\cal T}}
\def\CM{{\cal M}}
\def\CN{{\cal N}}
\def\CP{{\cal P}}
\def\CL{{\cal L}}
\def\CV{{\cal V}}
\def\CS{{\cal S}}
\def\CW{{\cal W}}
\def\BR{\IR}
\def\BZ{\IZ}
\def\BC{\IC}
\def\BM{\QM}
\def\BP{\IP}
\def\BH{\QH}
\def\BX{\QX}
\def\Bone{{\bf 1}}
\def\sym#1{{{\rm SYM}} _{#1 +1}}
\def\imp{$\Rightarrow$}
\def\IZ{\relax\ifmmode\mathchoice
{\hbox{\cmss Z\kern-.4em Z}}{\hbox{\cmss Z\kern-.4em Z}}
{\lower.9pt\hbox{\cmsss Z\kern-.4em Z}}
{\lower1.2pt\hbox{\cmsss Z\kern-.4em Z}}\else{\cmss Z\kern-.4em
Z}\fi}
\def\IB{\relax{\rm I\kern-.18em B}}
\def\IC{{\relax\hbox{$\inbar\kern-.3em{\rm C}$}}}
\def\ID{\relax{\rm I\kern-.18em D}}
\def\IE{\relax{\rm I\kern-.18em E}}
\def\IF{\relax{\rm I\kern-.18em F}}
\def\IG{\relax\hbox{$\inbar\kern-.3em{\rm G}$}}
\def\IGa{\relax\hbox{${\rm I}\kern-.18em\Gamma$}}
\def\IH{\relax{\rm I\kern-.18em H}}
\def\II{\relax{\rm I\kern-.18em I}}
\def\IK{\relax{\rm I\kern-.18em K}}
\def\IP{\relax{\rm I\kern-.18em P}}
\def\IX{{\bf X}}
\def\mod{{\rm mod}}
\def\ndt{\noindent}
\def\p{\partial}
\def\pab{\pb_{\bar A} }
\def\pb{{\bar \p}}
\def\pgp{\pb g g^{-1}}
\font\cmss=cmss10 \font\cmsss=cmss10 at 7pt
\def\IR{\relax{\rm I\kern-.18em R}}
\def\pbar{\bar{\p}}
\def\ker{{\rm ker\ }}
\def\cok{{\rm cok\ }}
\def\sqap{\sqrt{\ap}}
\def\S{{\Sigma}}
\def\f{\psi}
\def\ph{ {\hat p}}
\def\dd{ {\delta}}
\def\di{{ \widetilde D}}
\def\df{\dot{\psi}}
\def\pV{{p_\Vert}}
\def\RV{{R_\Vert}}
\def\s{{\sigma}}
\def\1{{\bf 1}}
\def\3{{\bf 3}}
\def\7{{\bf 7}}
\def\2{{\bf 2}}
\def\8{{\bf 8}}
\def\ec{{\bf 8_c}}
\def\ev{{\bf 8_v}}
\def\es{{\bf 8_s}}
\def\B{B}
\def\com#1#2{{ \left[ #1, #2 \right] }}
\def\h#1#2{{ h_{#1}^{ \{ #2 \} }  }}
\def\acom#1#2{{ \left\{ #1, #2 \right\} }}
 
\null
\vskip-2.0cm
{\hfill DAMTP-99-13}
\vskip 0.1cm
{\hfill hep-th/9903124}
\vskip 1.5 cm
 
\begin{center}   
 {\centerline {\large \bf Interconnections between type II superstrings,}} 
\vskip 0.3cm 
{\centerline {\large \bf M theory and ${\cal N} =4$ supersymmetric
Yang--Mills}}  
\vskip 1.0cm
{\centerline{ Michael B. Green}} \vspace{0.6cm} 
\medskip 
\centerline{\sl Department of Applied Mathematics and  
Theoretical 
Physics,} 
\centerline{\sl Silver Street, Cambridge CB3 9EW, UK} 
 \medskip
\vskip 0.3cm
 \end{center}
\vskip 3.0cm
\rm

\begin{quote} 

These lecture notes\footnote{\noindent
Extended version of lectures given at 
22nd Johns Hopkins Workshop (Gothenberg, August 20-22 1998);
`Quantum Aspects of Gauge Theories, Supersymmetry and 
Unification', TMR meeting (Corfu,  September  20-26 1998); Andrjewski 
lectures (Berlin, November 1-6 1998).} begin with a 
 review of the first nonleading contributions to the
derivative expansion of the M theory  effective action compactified on
a two-torus.  The form of these higher-derivative interactions is
shown to follow from  ten-dimensional 
type IIB supersymmetry as well as from one-loop quantum  corrections to
classical eleven-dimensional supergravity.  The detailed
information concerning D-instanton effects encoded in these terms is
related to the problem of evaluating the Witten index for $N$
D-particles in the type IIA theory.  Using the
AdS/CFT conjecture,  it also  leads to very specific predictions of
multi-instanton  contributions in  $\calN=4$ supersymmetric $SU(N)$
Yang--Mills theory in the limit of strong 't Hooft coupling.   

\end{quote}
\vskip 1cm
\normalsize

\newpage
\pagestyle{plain}
\setcounter{page}{1}

\def\baselinestretch{1.2}
\baselineskip 13 pt

\vskip 1.0cm
\renewcommand{\labelitemi}{}

\begin{itemize}

\item[1.]  Introduction.

\item[2.]  M theory on $T^2$ and type II superstrings on $S^1$.

\item[3.] Linearized supersymmetry and higher derivative terms.

\item[4.] Nonlinear supersymmetry constraints on terms of order ${\alpha'}^{-1}$.
 
\item[5.] The eleven-dimensional perspective.

\item{} 5.1 The $\calR^4$ term from one loop in eleven dimensions.

\item{} 5.2  One loop with sixteen dilatini.

\item[6.] Fourier series and properties of D-instantons.

\item[7.]  D-particle bound states and the Witten index.

\item{} 7.1 D-particles and D-instantons.

\item{} 7.2 The deficit  term.

\item[8.] The AdS/CFT correspondence and higher derivative terms.

\item{} 8.1 $\calN =4$ Yang--Mills fields and supercurrents.

\item{} 8.2  Generalities concerning higher derivative terms.

\item{} 8.3 Scalar $AdS$ Green functions and the instanton profile.

\item[9.] D-instantons and Yang--Mills instantons.

\item{} 9.1 The classical D-instanton in flat space and in \AdS5s5.

\item{} 9.2 Instantons in the limit of large $N$.

\item[A.] Derivative expansion of the string tree amplitude.

\item[B.] Modular covariant derivatives.

\item[C.] Some properties of Type IIB Supergravity

\item{} C.1 Spinors and gamma matrices.

\item{} C.2 The fields and their supersymmetry transformations.

\item[D.] Fermions in type IIB on $S^1$ and M theory on $T^2$.

\end{itemize}

\vfill\eject

\section{Introduction}
\medskip

\noindent{\it ``The incalculable in full pursuit of the ineffable''}\footnote{Apologies to Oscar Wilde.}
\smallskip

Over the  past few years a number of very interesting nonperturbative
aspects of string theory have emerged and are encapsulated in the term
`M theory'.  The precise meaning of this term is open to a multitude
of interpretations although  a common theme is that M theory
represents  
a framework for describing nonperturbative quantum gravity that
reduces to perturbative string theory or classical eleven-dimensional
supergravity in various limits.  
Although there have been  attempts to give a  microscopic definition
of M theory  it is not at all obvious that the correct concepts have 
yet been   discovered.  Nevertheless, certain fascinating features are
emerging that will surely be of lasting  significance.  Foremost among
these is the interplay between quantum gravity and Yang--Mills gauge
theory, which has been suggested by the structure of string
perturbation theory for many years but has recently come to the fore
in the realization of the `holographic' principle in the context of
the AdS/CFT correspondence.

The point of view followed in these notes is that it
is of interest  to unravel the extent to 
which features of the theory emerge purely as a consequence of its
very  large symmetries  and do not depend on the
microscopic model.
All present formulations of string theory and its M theory extensions
depend in some manner on the background.   Obviously this is not a
satisfactory state of affairs for a quantum theory of gravity.  In
practise, this means that the properties of the theory depend on 
the number of 
moduli of the particular background chosen.   The number of moduli
grows with the number of compact dimensions ---
there are no moduli in eleven dimensions while the ten-dimensional
type IIA theory has a single modulus (the dilaton) and the
ten-dimensional type IIB theory has two moduli (associated with the
complex scalar field). 
In lower dimensions the number of moduli increases further
and there is a corresponding increase in the size of the duality group.  
In order to avoid the possibility of terminological confusion the term
`M theory' will be restricted in these notes to apply to the
low energy  expansion of eleven-dimensional supergravity
compactified on  specified backgrounds\footnote{The ultimate fundamental 
theory is perhaps worthy of representation by an exquisite
third millenium 
\TeX\ symbol instead of a mundane latin letter}. 
\medskip

\noindent{\it Overview}
\smallskip

In section 2 we will begin with a  brief review of  
the classical  relationships
between    M theory compactified on a two-torus, $T^2$,  to nine
dimensions and the two type II superstring theories compactified on a
circle, $S^1$. 
These are configurations with maximal supersymmetry --- there are 32 unbroken 
components in the supercharges.  The nine-dimensional duality group is $\Sl2$ 
and there are three scalar moduli fields that parameterize the three-dimensional  coset space $\Sl
2 \backslash SL(2,\IR)/U(1) \times R$.  The scalars may be identified with 
the volume and the complex structure of $T^2$. 
The duality properties 
were  originally motivated \cite{witten,aspinwall,schwarz}
by comparison of the classical (Einstein--Hilbert)
eleven-dimensional action \cite{cjs} with the classical low energy
string type II string actions.    Furthermore, the BPS states of the
theories also coincide.  For example, there are states in which the
M2-brane (the membrane solution of eleven-dimensional supergravity)
is wrapped around the two-cycles of $T^2$.  Such states are
interpreted in the string theories as fundamental string states with
the string wound around the circle that forms the tenth dimension.
This is a simple  illustration of  
the more general network of dualities that relate all 
the different string perturbation theories and eleven-dimensional
supergravity \cite{hulltown,witten}.

The duality that is manifested by the low energy effective actions of
eleven-dimensional supergravity and type II string theories is
supposed to extend to the complete quantum theories.  We will consider
the expansion of the   quantum effective action  in powers of derivatives
on the fields.  Thus, in string theory language the  Wilsonian action
will be expanded in inverse powers of the string tension\footnote{Recall
that the
string tension $T_F$ has dimensions (length)${}^{-2}$ and 
$\alpha'= (2\pi T_F)^{-1}$ has dimension 
(length)${}^2$.}, 
\begin{equation}\label{effacs}{{\alpha'}^{-4} S = {\alpha'}^{-4} (S^{(0)} +
{\alpha'}  S^{(1)} + \ldots + (\alpha')^n  S^{(n)} + 
\ldots) ,}
\end{equation} %
where, by dimensional analysis,  each power of $\alpha'$ implicitly comes 
with two derivatives on bosonic fields whereas a fermionic field counts as a half power of a derivative, as usual.  
Obviously an  expansion such as (\ref{effacs}) 
is of limited value since it does not capture effects that are 
nonperturbative in $\alpha'$. 
The leading term, $S^{(0)}$,  contains the classical Einstein--Hilbert  
action and all the terms related by supersymmetry. We will use this
notation even in the case of the   type IIB theory where there is a
well-known difficulty since there is no globally well-defined action
for the self-dual five-form
field strength, $F_5$.  However, for the most part we will only use
the notion of an action as  a shorthand for
packaging the equations of motion, which are well-defined. For such
purposes a locally well-defined action, such as that in \cite{lechner,pasti},
is sufficient.    The first
nonzero terms 
beyond $S^{(0)}$ are those in  $S^{(3)}$  
which has a coefficient 
${\alpha'}^{-1}$.
A striking  example is the  topological term in 
type IIA ten-dimensional string theory,  
\begin{equation}\label{vafwiterm}{ {1\over (2\pi)^5 \alpha'}
\,\int d^{10} x \,  B\wedge\, X_8(R),}
\end{equation} %
where $X_8(R)$ is an eight-form made out of the curvature $R$ and $B$ is 
the two-form Neveu-Schwarz--Neveu-Schwarz (\NSNS) antisymmetric tensor potential.  This
term, with its  precisely determined overall coefficient,
can be discovered directly from a one-loop string
\cite{vafwit}  calculation in
which the stringy regulator plays a crucial r\^ole
in regulating an otherwise quadratically  divergent diagram.  It  can also
be discovered by considering the cancellation of chiral gravitational
anomalies in the type IIA 
five-brane.  These anomalies arise from the coupling of the
world-volume chiral fermions and self-dual antisymmetric three-form
field strength  to the pull-back of the  target
space gravitational field.  In the presence of a five-brane 
(\ref{vafwiterm})  transforms anomalously
under local diffeomorphisms and local Lorentz transformations in just
the correct manner to cancel the world-volume anomaly.  This is an
example of the `anomaly inflow mechanism' for cancelling local
anomalies \cite{callharv},  which  was the method used in \cite{duffmin} 
to derive the
generalization of the term (\ref{vafwiterm}) to eleven-dimensional
supergravity where it is the integral of an 
eleven-form, $C^{(3)}  \wedge X_8$.  
Both of these arguments for the presence of the term
(\ref{vafwiterm}) are outside the realm of perturbative quantum field theory --
the first method invokes a stringy cutoff while the second method requires 
consistency of nonperturbative solitonic chiral five-branes.  

Obviously (\ref{vafwiterm}) is not  the only higher derivative
term in the effective  action.  There will
be  terms with arbitrarily high numbers of derivatives as well as terms that are 
nonperturbative in $\alpha'$.   As we will see, the large amount of supersymmetry 
severely constrains the form of certain  classes of 
\lq protected' terms in the effective action.   
Whereas the classical type IIB
supergravity (the supersymmetrized Einstein theory) is invariant
under continuous SL(2,\IR) transformations the IIB superstring only
possesses a discrete (local) $\Sl2$.  This means that the quantum effective action 
only possesses this discrete symmetry which must be a symmetry order by order in $\alpha'$.

The first corrections to the Einstein--Hilbert term are of order ${\alpha'}^{-1}$ (${\alpha'}^{3}$
relative to the leading term).  One such term is a
specific contraction of four Riemann curvatures that only involves
the Weyl tensor components (recall that the Weyl tensor is the part
of the Riemann tensor that is trace-free).  In the type IIB theory 
this term, which will be denoted by $\calR^4$, 
must have the form (in string frame) \cite{greengut}, 
\begin{equation}\label{rfourf}{\int d^{10} x \sqrt{-g} e^{-\phi/2} f^{(0,0)}(\tau,\bar \tau) \calR^4,}
\end{equation} %
where $\tau$ is a complex scalar field, 
\begin{equation}\label{scaldef}{\tau \equiv  \tau_1 + i \tau_2 = C^{(0)} + i e^{-\phi},}
\end{equation} %
and  $C^{(0)}$ is the Ramond--Ramond (\RR) pseudoscalar field and  
$\phi$ is the type IIB dilaton field (which will sometimes be written 
as $\phi^B$ in order to distinguish it from the type IIA dilaton).
The power of $e^{-\phi/2}$ in (\ref{rfourf}) cancels after transformation 
to the Einstein frame, 
\begin{equation}\label{einstrans}{g \to g_{_{E}} e^{\phi/2},}
\end{equation} %
where $g_{{}_{E}}$ is the Einstein-frame metric.   
In this frame the curvature is invariant 
under $\Sl2$ so that the function 
$f^{(0,0)}(\tau,\bar\tau)$ must be a scalar under $\Sl2$ transformations which act as
\begin{equation}\label{slact}{\tau \to {a\tau + b \over c\tau + d},}
\end{equation} %
with integer $a,b,c,d:\ ad-bc=1$.   The notation  in (\ref{rfourf}) has been 
chosen with later generalizations in mind where  we will encounter 
modular forms, $f^{(w,\hat w)}(\tau,\bar\tau)$, where the superscripts $(w,\hat w)$ 
denote holomorphic and antiholomorphic weights.
The string coupling constant is proportional to $e^\phi$ when $\phi$
is constant.  It follows that the large $\tau_2 = e^{-\phi}$ expansion
of $f^{(0,0)}$ must reproduce the terms that are known in the
perturbative type IIB string expansion \cite{greengut,gv}.  It will also include
exponentially suppressed 
terms that correspond to the presence of arbitrary numbers of
D-instantons that will be discussed in subsequent sections.  

At the linearixed level the physical fields of IIB supergravity package
 together into a constrained  `on-shell' superfield  that
 is a function of a  $SO(9,1)$ Grassmann spinor, $\theta$ \cite{howest}.  In
 section 3 this will be used  to express the 
 $\calR^4$ term, together with many other interaction terms of the same
 dimension,  in terms of an integral over the sixteen components of
 $\theta$, which is half the total number of components of the two
 type IIB supersymmetries.  This makes manifest the fact that these
 terms of order ${\alpha'}^{-1}$ 
are related by supersymmetry at the linearised level.  
We will refer to terms such as these that are
 integrals over a fraction of the superspace as `protected' terms.
Generally these interactions violate the $U(1)$ R symmetry of the
classical type IIB theory.   For example, among
such terms is one of the form \cite{ggk} 
\begin{equation}\label{lamdef}{\int d^{10} x \sqrt{-g} e^{-\phi/2}
f^{(12,-12)}(\tau,\bar\tau) \lambda^{16},}
\end{equation} %
where $\lambda$ is the complex dilatino
which is a sixteen-component chiral SO(9,1) spinor.    
The nonholomorphic modular form, 
$f^{(12,-12)}$,  transforms with a phase under
$ \Sl2$ in a manner that cancels the phase
transformation of the dilatino.  The characteristic feature of these
terms is that  the linearized approximation is a good approximation  for 
 the D-instanton contributions but not for the perturbative
contributions of zero instanton number.  The sixteen components of $\theta$
are identified as the components of
supersymmetry that are broken in the 
D-instanton background \cite{greengut,ggk}.

The constraints imposed by the full nonlinear supersymmetry in the
type IIB theory will be the subject of section 4 where the terms
of order ${\alpha'}^{-1}$ in the derivative expansion
of the effective action of type IIB supergravity will
be deduced by    requiring the closure of the superalgebra at this order
\cite{greensethi}.  This is a kind of  Noether procedure that implements
supersymmetry up to any given order in $\alpha'$.  In this manner   terms 
we shall determine not only the terms in  $S^{(3)}$ but certain of the 
modified supersymmetry transformations that relate $S^{(3)}$   to the classical action $S^{(0)}$.
We will see
that supersymmetry requires that the functions of the
moduli $f^{(w,-w)}(\tau,\bar\tau)$ are eigenfunctions
of the Laplace operator on the Poincare upper-half plane with
specific eigenvalues that depend on $w$.    
Assuming that the theory is  invariant under $\Sl2$ 
determines the solutions uniquely.

We will see in section 5 that the  protected higher-derivative terms, 
such as $f^{(0,0)} \, \calR^4$  can be  deduced by consideration 
of one-loop Feynman diagrams in the eleven-dimensional 
theory compactified on $T^2$ \cite{ggv}.  Thus,  the $\calR^4$ term 
will be deduced by considering the one-loop  scattering of four
gravitons. This amplitude can be expressed as a sum over windings 
around the torus of the world-line of the particle circulating in 
the loop.   The dependence on the moduli of the torus emerges from the 
sectors of nonzero winding while the ultraviolet (UV) divergence is contained in 
the zero-winding sector.    Since the  supergravity 
field theory is not renormalizable these     UV 
divergences cannot be interpreted in the absence of  a microscopic 
definition of  M theory (just as string theory regulates the 
ultraviolet divergences in ten-dimensional supergravity).   However,
with any sensible regularization this  
term is proportional to the volume of the torus  and it disappears in 
the limit in which the torus has zero volume.  But this is the limit 
in which M theory is identified with the ten-dimensional type IIB  
theory.  We will find that the expression for the $\calR^4$ term that was 
deduced from  supersymmetry in section 4 is reproduced by this 
one-loop calculation.  More generally, for finite volume, there 
are volume-dependent terms that are ill-defined due to the UV divergence.  
However, if  the condition of T-duality, that relates the
type IIB and IIA string theories in nine dimensions is imposed, 
then the regularized coefficients 
are determined uniquely and there are no ambiguities.  Having fixed 
the value of the regularized divergence the eleven-dimensional limit 
may be recovered  with a specific finite  coefficient for the
$\calR^4$ term\footnote{
For related work  in toroidally compactified 
type II theories see   \cite{kiritsisb,kiritsisc} and in  
type I theories see
\cite{bacha,bachb,kiritsisa}.  $\Sl2$-invariant expressions for
higher-dimensional terms in the type IIB  
effective action have also been proposed in
\cite{russoa,russob,kehaa,kehab,berkova}. }.

In similar fashion the other terms of order ${\alpha'}^{-1}$ can be obtained 
from one-loop processes in M theory on $T^2$.  The interaction of sixteen 
dilatini, $\lambda^{16}$,  will also  be considered in detail in section 5.  
This involves the evaluation of a sixteen-gravitino amplitude in which the gravitini are polarized in
appropriate directions with respect to the two-torus so as to be interpreted 
as spin-1/2 dialtini in the IIB theory \cite{ggk}.
In this case there is no eleven-dimensional divergence at all and the loop 
amplitude is proportional to the modular form $f^{(12,-12)}$. 

The structure of these $o({\alpha'}^{-1})$ interactions will be analyzed   further
in section 6.  The  large $\tau_2$ (small coupling) 
expansion of any of these  terms is determined by writing  the  modular form, 
$f^{(w,-w)}(\tau,\bar\tau)$, 
as a Fourier series in powers of $e^{2\pi i \tau_1}$, which 
is a series of D-instanton terms.  The zero Fourier mode 
contains the perturbative contributions 
(powers of $\tau_2^{-1}$) which consist of the  tree-level 
and one-loop terms with   no higher-order power behaved terms.  The absence 
of perturbative terms beyond one loop points to  a perturbative  
non-renormalization theorem which is a consequence of the 
fact that the protected interactions can be expressed as integrals over 
half the superspace.\footnote{Some indirect arguments corroborate the validity of this `theorem'
\cite{antoniadis,berkovb,piolinea}.}   
The charge-$K$ D-instanton contribution is 
correlated  with a  phase factor  $e^{2\pi i K \tau_1}$. The significant 
$\tau_2$ dependence of the charge-$K$ D-instanton term 
is contained in the Bessel function $K_{-1}(2\pi  |K|\tau_2)$ which 
has a  large-$\tau_2$ expansion  that is suppressed by the exponential
factor $e^{-2\pi|K| \tau_2}(1 + o(\tau_2^{-1}))$.  The coefficient
of the leading charge-$K$ D-instanton contribution will be found to be (with suitable  normalization)
\begin{equation}\label{matform}{\sum_{m|K}{1\over m^2},}
\end{equation} %
where the notation indicates a sum over all the divisors of $K$.
On the other hand, an extension of ideas  in \cite{wittena} suggest that this 
D-instanton measure should be identified with the partition function of the $SU(K)$ 
`zero-dimensional matrix model',
that is the compactification to zero dimensions of
ten-dimensional supersymmetric Yang--Mills theory. 
In this manner we will be able to deduce the partition function for arbitrary $K$.

The $K$ D-instanton measure is related to the problem of
evaluating the Witten index for  
$K$ D-particles which will be the subject of section 7 \cite{ggone}. 
The chain of duality relations between string theory and 
eleven-dimensional supergravity requires that there be precisely
one threshold bound state of $K$ D-particles of charge
$K$   \cite{witten}.  This means that the Witten index for 
the $K$ D-particle system must be equal  to  one.      The case of two
D-particles ($K=2$) was considered in detail in \cite{yi,sethi}
where it was noted that the  Witten index has two contributions --
the `bulk' term is equal to the zero-dimensional matrix model
partition function while   the `boundary' term is  obtained 
from the region of moduli in which the two D-particles can be described as 
identical  free  particles.  
We will see how these ideas generalize to arbitrary $K$.  
The bulk term in the index is now given by the zero-dimensional
matrix model partition function for general $K$ (\ref{matform}).
Furthermore,   if it is assumed that   the
boundary term in the index is again  given by the free propagation of identical 
D-particles of any charge less than $K= mn$ then the total index is
equal to  one as expected.

In section 8 we turn to consider the effect of the higher derivative
terms on the conjectured equivalence of type IIB string theory in
\AdS5s5\ (five-dimensional anti de Sitter space $\times S^5$)  
and four-dimensional $\calN=4$ supersymmetric $SU(N)$  Yang--Mills theory
on the boundary of $AdS_5$ (with the $S^5$ appended). According 
to this correspondence \cite{maldacena} (see also \cite{gkp,wittads}) 
the higher-derivative interactions of order
${\alpha'}^{-1}$ in the type IIB theory should correspond to
contributions of  correlation functions of composite 
operators in the Yang--Mills theory of order $N^{-3/2}$
relative to the leading terms.  This leads to a prediction
of the precise form of the Yang--Mills instanton contributions
to classes of correlation functions that may be obtained from
the protected supergravity processes.  These exact expressions
exhibit the Montonen--Olive duality $\Sl2$ symmetry
\cite{montolive,sen} that is inherited
from the supergravity interaction terms discussed in the earlier sections.

Section 9 will focus on the correspondence between the D-instanton
contributions to IIB supergravity in \AdS5s5\  and the contribution of Yang--Mills
instantons in the boundary supersymmetric Yang--Mills theory.
The D-instanton is the stringy version of a classical solution to type IIB
supergravity that is concentrated around a point in euclidean
ten-dimensional space.  In the Einstein frame this solution is
one in which the metric is flat and $\nabla^2 e^\phi =0$ 
and the BPS condition relates the euclidean continuation of  
$C^{(0)}$  to $e^{-\phi}$.

We will illustrate the AdS/CFT correspondence by comparing the one-instanton 
contribution to the sixteen-dilatino amplitude in three different ways.     
Firstly (method (a)), we will extract the leading one D-instanton term in $f^{(12,-12)}$ which is the 
coefficient of the sixteen-dilatino vertex.
The amplitude with specified boundary data is obtained by contracting each 
leg of this vertex with a dilatino propagator from the vertex to a point on 
the boundary and   integrating the  position of the vertex in the bulk \AdS5s5\ space.  
The AdS/CFT correspondence identifies the boundary value of the dilatino as the source of a particular  fermionic
supercurrent   of the $\calN=4$
Yang--Mills theory.  
It is therefore of interest to compare the sixteen-dilatino 
amplitude   with the correlation function of sixteen of these supercurrents.  
The second method (method (b)) is the direct determination of the leading one-instanton term in
this correlator  
in $SU(2)$ $\calN=4$ Yang--Mills theory which we will see has the same form 
as the supergravity amplitude determined by method (a), up to an undetermined overall coefficient.  
The third method (method (c)), based on the semiclassical approximation to 
scattering in a D-instanton background, again gives the same expression.
The agreement between the functional form of the correlators  in $SU(2)$ 
Yang--Mills and the bulk amplitudes in IIB supergravity    might appear  
somewhat surprising since the  AdS/CFT correspondence is supposed to relate the leading $\alpha'$ piece of the 
$\calR^4$ term to the large-$N$ limit of $SU(N)$ Yang--Mills,  rather than 
to the $N=2$ case.  
 However, recent work on the large-$N$ limit of
the one-instanton contributions of the  
Yang--Mills theory \cite{doreyone} shows that all the $N$-dependence enters  
in the overall coefficients of these correlation functions  and their functional 
form is independent of $N$, thus explaining the agreement for $N=2$.  Furthermore, this overall
coefficient has the predicted $N$-dependence for large $N$.  This impressive 
agreement extends to the  behaviour of the charge-$K$ D-instanton contribution 
\cite{doreytwo,doreythree}  to  leading order
in the coupling constant.   This will also be described
(with few details) in  section 9.

\section{M theory on $T^2$ and string theory  on $S^1$}

The Einstein--Hilbert action of the eleven-dimensional theory has the form,
\begin{equation}\label{einshilb}{S_{EH} = {1\over 2 \kappa_{11}^2}
\int d^{11}x \sqrt {-G^{(11)}} R,}
\end{equation} %
where the overall normalization has been expressed in terms of the
coupling 
$\kappa_{11}\equiv (2\pi  l_P)^{9/2}$ and $l_P$ is the eleven-dimensional
Planck  length. The additional terms in the fully supersymmetric classical 
action of \cite{cjs} involve the three-form potential, $C^{(3)}$, and the 
gravitino, $\psi_{\mh}$  ($\mh,\nh = 0,1,\dots,9,11$).

It was argued  in \cite{witten}  
that eleven-dimensional M theory  compactified on a circle of radius $R_{11} l_P$
\footnote{We will use a convention in which capital letters denote
distances in eleven-dimensional Planck units while lower case letters
denote distances in string units.} should
be identified with ten-dimensional type IIA string theory with a
coupling constant, $g = e^{\phi^A}$ (where $\phi^A$ is the IIA
dilaton), that is related to $R_{11}$.
This is seen by parameterizing the eleven-dimensional metric,
$G_{\mh\nh}$, 
\begin{equation}\label{eleveng}{ ds^2=
G_{\mu\nu} dx^\mu dx^\nu + R_{11}^2 (dx^{11}- C^{(1)}_\mu dx^\mu)^2,}
\end{equation} %
where  $C_\mu^{(1)}$ 
is the \RR\ one-form potential of the type IIA
theory.  It is easy to check that with the  identifications
\begin{equation}\label{fundiden}{R_{11} = e^{2\phi^A/3},}
\end{equation} %
and 
\begin{equation}\label{metiden}{G_{\mu\nu} = e^{-2\phi^A/3} g^A_{\mu\nu}\left({l_P\over l_S}\right)^2,}
\end{equation} %
where  $g^A_{\mu\nu}$ is the string-frame metric of type IIA
string theory, the Einstein--Hilbert action (\ref{einshilb}) compactified on a circle 
becomes
\begin{equation}\label{ehtypea}{S_{EH} = {2\pi \over (2\pi l_S)^8} \int d^{10} x
\sqrt{- g^{A}} e^{-2\phi^A} R .} 
\end{equation} %
In these equations  $l_S$ is the string scale which is related to the fundamental 
string tension, $T_F$, by 
\begin{equation}\label{tendefs}{l_S^2 \equiv \alpha'  = {1\over 2\pi T_F}}
\end{equation} %
and  the curvature in (\ref{ehtypea}) is expressed in the string
metric. The overall coefficient in (\ref{ehtypea}) is often denoted by 
$(2\kappa_{10}^2)^{-1}$ where $\kappa_{10}^2 = (2\pi l_S)^8/4\pi$.
With these conventions the tension in a  Dp-brane is given by
\begin{equation}\label{tenp}{T_p = {1\over l_S}\, (2\pi l_S)^{-p}\, e^{-\phi^A}}
\end{equation} %
and so the tension of the 
D-string ($p=1)$ is $e^{-\phi^A} T_F$   \cite{polchtass}.   The identification of the masses of the Kaluza--Klein
modes of compactified M theory and the masses of the type IIA
D-particles  ($p=0$ in (\ref{tenp})) leads to the relation between the scales,
\begin{equation}\label{relmass}{{1\over R_{11} l_P} = {e^{-\phi^A}\over l_S},} 
\end{equation} %
or
\begin{equation}\label{relcoup}{l_P = e^{\phi^A/3} l_S,}
\end{equation} %
which expresses the eleven-dimensional Planck scale in terms of the string scale and coupling constant.

We will now generalize this description and obtain more insight by
making
use of the fact that in the
nine-dimensional theory
the $\Sl2$ symmetry of the IIB string theory can be interpreted as a
geometric symmetry of M-theory compactified on a torus
\cite{aspinwall,schwarz}.
Three scalar fields  arise from the compactification of the eleven-dimensional
theory (with coordinates labelled $x^0,\dots ,x^9,x^{11}$) on a
two-torus, $T^2$, oriented in the $(x^9,x^{11})$ directions so that
$G_{9\, 9}^2 = R_9^2$.
These scalars  correspond to the volume of the torus,
$\calV$, and the complex structure, $\Omega$.
Using the  ansatz  (\ref{eleveng}) for the eleven-dimensional metric we have 
\begin{equation}\label{met}{\sqrt{-G^{(11)}} = \sqrt{G^{(2)}} \sqrt {- G^{(9)}} 
= \calV \sqrt{-G^{(9)}},}
\end{equation} %
where $\calV=R_9R_{11}$
is the volume of $T^2$. The metric on the two-torus is 
\begin{equation}\label{mtorus}{ G^{(2)}_{IJ} = {\calV\over \Omega_2} 
\pmatrix{|\Omega|^2 & \Omega_1 \cr
\Omega_1 & 1 \cr}}
\end{equation} %
($I,J=9,11$) where the  complex
structure is given by
\begin{equation}\label{mcomp}{\Omega \equiv \Omega_1 + i \Omega_2  =  {G^{(2)}_{9\, 11} + 
i \sqrt{ G^{(2)}}  \over G^{(2)}_{11\, 11}} = C^{(1)} + i {R_{9} \over R_{11}}.}
\end{equation} %
In type IIA language the radius of the tenth dimension is identified as
\begin{equation}\label{radef}{r_A = R_9 {l_P\over l_S} = R_9 R_{11}^{1/2} =
\calV^{{3\over 4}} \Omega_2^{\quart}.}
\end{equation} %

Now T-duality on this circle can be used to determine the relation
with the type IIB theory.  This gives
\begin{equation}\label{rbdef}{r_B = {1\over r_A } =  R_9^{-1}
R_{11}^{-1/2},
\qquad e^{-\phi^B}= r_A e^{-\phi^A} = {R_9 \over R_{11}}}
\end{equation} 
and $C^{(0)} = C^{(1)}_9 = \Omega_1$.  
These equivalences can be stated in the form,
\begin{eqnarray}\label{idenstruc}\Omega   & = & 
\tau^A \equiv  C^{(1)}_9  + i {r_A} e^{-\phi^A} \nn \\ 
& = &  \tau^B \equiv  C^{(0)} + i
e^{-\phi^B} .
\end{eqnarray} %
Large diffeomorphisms of the torus are $\Sl2$ transformations under
which $\Omega$ transforms by 
\begin{equation}\label{omtrans}{\Omega \to {a\Omega + b \over c\Omega + d},}
\end{equation} %
with $a,b,c,d\in \IZ$ and $ad-bc=1$.

The type IIB Einstein--Hilbert action is obtained   by  compactification of the type IIA theory on a circle
of radius $r_A$ followed by T-duality,  leading to the type IIB
expression (now in the type IIB string frame), 
\begin{equation}\label{ehtypeb}{S_{EH} = {2\pi \over (2\pi l_S)^8}  \int d^{10} x 
\sqrt{-g^{B}} e^{-2\phi^B} R ,}
\end{equation} %
where $g^B_{\mu\nu}$ denotes the IIB string metric. 
Supersymmetry determines the remaining terms in the complete classical
action in any of these parameterizations, subject  to the caveat that the IIB 
theory does not have a globally well-defined
action because of the special features of self-dual antisymmetric 
tensor fields \cite{marcschwarz,wittfive}.  It follows from 
simple dimensional analysis that all the terms that only involve bosonic
fields have two derivatives.  A pair of fermionic fields is 
dimensionally equivalent to a derivative on a bosonic field, so there 
are terms with two fermion fields and a single derivative as well as 
terms with four fermionic fields and no derivatives.

In the following we will be  concerned with higher-derivative
corrections to the lowest-order action that necessarily arise in the 
quantum theory. Certain of these terms are known from tree-level and 
one-loop string perturbation theory.  For example, the exact expression for
the four-graviton scattering amplitude in either of 
the type II superstring theories has a simple expansion in powers of
$\alpha' = l_s^2$.    
The expression, which is reproduced in appendix A (A.1), has an overall kinematic
factor of $\tau_2^2  \tilde K$ where \cite{greenschwarzc},
\begin{equation}\label{kinematic}{\tilde K = t_8^{\mu_1\dots\mu_8}\, t_8^{\nu_1\dots\nu_8} 
\prod_{r=1}^4 \, \zeta_{\mu_r\nu_r}\, k^{(r)}_{\mu_r} \, k^{(r)}_{\nu_2},}
\end{equation} %
and  the eighth-rank tensor $t_8$ arises in open superstring amplitudes and 
is defined in \cite{greenschwarzd}.  The
factor, $\tilde K$, is eighth-order in the external momenta and is  
the linearized approximation to  $\calR^4$.  The leading term in the 
expansion of the amplitude in powers of $\alpha'$ in (A.1) summarizes the sum 
of the tree-level Feynman diagrams for on-shell four-graviton
scattering in either of the ten-dimensional low energy type II supergravity
theories. This consists of the sum of pole terms together with a four-graviton
contact interaction with two derivatives that arise from the
Einstein--Hilbert term. The next term, with coefficient $\zeta(3)$ is a contact interaction
with eight powers of momenta (contained in $\tilde K$) that arises from the
$\calR^4$ term (which was deduced in this manner in
\cite{grosswitten}
and from a four-loop sigma model term in \cite{grisaru}).  
The one-loop contribution to the four-graviton amplitude has coefficient 
$\tilde K$ with no dilaton dependence\footnote{Here and in the following we
will ignore the Gauss-Bonnet term which also arises in the one-loop amplitude and is
proportional to the product of two eight-dimensional Levi--Civita
symbols.}. It does not possess
massless particle poles  and its leading low energy contribution is
also to the $\calR^4$ term \cite{greenschwarzc}.  The effect of the
tree-level $\calR^4$ term on Calabi--Yau compactifications has
recently been reconsidered in \cite{ferrara,strominger} and its
effect on black hole solutions in \AdS5s5\ has been analyzed in \cite{gkt}
and \cite{theisen}.

After compactification to nine dimensions on a circle of radius $r_A$ in the
IIA units (or $r_B =1/r_A$ in the IIB units) the one-loop term gets
a simple $1/r_A^2$ (or $1/r_B^2$)  correction due to the winding of
the string around the compact tenth dimension.    
Putting these terms together with the appropriate coefficients leads to the 
expressions for the nine-dimensional $\calR^4$ term   in coordinates appropriate to the
type IIA, IIB or M-theory parameterizations \cite{gv},
\begin{eqnarray}
S_{R^4} & = & {1\over 3\cdot (4\pi)^7 l_S} \int d^{9} x \sqrt{-g^{A(9)}}\, 
\Rfour\, r_A\, \left[ 2 \zeta(3) (\tau^A_2)^2 + {2\pi^2 \over 3
}(1 + {1\over r_A^2}) + \cdots \right]\nn \\
& = &{1\over 3\cdot (4\pi)^7 l_S} \int d^{9} x \sqrt{- g^{B(9)}}\,
\Rfour\, r_B\, \left[2\zeta(3) (\tau^B_2)^2 + {2\pi^2\over 3 } (1 +
{1\over r_B^2}) + \cdots \right]\nn
\\
& = &{1 \over 3\cdot (4\pi)^7 l_P} \int d^9 x
\sqrt{- G^{(9)}}\, \Rfour\, R_{9} R_{11}
\left[2\zeta(3){1\over R_{11}^3}+ {2\pi^2\over 3 } +
{2\pi^2\over 3 R^2_{9}R_{11}} + \cdots \right], \nn\\
&&\label{mact} 
\end{eqnarray} 
where $ g^{A(9)}$, $ g^{B(9)}$ are the nine-dimensional metrics in
the IIA  and IIB theories.  An important feature of the first two expressions
is that they are related by T-duality.  This is trivial for the
leading terms but less so at one loop where the  winding
term of one of the type II theories 
(e.g., the $2\pi^2/3r_A^2$ term in the first expression) is
interchanged with the nonwinding term of the other (e.g., the
$2\pi^2/3$ term in the second expression).

The ellipsis in (\ref{mact}) indicate other terms that are yet to be determined. 
As we will see there are in fact no further perturbative terms (higher inverse 
powers of   $\tau_2$) but there will be
exponentially suppressed terms that are powers
of $e^{2\pi i \tau^A} = e^{2\pi i \tau^B} = e^{2\pi i \Omega}$.  Such
contributions are associated with D-instantons.  Generally, a $Dp$-brane is
a stringy soliton extended in $p\ge 0$ spatial directions while the
D-instanton corresponds to the value $p=-1$ and is associated with a 
point-like {\it space-time  event} rather like a small
Yang--Mills instanton.   The classical D-instanton is a solution of euclidean
type IIB supergravity in which  $e^{\phi^B}$ and $C^{(0)}$ have nontrivial
profiles while the other fields (including the Einstein-frame metric) are
trivial  \cite{ggp}.  The action for a D-instanton with charge $K$ is $ 2\pi|K|/g$, 
where $g$ is the asymptotic value of $\tau_2^{-1}$ so that the D-instanton 
contributions in (\ref{mact}) should be  characterized by terms proportional to 
\begin{equation}\label{dtyp}{e^{-2\pi(|K|/g \mp i K C^{(0)})},}
\end{equation} %
where the $-$ sign refers to a D-instanton and $+$ to an anti D-instanton. 
This classical configuration  will be reviewed in section 9 where we will also
see a direct correspondence between Yang--Mills instantons and D-instantons
in the context of the AdS/CFT conjecture.  
String scattering amplitudes in
a D-instanton background are described by world-sheets with boundaries on
which the target-space coordinates  satisfy Dirichlet boundary conditions
\cite{polchinst,greeninst,greengut}.  Such pointlike closed string
boundary conditions have a significant short distance effect on string 
scattering amplitudes -- a fact that led to the suggestion \cite{greenpoint}
that they may be relevant in the formulation of a string theory of hadrons.

An explicit calculation of the leading effect of a single D-instanton was
carried out in \cite{greengut} by considering  four-graviton scattering in
the presence of Dirichlet world-sheet boundaries.  This  established the
presence of the charge-one D-instanton term in   the ten-dimensional type
IIB effective action that arises in the $r_B \to \infty$ limit of
$S^{IIB}_{R^4}$.   However, in order to ensure the $\Sl2$ duality
invariance of the effective action the expressions in (\ref{mact}),  
there must be an infinite number of terms with D-instantons of arbitrarily
high charge.  In order to exhibit this  modular invariance it is instructive
to write the last line of (\ref{mact}) in terms of the complex
structure as
\begin{eqnarray}   S_{R^4} 
& =&  {1\over3\cdot (4\pi)^7 l_P}  \int d^9 x   \sqrt{- G^{(9)}} \,
\Rfour \nn\\
 & & \ \left\{  \calV ^{-1/2} \left[2\zeta(3) (\Omega_2
)^{3/2} + {2\pi^2 \over 3 } (\Omega_2 )^{ -1/2} +
\cdots \right] + {2\pi^2 \over 3 } \calV \right\}\nn  \\
& = & {1\over3\cdot (4\pi)^7 l_P}  \int d^9 x   \sqrt{- G^{(9)}} 
\Rfour \left\{ {2\pi^2 \over 3} \calV + \calV^{-1/2} 
f^{(0,0)}(\Omega,\bar\Omega) \right\}.\label{magain}
\end{eqnarray} 
We have here identified the expression in parentheses
with the expansion of an as yet undetermined  function, 
$f^{(0,0)}(\Omega,\bar\Omega)$,  that must transform as a scalar
under $\Sl2$ transformations of the complex structure, $\Omega$.  
This translates into a
function $f^{(0,0)}(\tau^B,\bar\tau^B)$ in the IIB theory and 
$f^{(0,0)}(\tau^A,\bar\tau^A)$ in the IIA theory.

The fact that this function must match the two known perturbative 
coefficients and that it must contain powers of $e^{2\pi i \tau^B}$ 
motivated the conjecture \cite{greengut} that the complete coefficient 
of the $\calR^4$ term in (\ref{rfourf})  is the modular function 
\begin{equation}\label{fzerdef}{f^{(0,0)}(\tau,\bar\tau) = \sum_{(m,n)\ne
(0,0)} {\tau_2^{3/2} \over |m+n\tau|^3 }\equiv 2 \zeta(3) \, E_{3\over 2}(\tau) ,}
\end{equation} %
where the notation $E_{3\over 2}$ denotes a nonholomorphic Eisenstein 
series (as defined, for example, in \cite{terras}).   The overall normalization 
in (\ref{fzerdef}) has been chosen so that the leading term for large $\tau_2$ is $2\zeta(3)\tau_2^{3/2}$. This  coincides 
with the tree-level expression in parentheses in the second 
line of (\ref{mact}) (after allowing for the extra factor of $\tau_2^{1/2}$ 
due to the transformation between the string frame and Einstein frame).
This expression for $f^{(0,0)}$ was further motivated by the 
study of the one-loop corrections to eleven-dimensional supergravity 
on $T^2$ \cite{ggv} (see section 5) and  will be derived from supersymmetry in section 4.

In the meantime it is worth noting that the perturbative contributions
to the  decompactification limits that give  the
ten-dimensional type IIA theories and  eleven-dimensional M-theory are
easily obtained from (\ref{mact}) and (\ref{magain}). 
In the ten-dimensional IIA limit, $r_A \to \infty$ in the first line in (\ref{mact}),  
the D-instanton terms vanish  exponentially since $\tau^A \sim   
ir_A e^{-\phi^A}\to  i \infty$, so $e^{2\pi i \tau^A} \to 0$  in this limit.   This leaves  only the perturbative 
terms.  If we now consider the further decompactification to 
eleven dimensions $e^{\phi^A} = R_{11}^{3/2} \to \infty$ the tree-level IIA contribution vanishes and 
the one-loop term gives a finite contribution.  This eleven-dimensional 
limit coincides with the limit $\calV \to \infty$ in  (\ref{magain}).   
In this limit the $\calV^{-1/2}$ term vanishes and the finite eleven-dimensional term  comes entirely from the term pr
oportional to $\calV$. 
It is important to note if there had been other terms of higher order 
in the type IIA coupling constant, $e^{\phi^A}= R_{11}^{3/2}$,  
the eleven-dimensional limit would have been singular order by order  in IIA perturbation theory  and the limit would not
 have been uniform. 
However, if $f^{(0,0)}$ is given by  the expression (\ref{fzerdef}) there are no further perturbative terms and the 
eleven-dimensional limit is simply
\begin{eqnarray}\label{mfin}
S_{R^4}  & = & {1 \over 18 \cdot (4\pi)^7 \cdot l_P^3}  \int d^{11} x  \sqrt{- G^{(11)}} \, 
\Rfour \nn\\
& = & {1\over 9\cdot 2^{12}\cdot  \pi^4\cdot  \kappa_{11}^{2/3}}
\int d^{11} x  \sqrt{- G^{(11)}} \,
\Rfour .
\end{eqnarray} %
As argued in \cite{gv} the coefficient  of this 
term is in precise agreement  with
supersymmetry which relates the $\calR^4$ term to the
eleven-form $C^{(3)} \wedge X_8$. 

\section{Linearized supersymmetry and higher derivative terms}

The existence of a large number of interactions in the IIB theory that are
of the same dimension as  the ${\cal R}^4$ interaction can be motivated
very simply by using
linearized supersymmetry.  This can be implemented by
packaging the physical fields or their field strengths into a constrained
superfield $\Phi(x^\mu - i\bar \theta\gamma^\mu \theta,\theta)$ where $\theta^a$  ($a=1, \dots,16$)
is  a complex Grassmann coordinate that transforms as a Weyl spinor of
$SO(9,1)$.  This superfield is taken to satisfy the 
holomorphic condition \cite{howest},
\begin{equation}\label{holocon}{D^* \Phi=0,} 
\end{equation} %
and the constraints,
\begin{equation}\label{onshell}{D^4  \Phi =  D^{*4} \Phi^*,}
\end{equation} %
where
\begin{equation}\label{covderiv}{
D_A = {\partial \over \partial \theta^A} +2i (\gamma^\mu  \theta^*)_A
\partial_\mu, \qquad   D^*_A = - {\partial\over \partial\theta^{*A}}}
\end{equation} %
are the holomorphic and anti-holomorphic covariant
derivatives that anticommute with
the rigid supersymmetries
\begin{equation}\label{susys}{
Q_A ={\partial \over \partial \theta^A},
\qquad   Q_A^* = - {\partial \over \partial
\theta^{*A}} +  2i
(\bar \theta \gamma^\mu  )_A \partial_\mu .}
\end{equation} %
The  constraints (\ref{holocon}) and (\ref{onshell}) ensure that the field $\Phi$ has an expansion in powers of $\theta$ (but not $
\theta^*$), that terminates after
the
$\theta^8$ term  and contains  the 256 fields in an `on-shell' supermultiplet \footnote{We are 
using the usual convention that $\gamma^{\mu_1 \dots \mu_p}$ is the 
antisymmetrized product of $p$ gamma matrices, normalized so that 
$\gamma^{\mu_1\dots \mu_p} \equiv \gamma^{
\mu_1} \dots \gamma^{\mu_p}$ when $\mu_1 \ne \dots \ne \mu_p$.},
\begin{eqnarray}
\Phi & = & \tau_0(1 + \Delta) \nn\\
& = & \tau_0(1  + a   + i \bar \theta^* \lambda +  \hat G_{\mu\nu\rho}
\bar\theta^* \gamma^{\mu\nu\rho} \theta +  
i\bar\theta^*\gamma^{\mu\nu\sigma}\theta\, 
\bar\theta^*\gamma_{\nu}\partial_{\sigma}\psi_{\mu}\nn\\  
& + &    
\calR_{\mu\sigma\nu\tau}
  \bar\theta^* \gamma^{\mu\nu\rho} \theta \bar\theta^*
\gamma^{\sigma\tau}_{\ \ \ \rho} \theta  +  \partial_\mu 
\hat F_{5\, \nu\rho\sigma\tau\omega}\, \bar\theta^* \gamma^{\mu\nu\rho} 
\theta \bar\theta^*
\gamma^{\sigma\tau\omega} \theta +
 \cdots  + \theta^8 \partial^4 \bar\tau)\nn \\
& \equiv & \sum_{r =0}^8 \Phi^{(r)} \theta^r ,\label{expphi}
\end{eqnarray} %
where $\tau_0 \Delta$ is the linearized fluctuation
around   a flat background with a constant scalar, $\tau_0 = \tau - \tau_0 a =
C^{(0)}_0 +i g^{-1}$.  
The normalizations that will not concern us here.
The  symbol  $\hat G_{\mu\nu\rho}$
($\mu,\nu,\rho = 0, \ldots, 9$) denotes the `supercovariant' combination of
$G$ and fermion bilinears defined in appendix C, where 
 $G_{\mu\nu\rho}$ and $G_{\mu\nu\rho}^*$ are complex combinations of the field
strengths
of the \RR\ and \NSNS\ two-form potentials.
The $\theta^4$ terms are  $\calR$,  the
Weyl curvature, 
and  $\hat F_{5\, \rho_1\cdots\rho_5}$, which is the field strength of the
fourth-rank \RR\ potential (where the hat again denotes the supercovariant 
combination defined in appendix C).   The fermionic field $\lambda$ 
is the dilatino and $\psi_\mu$ is the gravitino.  The gamma matrices with world indices are defined 
by
$\gamma^\mu = e^\mu_m \gamma^m$, where $m=0,\cdots,9$ is the $SO(9,1)$
tangent-space index and $e^\mu_m$ is the inverse zehnbein.
A barred Weyl spinor, such as $\bar\theta$, is defined by
\begin{equation}\label{bardef}{\bar\theta_a \equiv \theta^*_b (\gamma^0)_{ba}.}
\end{equation} %
The terms  indicated by $\cdots$ in (\ref{expphi}) 
fill in the remaining members of the ten-dimensional $N=2$ chiral
supermultiplet, comprising (in symbolic notation)   $\partial
\psi_{\mu\nu}^*$,  $ \partial ^2 G^*_{\mu\nu\sigma}$,  $\partial^3\lambda^*$
and $ \partial^4\tau^*$.

The $U(1)$ R-symmetry  charge $u_r$  of any of the   component 
fields $\Phi^{(r)}$ in the expansion (\ref{expphi}) can easily be determined 
since they are correlated with the powers of $\theta$. Assigning a
charge $-1/2$ to  $\theta$ and an overall charge $2$ to the superfield leads
to the charge for the field with $r$ powers of $\theta$,
\begin{equation}\label{uchar}{u_r = 2 - {r\over 2}}
\end{equation}
(as in \cite{schwarzwest,schwarza}).  For example,   
$u_{\partial \tau} = 2$; $u_{\lambda} = 3/2$; $u_{G}=1$; $u_{\psi} = 1/2$; $u_{R} = u_{F_5} =0$; $\dots$.

Although the linearized theory cannot capture the full structure of the terms in
the effective action it can be used to relate various terms in
the limit of weak coupling, ${\rm Im}\ \tau_0 = g^{-1}\to \infty$ (where $g=e^{\phi_0}$ is the string 
coupling constant).  The linearized approximations to the complete 
interactions are those that
arise by integrating a function of $\Phi$ over the sixteen components of $\theta$,
\begin{equation}\label{actdef}{
S^{(3)}_{linear} = \int d^{10}x  d^{16} \theta \,   F[\Phi] +\ {\rm c.c.},}
\end{equation} %
which is manifestly invariant under the rigid supersymmetry
transformations, (\ref{susys}).
The various  component interactions contained in (\ref{actdef}) are obtained
\ from the $\theta^{16}$ term in the expansion,
\begin{equation}\label{expands}{
F[\Phi] = F(\tau_0) + \Delta {\partial \over \partial \tau_0}
F(\tau_0) + \half \Delta^2 \left( {\partial \over \partial \tau_0}\right)^2  F(\tau_0)   +
\cdots.}
\end{equation} %
Using the expression for  $\Delta$ in (\ref{expphi}) and substituting into (\ref{actdef})
leads to all the possible interactions at this order in $\alpha'$ \cite{ggk,kehab},
\begin{eqnarray}\label{nonpert} 
S^{(3)} & = & 
\int d^{10}x\, \det\, e\, \left(
f^{(12,-12)} \lambda^{16} + f^{(11,-11)} \hat G \lambda^{14} + \ldots
\right .\nn \\
& &  \left.
+ f^{(8,-8)} \hat G^8 +\ldots + f^{(0,0)} \calR^4 + \ldots +
f^{(-12,12)} \lambda^{* \, 16}\right),
\end{eqnarray} %
where  $\det\, e = \det\, e_\mu^m$ is the determinant
of the zehnbein. 

The $\calR^4$ interaction  comes from the $\Delta^4$ term
while the $\lambda^{16}$  comes from the $\Delta^{16}$ term. 
We see, in particular, that the tensor structure of the contractions 
between the four Riemann tensors in the $\calR^4$ term is summarized by the Grassmann integration
\cite{nilsson},
\begin{equation}\label{rfourdef}{R^4=\int d^{16}\theta (R_{\theta^4})^4,}
\end{equation} %
where
\begin{equation}\label{rthetdef}{R_{\theta^4} \equiv \bar \theta
\gamma^{\mu\nu\sigma} \theta \bar \theta \gamma^{\rho\tau}{}_{\sigma} \theta \, R_{\mu\nu\rho\tau}.}
\end{equation} %
It follows from a standard Fierz transformation that all possible
contractions of the Riemann tensor appearing in $R_{\theta^4}$ vanish,
so that only the trace-free part --- the  Weyl tensor --- survives.

The superscripts that label the coefficients $f^{(w,-w)}$ are related
to the violation of the $U(1)$ charge.  Thus, the linearized form of 
the general term in (\ref{nonpert}) contains a product of $p$ fields, 
\begin{equation}\label{genterm}{
\int d^{10}x\, \det\, e\, f^{(w,-w)} \prod_{k=1}^p \Phi^{(r_k)},}
\end{equation} %
which violates the $U(1)$ charge by
\begin{equation}\label{uviol}{2w =  \sum_{k=1}^p u_{r_k} = 2p - 8}
\end{equation} %
units, where we have used (\ref{uchar}) and the fact that the total power of
$\theta$ must be $\sum_k r_k = 16$.  For example the $\calR^4$ term ($w=0$) 
conserves the $U(1)$ charge while the 
$\lam16$ term ($w=12$) violates the $U(1)$ charge by $24$  and
there are many other terms that violate the charge by any even number.

In the linearized approximation,  $g\to 0$ ($\tau_2 \to \infty$),  
the coefficients $f^{(w,-w)}$ are constants that
are related to each other by use of the Taylor expansion, (\ref{expands}).
{}For example, the $\calR^4$ term has coefficient $\partial_{\tau_2}^4F$
while the $\lambda^{16}$ term has coefficient $\partial_{\tau_2}^{16} F$ so
that, at the linearized level,
\begin{equation}\label{linrels}{f^{(12,-12)} \sim \left( \tau_2 {\partial \over
\partial \tau_2}\right)^{12} f^{(0,0)},}
\end{equation} %
where for the moment we are not concerned about the overall constant.
In writing this we have used the fact that
the linearized approximation  is valid only if  the inhomogeneous term
in the modular covariant derivative, $\calD$ (defined in appendix B), is negligible which requires that 
\begin{equation}\label{validl}{2\tau_2 \partial_{\tau_0} f^{(w, - w)}   >> w  f^{(w, - w)}}
\end{equation} %
since only in this case does the modular covariant derivative reduce to the ordinary derivative.   
This inequality is obviously not satisfied by terms in the expansion of $f^{(w,- w)}$ that are
powers of $\tau_2$, such as the perturbative tree and one-loop terms in (\ref{mact})
(with $\tau_2 \to \tau_0$). However, when acting on a factor  such as 
$\tau_0^ne^{ - 2\pi |K|\tau_0}$ (where $n$ is any constant) 
which is  characteristic of a  charge-$K$ D-instanton, the
inhomogeneous term may be neglected in the limit  $\tau_0\to \infty$ and the  covariant
derivative linearizes. Therefore, a linearized superspace
expression such as  (\ref{actdef}) should  contain the exact  leading
multi-instanton contributions to the    $\Rfour$ and related   terms.
These  leading  instanton terms arise by  substituting  the expression
\begin{equation}\label{leadinst}{F_K =  c_K e^{2\pi i |K| \Phi}}
\end{equation} %
into  (\ref{actdef}). 

In the exact theory that will be considered in detail in the next section,
the $\Sl2$ symmetry of the IIB theory requires that the   $f^{(w,-w)}(\tau,\bar\tau)$ are modular forms with
holomorphic and anti-holomorphic weights as indicated in the
superscripts.  In that case the  derivative $\tau_0\partial_{\tau_0}$ in (\ref{linrels}) will be
replaced by the  covariant derivative, $D_w$,  that maps a modular form 
of weight $(w, \hat w)$  into a modular form of weight
$(w+1,\hat w-1)$\footnote{Appendix B summarizes some relevant properties
of modular covariant derivatives and laplacians.},
\begin{equation}\label{covarderiv}{D_w F^{(w,\hat w)} = i\left(\tau_2 {\partial
\over \partial \tau} - i {w\over 2} \right) F^{(w+1,\hat w-1)} = F^{(w+2,\hat w)}.} 
\end{equation} %
The covariant version of (\ref{linrels}) is
\begin{eqnarray}\label{ftwel}f^{(12,-12)}  (\tau,\bar\tau) & = & D ^{12} f^{(0,0)}(\tau,\bar\tau)\nn\\
& \equiv & D_{11} D_{10} \dots D_0 f^{(0,0)}(\tau,\bar\tau).
\end{eqnarray} %
We will take this relation to define the relative normalizations of $f^{(0,0)}$ and $f^{(12,-12)}$.

\section{Nonlinear supersymmetry constraints on terms of order 
${\alpha'}^{-1}$}

In this section, which is based on \cite{greensethi} where further details may be 
found,  the constraints imposed by the full nonlinear supersymmetry of the 
type IIB theory will be used to derive the detailed expressions for all the terms in $S^{(3)}$ --- 
the  terms in the derivative expansion of the action of the type IIB theory that are of order ${\alpha'}^{-1}$.

The procedure will be to impose supersymmetry order by order in $\alpha'$ by expressing the  supersymmetry transformations on an arbitrary field $\Psi$  as the series,
\begin{equation}\label{susyexp}{\delta_\epsilon \Psi
=\left( \delta^{(0)} + \alpha' \delta^{(1)}
+\ldots + {\alpha'}^n \delta^{(n)} + \ldots \right)\Psi,}
\end{equation} %
while the effective action has the expansion (\ref{effacs}).
In principle, the  action can be constructed by imposing  the conditions,
\begin{equation}\label{noethn}{ \left( \sum_{m=0}^r {\alpha'}^m \delta^{(m)} \right) \sum_{n=o}^r
{\alpha'}^n S^{(n)}  = 0 ,}
\end{equation}
order by order in $\alpha'$.   
In addition, it is important to impose closure of the supersymmetry algebra 
on all the fields.  In theories of this type the supersymmetry algebra only 
closes on a field $\Phi$ up to the equations of motion and also includes specific local symmetry transf
ormations, $\delta_{\rm local} \Phi$. The generic structure of the superalgebra is therefore of the form,
\begin{equation}\label{supalgv}{(\delta_1 \delta_2 -
\delta_2\delta_1) \Phi =   \xi^\mu  D_\mu\Phi + {\Phi\ \rm equations\ of \ motion}
+ \delta_{\rm local} \Phi,}
\end{equation} %
where the first term is the usual translation term with parameter
\begin{equation}\label{xidef}{\xi^\mu = -2\, \Im\, \bar\epsilon_2 \gamma^\mu \epsilon_1.}
\end{equation} %

Imposing the supersymmetry 
conditions (\ref{noethn}) together with closure of the superalgebra
(\ref{supalgv})  simultaneously 
determines both the action and the supersymmetry transformations on the fields.
The ${\alpha'}^0$ ($n=m=0$) term in (\ref{noethn}) corresponds to the supersymmetry 
of the classical theory.  These terms in the action and in the supersymmetry 
transformations were determined in \cite{schwarza}, ignoring terms
that are quartic in fermion fields\footnote{As 
stated in the introduction, it is really the equations of motion 
rather than the action that will be considered in the following.}.  Some of 
these zeroth order supersymmetry transformations and action are reviewed in appendix C. One particular 
term quartic in the fermion fields is needed in the following 
(the term $L_1^{(0)}$ below) with a coefficient that  is derived in \cite{greensethi}.

There are no $n=1$ or $n=2$ terms at tree-level
or one-loop, and there  is strong evidence that supersymmetry precludes 
the presence of any such terms in  (\ref{effacs}). We will {\it assume} this to be the case so that
the first corrections are the terms in $S^{(3)}$, which are of order
$(\alpha')^{3}$ relative to $S^{(0)}$. These terms are eighth order in
derivatives.

The   procedure of imposing supersymmetry and closure  can become very complicated 
unless one makes a judicious choice of starting point.  We will start by 
selecting two special terms in $L^{(3)}$ that involve sixteen fermionic 
fields\footnote{Our notation is chosen so that $(\alpha')^{n-4}\int d^{10} x L^{(n)} = 
S^{(n)}$.}.  The choice is motivated by the fact that these two terms  
mix with each other and with no other terms  under the classical supersymmetry 
transformations.  The two terms are the sixteen-fermion terms contained in 
\begin{eqnarray}\label{starti} L^{(3)}_1 & = & {\det\, e} \,  \left(
f^{(12,-12)}(\tau,\bar\tau) \lambda^{16} + f^{(11,-11)}(\tau,\bar\tau)
\hat G \lambda^{14}\right)\nn\\ 
& = & {\det\,e}\, \left( f^{(12,-12)} \lambda^{16} - 3\cdot 144
f^{(11,-11)} \, \left(\lambda^{15} \gamma^\mu \psi_\mu^*\right) + \ldots
\right),
\end{eqnarray} %
where the ellipsis represents other terms in $\hat G$ which do not
affect the subsequent argument\footnote{The overall normalization of the 
action does not affect the arguments of this section.}
(details of the spinor algebra that leads to the precise coefficients 
in this and following equations can be found in appendix C and \cite{greensethi}).

{}From  the lowest order supersymmetry transformations given in appendix C 
we obtain terms that transform  (\ref{starti})  into  $\det \, e\, \lambda^{16}\,
\psi_\mu^*\,\epsilon$,
\begin{equation}\label{varone}{\delta_1^{(0)} L^{(3)}_1 = -i \, {\det\,e}
\, \left(\bar \epsilon^* \gamma^\mu \psi_\mu^*\right)\, \lambda^{16}\,
\left( 8 f^{(12,-12)} + 6 \cdot  144\, D_{11} f^{(11,-11)}\right),}
\end{equation} %
where we have only kept terms proportional to  $\lambda^{16}\,
\psi_\mu^*\,\epsilon$. It is important to check whether there could
also be a contribution of the same form as (\ref{varone}) arising from a
$(\alpha')^3\delta^{(3)}$
variation of the fields  in the lowest order action $S^{(0)}$.
However, it is easy to see by inspection
that no term with $\lambda^{16} \psi^*_\mu$  can arise from the
variation of any term in $S^{(0)}$. This means that we must require $\delta_1^{(0)} L^{(3)}_1 =0$,
which implies that
\begin{equation}\label{constwo}{D_{11} f^{(11,-11)} =  -{4\over 3 \cdot  144}
f^{(12,-12)}.}
\end{equation} %
This condition is consistent with the modular weights assigned to the
functions  $f^{(w,-w)}$.

We  now consider the term in the variation of (\ref{starti}) that is proportional to  $ \det\, e\, \lambda^{16} \,
\lambda^*\,\epsilon^*$,
\begin{equation}\label{vartwo}{\delta_2^{(0)} L^{(3)}_1    = - 2i\, {\det\, e}
\, \lambda^{16}  (\bar\epsilon \lambda^*)
\left(\bar D_{-12}   f^{(12,-12)} +
3\cdot 144\cdot {15\over 2} f^{(11,-11)}   \right) + \ldots,
}
\end{equation} %
where we have made explicit only the terms containing $\lambda^{16} \lambda^*\epsilon^*$.
In this case, there is another contribution of the same form as 
$\delta_2^{(0)} L^{(3)}_1$ that arises from the $(\alpha')^3 \delta^{(3)}$
variation of  terms in the lowest order IIB  Lagrangian $L^{(0)}$.   Even
though the complete set of interactions in the classical theory is not
tabulated explicitly in the literature (it is implicit in the superspace formulation \cite{howest}),
it is easy to convince oneself that
the only possible term that can vary into $\delta_2^{(0)} L^{(3)}_1$
is a term of the form,
\begin{equation}\label{nonterm}{L^{(0)}_1 = {1\over 256} \, \det\, e\,
\bar \lambda \gamma^{\mu\nu\rho} \lambda^*\, \bar \lambda^*\gamma_{\mu\nu\rho} \lambda ,}
\end{equation} %
which is the unique tensor structure
containing $\lambda^2 \lambda^{*2}$.  The coefficient of this term  was determined by the lowest order
 supersymmetry transformations in \cite{greensethi}  by considering the 
mixing of $L_1^{(0)}$ with other terms in the classical action.

We see that $L_1^{(0)}$ can vary into the same form as
$\delta_2^{(0)} L_1^{(3)}$ if we assume a higher-order variation of $\lambda^*$ of the form,
\begin{equation}\label{newvar}{\delta^{(3)} \lambda^*_a =- {1\over 6}\,i
 g(\tau,\bar\tau)\,  (\lambda^{14})_{cd}
(\gamma^{\mu\nu\rho} \gamma^0)_{dc} \, (\gamma_{\mu\nu\rho} \epsilon^*)_a,}
\end{equation} %
where $g(\tau,\bar\tau)$ is another  function to be determined.  Substituting in (\ref{nonterm})
gives a contribution,
\begin{eqnarray}\label{threelone}
\delta^{(3)} L^{(0)}_1 & = & -{1\over 768}\, i\, {\det\, e}\,
g(\tau,\bar\tau)\, \bar\lambda
\gamma^{\mu\nu\rho}\gamma_{\rho_1\rho_2\rho_3} \epsilon^*\,
(\lambda^{14})_{cd} (\gamma^{\rho_1\rho_2\rho_3} \gamma^0)_{dc}\, \bar
\lambda^* \gamma_{\mu\nu\rho} \lambda\nn\\
& = & - 30\, i\,  {\det\, e} \, g(\tau,\bar\tau) \lambda^{16}
(\bar\epsilon \lambda^*).
\end{eqnarray} %
Comparing with (\ref{vartwo}) we see that in
order for the total contribution to $\delta L_1$ to vanish at order
$(\alpha')^3$, there must be a linear relation between the function
$g$ and the functions $f^{(11,-11)}$ and $\bar D_{-12} f^{(12,-12)}$,

\begin{equation}\label{linrel}{\bar D_{-12} f^{(12,-12)} \, + 3\cdot 144 \cdot {15\over 2} \,
f^{(11,-11)} + 15   \,g =0.}
\end{equation} %

At this stage we have two equations ((\ref{constwo}) and (\ref{linrel})) relating the three unknown functions, 
$f^{(12,-12)}, f^{(11,-11)}$ and $g$.  
A further constraint on these functions is obtained by requiring the closure of the superalgebra.
We need to  consider closure of the supersymmetry transformations
on the field $\lambda^*$.  First, in the classical theory (and keeping only the terms linear in 
$\lambda$ derivatives) it was found in  eq. (4.5) of \cite{schwarza} that
\begin{eqnarray}(\delta^{(0)}_1 \delta^{(0)}_2 -
\delta^{(0)}_2\delta^{(0)}_1) \lambda^*  & = &  \xi^\mu  D_\mu
\lambda^* -{3\over 8} i [\bar\epsilon_2 \gamma^\rho \epsilon_1 -
(1\leftrightarrow 2)] \gamma_\rho \gamma^\mu D_\mu \lambda^*\nn\\
& - & {1\over 96}i [\bar \epsilon_2 \gamma^{\rho_1\rho_2 \rho_3}
\epsilon_1 - (1\leftrightarrow 2)] \gamma_{\rho_1\rho_2\rho_3}
\gamma^\mu D_\mu \lambda^*\nn  \\
& & \dots .\label{susyclos}
\end{eqnarray} %
This has the form of (\ref{supalgv}). The first term on the right-hand-side is of the form expected for the
commutator of two supersymmetry transformations.  The other terms that 
have been exhibted explicitely are proportional to
the linear terms in the $\lambda^*$ equation of motion.
Many  other terms that are not needed  are indicated by $\dots$.  
These terms  contribute to the   commutator to complete the low-energy
$\lambda^*$ field equation  as well as generating local
transformations of $\lambda^*$ \cite{schwarza}.

The higher order terms in  $L^{(3)}$ modify the equations
of motion and this should also be apparent by considering the closure of
the algebra.  Therefore, we now consider
terms that enter at order ${\alpha'}^3$ from the commutator of a $\delta^{(0)}$
with a $\delta^{(3)}$.  More precisely, we shall consider terms in the
commutator involving only $\epsilon_2^*$ and $\epsilon_1$,
\begin{eqnarray} \left( \delta^{(0)}_{\epsilon_1}
\delta^{(3)}_{\epsilon^*_2}  \right.&  -&    \left. \delta^{(3)}_{\epsilon_2^*}
\delta^{(0)}_{\epsilon_1}\right) \lambda^*_a\nn\\
&=& - {1\over 3}\left( \tau_2 {\partial  \over \partial \tau}
- i {45 \over 8}  \right)i\, g \, (\bar \epsilon_1^*
\lambda)( \lambda^{14})_{cd}(\gamma^{\mu\nu\rho}\gamma^0)_{dc}
(\gamma_{\mu\nu\rho}  \epsilon_2^*)_a\nn \\
& = &\, 32\,  D_{11}g \, \lambda^{15}_b\,  \left[{3\over 8}
\bar\epsilon_2 \gamma^\mu \epsilon_1
(\gamma_\mu)_{ba} + {1\over 96} \bar \epsilon_2
\gamma^{\mu\nu\rho} \epsilon_1 (\gamma_{\mu\nu\rho})_{ba}\right] \, +
 \, \delta_{\hat\epsilon} \lambda^* .\nn\\
&&\label{termone}
\end{eqnarray} %
In the last line, we have
separated a local supersymmetry transformation, $ \delta_{\hat\epsilon} \lambda^*$, 
which is to be identified with a supersymmetry transformation of the form (\ref{newvar}) with a particular
field dependent coefficient, $\hat \epsilon = {i\over 4} \epsilon_2^* \, (\bar\epsilon_1^* \lambda)$.
 
Combining (\ref{susyclos}) and (\ref{termone}) 
(including the powers of $\alpha'$) we
see that in order for the right-hand side of the commutator to vanish
the $\lambda^*$ field equation must be of the form,
\begin{equation}\label{propr}{ i \gamma^\mu D_\mu \lambda^* - (\alpha')^3\,  32\, D_{11}g\,
 \lambda^{15} + \dots =0,}
\end{equation} %
where the ellipsis indicates terms with different structure that we
have not considered. This equation has to be identified with the appropriate sum of 
terms in the $\lambda^*$ equation of motion that  is obtained by varying the action 
with respect to $\lambda$. At the same order in $\alpha'$ this is given by,    
\begin{equation}\label{lammot}{ i\gamma^\mu D_\mu \lambda^*  -  (\alpha')^3  
f^{(12,-12)}\, \lambda^{15} + \dots=0,}
\end{equation} %
where we have  only made explicit the term that is proportional to
$\lambda^{15}$.
Comparing (\ref{propr}) and (\ref{lammot}) gives the relation,
\begin{equation}\label{neweqm}{32 D_{11} g =  f^{(12,-12)}.}
\end{equation} %
Substituting (\ref{neweqm}) into (\ref{constwo}) gives,
\begin{equation}\label{grelf}{ g=  - {3\cdot 144 \over 128}   f^{(11,-11)}.}
\end{equation} %
There is no ambiguity in this relation between $g$ and $f^{(11,-11)}$ because 
there is no solution to  $D_{11} g =0$.
Substituting (\ref{grelf})  into (\ref{linrel})   gives,
\begin{equation}\label{twelveg}{\bar D_{-12} f^{(12,-12)} = 3\cdot 144 \left(-{15\over 2} +
{45\over 64}  \right) f^{(11,-11)}.}
\end{equation} %
The two simultaneous  first-order differential equations, (\ref{twelveg}) and (\ref{constwo})
are simply reduced to the independent second-order eigenvalues equations,
\begin{equation}\label{laplacee}{ \nabla^2_{(-)\, 12}f^{(12,-12)} =
4 D_{11} \bar D_{-12}  f^{(12,-12)}=\left(-  132 + {3\over 4}\right) f^{(12,-12)}}
\end{equation} %
and
\begin{equation}\label{laplaceeb}{ \nabla_{(+)\, 11}^2 f^{(11,-11)} \equiv  4 \bar
D_{-12} D_{11} f^{(11,-11)} =\left(-  132 + {3\over 4}\right) f^{(11,-11)}, }
\end{equation} %
where the laplacians are defined in appendix B.   

By applying $\bar D^{12}$ to (\ref{laplacee}) we can define the   function,
\begin{equation}\label{fzert}{\tilde f^{(0,0)}(\tau,\bar\tau) \equiv \bar D^{12} f^{(12,-12)}(\tau,\bar\tau).}
\end{equation} %
Applying $D^{12}$ to this equation and using (\ref{laplacee}) together with  properties of 
covariant derivatives in appendix B leads to 
\begin{equation}\label{ftwelag}{f^{(12,-12)}  (\tau,\bar\tau) = k D ^{12}\tilde  f^{(0,0)},}
\end{equation} %
where $k$ is an irrelevant constant.  Therefore,  from (\ref{ftwel}), we can make the identification
\begin{equation}\label{fidenb}{\tilde f^{(0,0)} = k^{-1} f^{(0,0)}.}
\end{equation} %
Putting these relations together we see that the Laplace eigenvalue equation for $f^{(12,-12)}$ 
(\ref{laplacee}) implies
\begin{equation}\label{lapzero}{\nabla_0^2  f^{(0,0)}(\tau,\bar\tau)\equiv  4\tau_2^2
{\partial \over \partial \tau}{\partial \over \partial \tau^*}  f^{(0,0)}(\tau,\bar\tau) = 
{3\over 4} f^{(0,0)}(\tau,\bar\tau) .} 
\end{equation}%
It is an important fact that we will not prove here (see, for example,
exercise 2 in section 3.5 of \cite{terras}) that  (\ref{lapzero})  has the  unique solution 
(\ref{fzerdef}) provided   it  is assumed to transform as a scalar under
$\Sl2$ and has polynomial growth as $\tau_2 \to \infty$ (which is
the perturbative regime). 

Substituting (\ref{lapzero}) in (\ref{ftwelag}) and (\ref{fidenb}) determines $f^{(12,-12)}$ to be
\begin{equation}\label{ftwelvedeff}{f^{(12,-12)}(\tau,\bar\tau) = {1\over 2^{12}} 
{\Gamma\left({27\over 2}\right) \over \Gamma\left({3\over 2}\right)}
\sum_{(m,n)\ne (0,0)}  \tau_2^{3/2}{(m+n\bar \tau)^{24}\over |m+n\tau|^{27}}.}
\end{equation} %
Similarly, (\ref{laplaceeb}) gives a unique expression for the modular form $f^{(11,-11)}$. 
The expressions for all the coefficients, $f^{(w,-w)}$,  in (\ref{nonpert}) should also emerge from a 
more detailed application of the Noether procedure that  
considers all the possible mixing of terms in $S^{(3)}$
with arbitrary $U(1)$ charges. These are given by applying $w$ covariant derivatives to  
$f^{(0,0)}$ which results in the expression
\begin{eqnarray}\label{eisendef}
f^{(w,-w)}(\tau,\bar\tau) & = &   {1\over 2^{w}} 
{\Gamma\left(w+ {3\over 2}\right) \over \Gamma\left({3\over 2}\right)}
\sum_{(m,n) \ne (0,0)}  {\tau_2^{3/2} \over |m+n\tau|^3 }
\left( {m + n \bar \tau\over m+ n \tau}\right)^w \nn \\
& = &   {1\over 2^{w}} 
{\Gamma\left(w+ {3\over 2}\right) \over \Gamma\left({3\over 2}\right)}
\sum_{(m,n) \ne (0,0)}  \tau_2^{3/2}  {(m + n \bar \tau)^{2w}
\over |m+ n\tau|^{2w+3}}. 
\end{eqnarray} %
This transforms under $\Sl2$ transformations by a phase
\begin{equation}\label{tranpp}{f^{(w,-w)}(\tau,\bar\tau) \to \left({c\tau + d \over c\bar\tau + d}\right)^w  
f^{(w,-w)}(\tau,\bar\tau) ,}
\end{equation} %
as expected for a $(w,-w)$ modular form.

The Noether procedure rapidly escalates in complication when applied at 
higher orders in the derivative expansion and it is probably not a
practical procedure for terms with many more derivatives.  It may just 
about be feasible to deduce terms of order $\alpha'$ that are of order 
${\alpha'}^5$ beyond the Einstein--Hilbert term.  For example, this may be a method of  
determining the higher derivative  terms suggested in \cite{berkova} which were motivated 
by exactly known perturbative contributions \cite{berkovafd,oogvaf}.  

The dimensions of the terms of yet higher order in $\alpha'$  suggested in \cite{russoa,berkova}  
are so high that it is not obvious why they should be protected by supersymmetry.  
Nevertheless, they probably are protected based on the evidence for the
exact form of these terms  presented in these references.

There is also a suggestion \cite{kiritsisb,piolobers} 
for how the structure of the protected $o({\alpha'}^{-1})$ terms generalizes to compactifications 
to lower dimensions, where there are more moduli.  The suggestion is that the functions $f^{(w,-w)}$ 
are generalized Eisenstein series for the appropriate du
ality groups.  

\section{The  eleven-dimensional perspective}

We shall now review the  manner in which the $o({\alpha'}^{-1})$ terms in the
type II string actions can be deduced from eleven-dimensional supergravity.
This was the subject of \cite{ggv} where it was pointed out that
the complete $\calR^4$  term arises  from the  one-loop quantum
contribution to four-graviton scattering in  eleven-dimensional
supergravity compactified on $T^2$. Similarly, the $\lambda^{16}$ interaction can be deduced from the
one-loop scattering amplitude of sixteen gravitini in eleven-dimensional
supergravity \cite{ggk}.  The type IIB dilatino, $\lambda$, is
identified with the gravitino of eleven-dimensional supergravity with its  polarizations  
in appropriate directions.  
These calculations and generalizations were reviewed in 
\cite{greenmult}  where some further details may be found, 
including a discussion of higher derivative contributions  
(which were also the subject of \cite{russoa}).

\subsection{The $\calR^4$ term from one loop in eleven dimensions}

We will begin with the $\calR^4$ term which will be determined from the scattering amplitude
for four   gravitons at one loop in eleven-dimensional supergravity \cite{ggv}.   
A fully covariant calculation would involve the sum of loop
diagrams with the various component fields (the graviton,
$g_{\mh\nh}$, the third-rank antisymmetric potential,
$C^{(3)}_{\mh\nh\rh}$, and the gravitino, $\Psi_\mh^a$)
and their ghosts  circulating around the loop. 
An efficient method for summing all these contributions
is to use a manifestly supersymmetric formalism based on the
eleven-dimensional point particle \cite{ggv} 
(details of the superparticle 
formalism outlined below have yet to appear \cite{ggktwo}).

This method is modeled on the method used to construct string 
one-loop amplitudes in the light-cone gauge which
is the parameterization in which
\begin{equation}\label{xplus}{X^+ (t)  = p^+ t + x^+,} 
\end{equation} %
where $t$ labels the world-line of the superparticle,
$X^\pm(t) = (X^0(t) \pm X^8(t))/\sqrt 2$ and $p^+$ is a constant
(the directions $x^9, x^{11}$ will continue to be the $T^2$
directions).  The physical states are described in terms of the
transverse $SO(9)$ representations (instead of the $SO(8)$ that
is relevant for the transverse space in the case of the superstring).
The quantum states are constructed in terms of fermionic sixteen-component
spinors, ${\cal S}^A$ ($A=1, \dots, 16$), satisfying the anticommutation 
relations,
\begin{equation}\label{calssa}{\{S^A, S^B\} = \delta^{AB}.}
\end{equation} %
The 32 components of the supercharges decompose the components that 
are realized linearly in the light-cone gauge,
\begin{equation}\label{linq}{Q^A = \sqrt p^+ \calS^A,}
\end{equation} %
and those that are nonlinearly realized,
\begin{equation}\label{nonlinq}{\tilde Q^A = {1\over \sqrt p^+} (\Gamma \cdot \dot X \, \calS)^A,}
\end{equation} %
where $\Gamma^I_{AB}$ are the $16 \times 16$ $SO(9)$ gamma matrices and
$X^I(t)$ are the transverse coordinates of the superparticle.

The vertex operators that describe the linearized interactions 
can be derived by imposing the conditions
that the $g$,  $\psi$ and $C^{(3)}$ vertices transform
into each other under supersymmetry in the appropriate manner \cite{ggktwo} which
mimics the way in which light-cone superstring vertices can be derived. 
The resulting light-cone  graviton vertex operator is given by
\begin{equation}\label{gravert}{V_g^{(r)} =  \zeta^{(r)}_{IJ}  (\dot X^I \dot X^J -2 \dot X^I
S\gamma^{JL} S k^L +2  S\gamma^{IL} S k^L S\gamma^{JM} S k^M) e^{i
k\cdot X},}
\end{equation} %
where, as usual in the light-cone gauge operator formalism, 
a special frame has been chosen in which $k^+ =0$ and
$\zeta^{(r)}_{IJ}$ is the transverse polarization tensor of the
eleven-dimensional graviton ($I = 1, \cdots, 7, 9,11$). 
This expression reduces to the graviton
vertex in the   type IIA superparticle theory when reduced on a circle to
ten dimensions \cite{ggktwo}.  In addition to the cubic graviton  vertex 
higher-order contact interactions are 
necessary in order to reproduce general gauge-invariant amplitudes.  
Such contact terms compensate for short distance singularities in
products of $V_g^{(r)}$'s that involve factors such as
$\langle \dot X(\tau)  \dot X(\tau') \rangle$ but are not needed for the  amplitudes we are interested in.

The loop amplitude compactified on a two-torus is given by
\begin{equation}\label{loopstruc}{A_{R^4} ={1\over \calV} \int \prod_{r=1}^4  dt_r \int d^9p 
\sum_{m,n} \tr_{_{S}}( V_g^{(1)}(k^1) V_g^{(2)}(k_2) V_g^{(3)}(k_3) V_g^{(4)} (k_4)),}
\end{equation} %
where $\tr_{_{S}}$ denotes the trace over the fermionic variables.
The integers $m,n$ are the Kaluza--Klein charges in the $T^2$ directions $(x^9, x^{11})$.
For convenience we will only consider the situation in which all the external polarizations and momenta are 
oriented in the directions $x^1,\dots,x^7$ which are transverse to the light-cone as well as the $T^2$ directions. 
The fermionic trace is saturated by the term with sixteen $\calS$ factors, which picks out the last term 
in (\ref{gravert}) (the term with four $\calS$'s) so that no contractions of   $\dot X$'s in the prefactor of the vertices.  
This picks out an overall kinematic factor which can be shown to be the same as the overall factor in the ten-dimensional  
one-loop superstring amplitude defined in (\ref{kinematic}) and can be expressed in a  compact manner as an 
integral over a sixteen-component Grassmann spinor, as described in section 3.

After extracting the overall kinematic factor the remaining bosonic factor
is simply the loop amplitude for the four-point amplitude in scalar $\phi^3$
field theory.   The full momentum dependence of this amplitude was discussed
in  \cite{russoa,greenmult}.  Expanding in a power series in the momenta
leads to higher derivatives acting on the fourth power of the curvature. 
However, here we are only interested in the leading momentum dependence. 
Therefore, after extracting the kinematic factor the external momenta can
be set equal to zero in the remainder of the expression.  The result can
be written as
\begin{equation}\label{afourf}{A_{R^4} ={1\over \calV} \tilde K \int_0^\infty dt t^3 \int d^9p
\sum_{l_1,l_2} \exp\left(- t(p^2 + G^{IJ} l_Il_J)\right), }
\end{equation} %
where $G^{IJ}$ is the inverse metric on $T^2$ and $l_I = (m,n)$ 
($I,J =1,2$).  Substituting for the inverse metric, 
\begin{equation}\label{metinv}{G^{IJ} l_Il_J = {1\over \calV}{|m+n\Omega|^2\over \Omega_2},}
\end{equation} 
and performing a double Poisson resummation converts the sum over Kaluza--Klein 
charges ($m,n$) to a sum over the windings ($\hat m,\hat n$)
 of the world-line of the particle circulating in the loop around the
two independent cycles of $T^2$.   After the 
straightforward gaussian integration the result is
\begin{eqnarray}\label{afourwin}\calV A_{R^4}  & = & \tilde K \calV \int_0^\infty d\hat t \hat t^\half 
\sum_{\hat m,\hat n} e^{-\calV \hat t {|\hat m + \hat n \Omega|^2\over\Omega_2}}\nn\\
& = & \tilde K \calV C + \tilde K \calV^{-\half} \sum_{(\hat m,\hat n) \ne(0,0)} 
{\Omega_2^{3/2} \over |{\hat m} + {\hat n} \Omega|^3}\\
& = & \tilde K\left\{\calV C+   \calV ^{-1/2} \left[2\zeta(3) (\Omega_2
)^{3/2} + {2\pi^2 \over 3 } (\Omega_2 )^{ -1/2} +
\cdots \right]\right\},\nn
\end{eqnarray} %
where $\hat t = t^{-1}$.   The cubic ultraviolet divergence  
in (\ref{afourwin}) is contained  in the  zero winding term, $\hat m=\hat n=0$, which 
has the divergent coefficient $C$.   The second equality in (\ref{afourwin})
gives the first two
terms in the expansion for large $\Omega_2$ (this expansion will
be discussed in detail in section 6).  The amplitude (\ref{afourwin}) comes from 
a term in the effective action of the form (\ref{magain}) with $f^{(0,0)}$ again of the form (\ref{fzerdef}).
As remarked earlier, in the limit $\calV\to 0$ the regularized $\calV C$ term vanishes and the  amplitude has a
finite limit in type IIB coordinates, i.e., $r_B\to \infty$ with 
$e^{\phi^B}$ held constant. Substituting $\Omega$ by $\tau = C^{(0)}
+ i e^{-\phi}$ leads to precisely  the same 
expression for the $\calR^4$ term in the type IIB theory that we deduced earlier from supersymmetry.

It is easy to argue on dimensional grounds that the 
finite part of the  coefficient of the $\calR^4$ term can only come from   one loop  in eleven
dimensions.  Higher loops lead to finite  contributions that have extra
derivatives acting on $R^4$ \cite{greenmult}.  However, the systematics of the
divergent parts is more obscure and the coefficient $C$ has
contributions from divergences of diagrams with 
arbitrary numbers of loops. These divergences must be 
regularized by  microscopic effects in the eleven-dimensional theory
that lie outside the realm of perturbation theory.
Although we do not know how to fix the regularized value of $C$ 
directly it is
precisely determined by symmetry considerations.  Keeping $\calV$
nonzero and comparing the
leading terms of the large-$\Omega_2$ expansion in 
(\ref{afourwin}) with (\ref{magain}) shows that these expressions only agree if
$C$  has the value 
\begin{equation}\label{cval}{C= {2\pi^2\over 3}.}
\end{equation} %
We saw that the coefficients in (\ref{magain}) are consistent with T-duality
that relates the IIA and IIB string perturbation expansions in nine 
dimensions. 
This means that only for the value of $C$ in (\ref{cval})  is 
(\ref{afourwin}) consistent with T-duality.  This leads to the same value of  $C$ as 
the argument based on supersymmetry that was mentioned at the end of section 2.
 
\subsection{One loop with sixteen dilatini}

In a similar manner all the interactions that are related by linearized supersymmetry 
to the $\calR^4$ term can be deduced by considering one-loop processes in eleven-dimensional 
supergravity on $T^2$ \cite{ggk}.  As we saw in section 3 the generic terms
 of this type violate the $U(1)$  R-symmetry charge of the type IIB theory  that is conserved 
classically.  An extreme example of this is the interaction between sixteen  dilatini.  
Since  the  dilatino $\lambda$ carries a $U(1)$ charge of $3/2$  the inte
raction $\lambda^{16}$ violates 24 units of charge.

As in the case of four-graviton scattering the calculation of the general 
loop amplitude that describes the scattering of  sixteen external dilatini is
facilitated by using the light-cone superspace description of  the eleven-dimensional
super-particle. However, the $\lambda^{16}$ term  of
interest can be extracted from the  zero-momentum process and 
this can be calculated without using the general gravitino vertex operators.
The  zero-momentum loop amplitude has the
structure (ignoring an overall constant)
\begin{equation}\label{sixteenl}{A_{\lam16}   ={1\over \calV} \sum_{m,n}  \int d^9 p  \int
{d\tau\over \tau}
\tau^{16} \tr_S( V_\lambda^{(1)}
(0)\cdots V_\lambda^{(16)} (0))   e^{ - \tau (p^2+{1 \over \calV\Omega_2}
|m+n\Omega|^2)},}
\end{equation} %
where  $V_\lambda^{(r)}(0)$ is the zero-momentum vertex for the
$\lambda$
component of the $r$th gravitino. The trace in the above
expression is over the fermionic operators in the vertices.

The fermions of the IIB theory compactified on $S^1$, the complex fields $\psi$ and $\lambda$,  
are related to specific projections of the eleven-dimensional gravitino of M theory on $T^2$.  
These relations are described in appendix D (based on  \cite{ggk}).
It follows that  the 
vertex $V_{\lambda}$ is a  specific projection of the vertex operator for the eleven-dimensional gravitino, $V_\Psi$.
The gravitino vertex can again be deduced from eleven-dimensional supersymmetry in the  light-cone gauge \cite{ggk}.  
The zero-momentum vertex is very simply expressed in
terms of the 32-component
eleven-dimensional supercharge, $Q$, by the covariant expression
\begin{equation}\label{gravivertss}{V_\Psi = \bar\zeta^*_\mh  Q  p^\mh,}
\end{equation} %
where $\zeta_\mh$ is the zero-momentum wave function, $p_\mh$ is
the momentum of the particle circulating around the
loop and $\bar\zeta^*_\mh \equiv \zeta_\mh \Gamma^0$ (with no complex
conjugation). Since the zero momentum vertex  is independent of the
proper time around the loop  the volume of integration over
the proper times of the vertices gives the factor of $\tau^{16}$ in (\ref{sixteenl}).

We now want to consider the components of $\Psi_\mh$ containing
$\lambda$.  Inverting the relations in equation (D.12) in appendix D gives,
\begin{equation}\label{psilam}{
\Psi_z = \calP_{\bar z} \Gamma_z \chi + \calP_z \Gamma_{\bar z}\lambda, \qquad
\Psi_{\bar z} = \calP_z \Gamma_{\bar z}  \chi^* +  \calP_{\bar z}
\Gamma_z\lambda^*.}
\end{equation} %
\ from which it follows that
\begin{equation}\label{newlam}{
V_\lambda=-\bar \lambda^* \Gamma_{\bar z}  \calP_{\bar z}  Q p_{\bar
z} = -\bar\lambda^*\Gamma_{\bar z} q p_{\bar z} ,}
\end{equation} %
where $q = \calP_{\bar z}  Q$ is a projected supercharge that
satisfies the anticommutation relations
\begin{equation}\label{commsq}{
\{q^A, q^B\} = \calP^{AC}_{\bar z} \calP^{BD}_{\bar z} \{Q^C, Q^D\} =
\Gamma_z^{AB} p_{\bar z}.}
\end{equation} %
The momentum dependence can be scaled out  by changing to   $\hat
q^A = q_A (p_{\bar z})^{-\half} $, which satisfies $\{\hat q^A, \hat q^B\} =
\Gamma^{AB}_z$.  In the chirally projected subspace  $\Gamma^{AB}_z
\propto \delta^{AB}$ and the commutation
relation does not depend on $\bar z$.

Substituting  the vertex  (\ref{newlam})  into the loop amplitude
(\ref{sixteenl}) (contact terms are again  not
needed in this  spin-$\half$ process)
and integrating over   the loop momentum gives an expression of the form
\begin{equation}\label{resf}{
\calV A_{\lam16}  = \hat K \calV^{-\half} f^{(12,-12)} (\Omega, \bar \Omega),}
\end{equation} %
where the kinematic prefactor,
\begin{equation}\label{kinems}{
\hat K = \bar\lambda^{(1)*}_{A_1} \cdots \bar
\lambda^{(16)*}_{A_{16}}  \tr (
\hat q^{A_1} \cdots \hat q^{A_{16}}),}
\end{equation} %
is manifestly antisymmetric under permutations of the (commuting)
fermion wave functions  due to
the anticyclic property of the trace.  It  can be rewritten (up to
an overall constant) as
\begin{equation}\label{kinfac}{
\hat K = \bar\lambda^{*}_{A_1} \cdots  \bar
\lambda^{*}_{A_{16}}  \varepsilon_{16}^{A_1 \dots A_{16}},}
\end{equation} %
where $\varepsilon_{16}$ is the rank sixteen epsilon  tensor with spinor indices.

The rest of the expression (\ref{resf})  depends on the
circumference $r_B$ through the factor $\calV^{-\half}$.  The
$\Omega$ and
$\bar \Omega$ dependence  comes from the  loop integration and is
given by
\begin{eqnarray}\label{resultintegrals}
f^{(-12,12)} (\Omega, \bar \Omega)  & = & c \sum_{m,n} \int {dt\over t} t^{23/2}
\left({1\over \sqrt{\Omega_2}} (m+n\bar \Omega)\right)^{24}
\exp\left(-t {1\over {\Omega_2}}|m+n\Omega|^2\right)\nn\\ 
& = & {c\over \sqrt{\Omega_2}}\Gamma (23/ 2)\sum_{m,n}
{ (m+n\bar \Omega)^{24} \over | m+n \Omega|^{23}} 
\end{eqnarray} %
(where $c$ is an overall constant).
This  expression   is superficially cubically
divergent due to the sum over the Kaluza-Klein charges.    However,
this does not take into account the phase dependence which, for
generic $\Omega$, can lead to a cancellation between the growing terms
in the sum. Indeed,
for asymptotically large $m$ and $n$ the sum over discrete
momenta $p_z$ and $p_{\bar z}$ can be replaced by integrals and the
result is proportional to
\begin{equation}\label{contin}{
\int dp_z dp_{\bar z} p_{\bar z}^{24} (p_z p_{\bar z})^{-23/2},}
\end{equation} %
which vanishes if $|p_z|$ is regularized. Equivalently, it vanishes if the phase
integration is carried out before the integration
over $|p_z|$. This is the regularization prescription that will be
used in the following, motivated by the fact that it leads to an $\Sl2$ 
invariant answer.

In order to extract the finite result it is useful, as before,  to
use a double
Poisson summation to transform from the discrete momentum sum to a
sum over
windings of the loop  around the torus.  Writing
\begin{equation}\label{doubpois}{f^{(12,-12)} (\Omega,\bar\Omega) ={c\over \sqrt{\Omega_2}}
\left({\partial \over \partial\alpha}\right)^{24}
\sum_{m,n} \int {dt \over t} t^{23/2}\left.
e^{-t|m+ n\Omega|^2 + \alpha (m+ n\bar \Omega)}\right|_{\alpha =0},}
\end{equation} %
the Poisson resummation equates this to the manifestly finite expression,
\begin{eqnarray}\label{finalfin}
f^{(-12,12)} (\Omega,\bar\Omega)  & = &  {c\over \sqrt{ \Omega_2}}
\left({\partial \over \partial \alpha}\right)^{24}
\!\!\sum_{\hat m,\hat n}{\pi \over
\Omega_2}\nonumber\\
&&\qquad \int \!\!\!{dt \over t} t^{21/2}\left.\!\!\exp\left(\!\! -{\pi^2 \over
t \Omega_2^2} |\hat n \!\!+\!\! \hat m \Omega|^2 \!\!+\!\! {i\pi \alpha \over t
\Omega_2}(\hat n  + \hat m\bar \Omega)\!\!
\right)\right|_{\alpha=0}\nn\\ 
& = &  {c \over \pi^2 } \Gamma (27/ 2) \Omega_2^{3/2} 
\sum_{(\hat m,\hat n)\ne (0,0)} {(\hat n + \hat m\bar \Omega)^{24} \over
|\hat n + \hat m \Omega|^{27}}. 
\end{eqnarray} %
where $\hat m$ and $\hat n$ are winding numbers.
 
After translating from the M-theory coordinates ($\Omega$, $\calV$) to the string-frame  IIB
coordinates  ($\tau$, $r_B$) the expression (\ref{finalfin}) is identical to
(\ref{ftwelvedeff}) up to an overall normalization that can be absorbed in $c$.  The  amplitude
$A_{\lam16}$ corresponds to the  $\lambda^{16}$ term in the IIB effective action,  (\ref{nonpert}),  in the limit, $r_B\to
\infty$, of
ten decompactified dimensions.  In fact, in this case there is no divergent term at all even at finite $r_B$.  
The eleven-dimensional limit of $A_{\lam16}$ is not only finite but it vanishes.
The nine-dimensional effective action in the IIB parameterization   contains the term
\begin{equation}\label{beffamp}{S^{(3)}_{\lam16} = {1\over 3\cdot (4\pi)^7 l_S} \int d^9 x\,r_B\, \det\, 
e\,e^{-\phi/2}\, \lambda^{16} f^{(12,-12)} (\tau,\bar\tau).}
\end{equation} %
This expression has been written in the  string frame. The transformation  from    
Einstein frame to string frame includes the rescaling  $\lambda \to \lambda e^{-\phi/8}$ 
in addition to the rescaling of the zehnbein, $ e_\mu^a \to  e_\mu^a e^{\phi/4}$.

\section{Fourier series  and properties of D-instantons} 

We have seen that the terms in the ten-dimensional type IIB effective action 
of order ${\alpha'}^{-1}$ have the form (in string frame) \cite{ggIII}, 
\begin{equation}\label{termss}{{1\over \alpha'} \int d^{10}x \det e\,  e^{-\phi/2}f^{(w,-w)}(\tau,
\bar\tau)\calO_{\{u_r\}},}
\end{equation} %
where    $\calO_{\{u_r\}}$ is a function of the fields that reduces, at the linearized 
level to a monomial of $p$ fields with $U(1)$
R-symmetry charges $u_1,u_2, \dots u_p$ (see (\ref{genterm})).  The total $U(1)$ charge of
the interaction is $2w$ where $w = p-4$ (see (\ref{uviol}))
and the  coefficient $f^{(w,-w)}$ is  given by (\ref{eisendef}).
The coefficient functions  can be expressed as Fourier series,
\begin{equation}\label{fourone}{f^{(w,-w)}= 
\sum_{K=-\infty }^\infty \calF_{K,w}\, e^{2\pi i K
\tau_1},}
\end{equation} %
where the Fourier coefficients, $\calF_{K,w}$, with nonzero $K$ are
the D-instanton terms when $K>0$  and   anti D-instanton terms   when $K<0$.   
This expansion is readily derived  by a generalization of the
method that leads to the analogous expansion of the Eisenstein series,
$E_s$ (see, for example, \cite{terras}).  First one performs the sum over
all the terms in (\ref{eisendef}) with $n=0$ and  $m\ne 0$.  This leads to the `tree-level'
term of the form  $2\zeta(3) e^{-2\phi}$. For the  terms with $n\ne 0$
one may perform a
Poisson resummation  that converts the sum over $m$ into a sum over a
conjugate integer, $\hat m$.  The  sum over all $n\ne 0$ with  $\hat
m=0$ gives the `one-loop' term that is independent of $\tau_2$ while
the remaining sum (over all $n\ne 0$, $\hat m\ne 0$) gives the
D-instanton contributions  with $K = \hat m \, n$.

The result is \cite{ggIII,kehaa}
\begin{equation}\label{expadef}
{f^{(w,-w)} = 
2\zeta(3)\, \tau_2^{{3\over 2}} + {2\pi^2 \over 3}\, \tau_2^{-{1\over
2}} c_w +  
\sum_{K =1}^\infty \left( \calF_{K,w} e^{2\pi i K\tau_1} + 
\calF_{K,-w} e^{-2\pi i K \tau_1} \right),}
\end{equation} 
where 
\begin{equation}\label{cpdef}{c_w = (-1)^w {\pi \over 4} {1 \over
\Gamma({3\over 2} + w) 
\Gamma({3\over 2}  -w)}.}
\end{equation} %
The first two terms in (\ref{expadef}) have the interpretation of the  tree-level and 
one-loop string terms while the instanton and anti-instanton terms are contained in
\begin{equation}\label{znpddef}{\calF_{K,w} 
 = 4\pi^{1/2} \,  (2\pi |K|)^{1/2}\, Z_K\,
\sum_{k=0}^\infty{c_{w, k}\over (2\pi K\tau_2)^{k-w}}
e^{2\pi i K\tau_2},}
\end{equation} %
where
\begin{equation}\label{ckrdef}{c_{w,k} = {(-1)^w \over 2^{k-w} k!} {\Gamma(3/2) \over 
\Gamma(-w + 3/2)} {\Gamma (k - w-1/2)\over \Gamma(- k+ w -1/2)}}
\end{equation} %
and
\begin{equation}\label{mudef}{Z_K  = \sum_{\hat m|K} {1 \over \hat m^2}.}
\end{equation} %
The infinite series  in (\ref{znpddef}) 
begins with the power $\tau_2^w$ for a  D-instanton  $(K>0$)   
while the series of corrections to the anti D-instanton ($K<0$) starts with the power $\tau_2^{-w}$.

A particularly simple example of this formula is the expansion of the 
coefficient $f^{(0,0)}$  of the $\calR^4$ term which  is given by
\begin{eqnarray}\label{newsum}  f^{(0,0)} (\tau, \bar \tau) & = &
2\zeta(3)\, \tau_2^{{3\over 2}} +{2\pi^2\over3} \, \tau_2^{-{1\over
2}} +  8\pi \tau_2^{{1\over 2}} \sum_{K\ne 0} |K|\, Z_K \,
K_{-1}(2\pi |K|\tau_2) e^{2\pi i K\tau_1}
\nn\\
& = & 2\zeta(3)\, \tau_2^{{3\over 2}}+ {2\pi^2\over 3} \, \tau_2^{-{1\over
2}}  + 4 \pi \sum_{K=1}^\infty |K|^{1/2} Z_K 
\\
& \times & \left(e^{2\pi i K
\tau} + e^{-2\pi i K \bar \tau} \right) \left(1 + \sum_{k=1}^\infty
(4\pi K\tau_2)^{-k} {\Gamma( k -1/2)\over \Gamma(- k -1/2) k!} \right),\nn
\end{eqnarray} %
where the asymptotic expansion of the Bessel function
$K_{-1}(2\pi|K|\tau_2)$ (which is useful for small coupling, or large
$\tau_2$) has been used in the second equality.

The expressions (\ref{expadef}) and (\ref{newsum}) reproduce precisely the perturbative tree-level and one-loop
terms  and  demonstrate that there is  a  perturbative
non-renormalization theorem   beyond one loop. The absence of higher-order perturbative
corrections must be related to the fact that the terms in (\ref{termss}) are given by 
integrals over
half the on-shell superspace although a direct proof of the absence of higher-order 
perturbative corrections has not been given (see however \cite{berkovb}).   
The non-perturbative terms  in (\ref{newsum}) are contributions of
D-instantons (and anti-D-instantons) of arbitrary charge $K$ together with an infinite
sequence of perturbative fluctuations around each instanton configuration.

The D-instanton sum has a simple origin from the point
of view of the loop integrals considered in section 5.  There, the
circulating particle can be considered to be a Kaluza--Klein mode  of
charge $n$ of  M theory compactified on the first circle of circumference $R_{11}$.
{}For $n\ne 0$ this is a charge-$n$ D-particle.  The
world-line of this D-particle winds $\hat m$ times around the second
circle of radius $r_A$ (in IIA string units).  
The action for such a configuration is given
by the Dirac--Born--Infeld action for the D-particle and is equal to 
\begin{equation}\label{acta}{2\pi  |\hat m n|(r_Ae^{\phi^A} \mp i C^{(1)})}
\end{equation} %
(where the $\mp$ distinguishes instantons and anti-instantons). 
After performing a T-duality on the euclidean circle 
\cite{greengut} the real part of this expression is identified with the
charge-$K$ D-instanton  action, 
\begin{equation}\label{actb}{S_{K=\hat m n} = 2\pi |K| e^{\phi^B} }
\end{equation} %
and the \RR\ potential $C^{(1)}$ is identified with $C^{(0)}$. 

{}From (\ref{znpddef}) we see that  
the leading contribution to the $N$ D-instanton contribution in  
$\calF_{K, w}(\tau,\bar \tau)$ is  
\begin{equation}\label{coeform}{\tau_2^{{1\over 2}} \,\calF_{K,w} \sim   Z_K S_K^{-7/2 + p} 
e^{ - (S_K - 2\pi iK C^{(0)})} \left(1+ o(Ke^{\phi})\right),}
\end{equation} %
where $p=w+4$ and we have factored out the measure factor $Z_K$ (\ref{mudef}), 
which depends on $K$ but not on $S_K$ and is normalized so that $Z_1 =1$. 
The  power  $-7/2$ in (\ref{coeform}) arises from the combination of ten
bosonic zero modes and
sixteen  fermionic zero modes while the factor of $S_K$ for each external state comes from the 
normalization of the external states.  

\section{D-particle bound states and the Witten index}

One of the central planks of the web of dualities that relate string theory and M theory is the 
identification of D-particles (D0-branes) of type IIA 
superstring theory with Kaluza--Klein modes of eleven-dimensional supergravity compactified on 
$S^1$ \cite{witten}.  There  is a single Kaluza--Klein mode of a given charge, $N$, with mass 
$N$ times the mass of charge-one D-particle.  This implies that there 
must be precisely one threshold bound state (a bound state with zero binding energy) of 
$N$ D-particles.  The interaction of $N$ D-particles is described by the `one-dimensional 
matrix model' ---  a quantum mechanical system that is isomorphic to the
dimensional reduction of ten-dimensional
$U(N)$  supersymmetric Yang-Mills   theory to one (time) dimension \cite{wittena}.   
It is notoriously difficult to analyze the possible
 threshold bound states of this theory.  
One approach used in \cite{yi,sethi} in the case $N=2$ is to demonstrate that the Witten 
index is equal to one.  This shows that there is at least one bound state.  
Here we will show   how  the results 
of the last section   indicate that the Witten index is equal to one for arbitrary $N$ (this section is based on \cite{ggone})
\footnote{See \cite{porrati} for a very different approach to the threshold 
bound state problem}.

\subsection{D-particles and D-instantons} 

The Witten index for the system of $N$ D-particles
was defined in \cite{yi,sethi} by
\begin{eqnarray}I^{(N)} & = & \int d^9x \lim_{\beta\to \infty} \tr (-1)^F
e^{-\beta H}(x,x)\\
& = & \lim_{R\to \infty} \lim_{\beta\to 0} \left\{
\int_{|x|<R}d^9x\;\tr(-1)^F e^{-\beta H}\right.  \nn\\
&&\quad  + \left. {1\over2} \int_{|x|=R}d^9x \int_\beta^\infty d\beta^\prime\; 
\tr e_n (-1)^F Q e^{-\beta^\prime H}\right\},\label{indexa}
\end{eqnarray} %
where the first term defines a bulk contribution $I^{(N)}_{bulk}$ (which
in general is not an integer) and the second term a deficit contribution $I^{(N)}_{def}$  
(which corrects the bulk term).   The separation  of the integer  $I^{(N)}$
into these two parts is dependent on the order of limits.  We shall always consider the
infinite volume limit to be taken before the limit $\beta \to 0$.

The trace in (\ref{indexa}) is defined over gauge invariant states of the
Hilbert
space, which may be written as a trace over all  states  if a
projector
onto gauge
invariant states (states satisfying the constraints of Gauss' law, $C|\, \rangle =0$) is inserted.  
For the bulk term this gives
\begin{eqnarray}\label{fulltrace}{ I^{(N)}_{bulk}={1\over Vol(SU(N)/Z_N)} \lim_{\beta\to 0}\int_{SU(N)} d\eta \int
d^9x \; \tr(-1)^F e^{i \eta C}e^{-\beta H}(x,x),}
\end{eqnarray} %
where $\eta^a$ is the gauge parameter. 
Using the heat kernel approximation for the propagator in the
hamiltonian  $H$, which is valid in
the limit $\beta\to 0$, it was observed in \cite{yi} and \cite{sethi} that the exponent
of the bulk part
of the index can be written in a $SO(10)$ invariant form. This is
possible because the
parameter $\eta$ turns into a tenth bosonic coordinate, $x^9$. 

The  resulting integrals are seen to be equivalent to the $SU(N)$ integral  in
the partition function for the {\it zero}-dimensional matrix model.
This is the model that is obtained by compactification 
of supersymmetric Yang--Mills theory from ten (euclidean)
dimensions to zero dimensions 
that describes dynamics in the presence of $N$ D-instantons 
\cite{wittena,ikkt,periwal}.  The situation with D-instantons
is somewhat distinct from that of other Dp-branes  since $p=-1$
and  the world-volume reduces to a space-time point. 
The configuration space of $N$ D-instantons is determined by bosonic and
fermionic $U(N)$ matrices $A_\mu$ and $\psi_\alpha$, where $\mu$
and $\alpha$ denote  $SO(10)$
vector and spinor indices, respectively. The
reduction of the supersymmetric Yang--Mills action to a point is given by
\begin{equation}\label{DinstSYM}{S=  {1\over 4}
\tr([A_\mu,A_\nu]^2)+{i\over2}\tr(\bar{\psi} \Gamma^\mu [A_\mu,\psi]).}
\end{equation} %

The `center of mass' degrees of freedom  are
associated with
the element of the Cartan subalgebra  of $U(N)$ proportional to the
unit matrix  and do not appear
in the action (\ref{DinstSYM}).  They do, however,  enter as shift  
symmetries in the
supersymmetry transformations,
\begin{eqnarray}\label{susytrafo} & & \delta A_\mu = i \bar{\eta}\Gamma_\mu \psi,\nn \\
& & \delta \psi = [A_\mu,A_\nu]\Gamma^{\mu\nu}\eta + {\bf 1}
\,\epsilon. 
\end{eqnarray} %
The sixteen-dimensional spinor $\eta$ parameterizes the dimensionally
reduced
supersymmetry of the $SU(N)$ YM theory and the spinor $\epsilon$
acts as a constant
shift on the fermions in the `center of mass' degrees of freedom.
Hence the U(1)
part of $U(N)$ fermionic fields $\psi$ plays the role of the
sixteen  fermionic collective coordinates associated with the charge-$N$ D-instanton.

After factoring out the centre of mass coordinates the $N$ D-instanton
partition function is given by
\begin{equation}\label{integSYM}{Z = \int d^{10}y \int
d^{16}\epsilon\,  Z_N    ,}
\end{equation} %
where
\begin{equation}\label{zdef}{Z_N = {1\over Vol(SU(N)/Z_N)} \int_{SU(N)} d\psi dA \exp(-S_{SYM}[A,\psi]).}
\end{equation} %
The integration over $y^\mu$ and $\epsilon^A$  is the integral
over the overall bosonic and fermionic collective coordinates of  the
collection of D-instantons  of total  charge $N$.   The integration
measure in (\ref{zdef}) is the group invariant Haar measure which defines  the
appropriate normalization (as discussed in \cite{nicolaia,kosvan}). 
A non-zero value for
$Z_N$ can only arise from configurations in which there are
no extra fermionic
zero modes (in addition to  $\epsilon$) which is characteristic of
a single
multiply-charged D-instanton.  
According to \cite{wittena} the partition function of the SU(N)
zero-dimensional matrix model should be  
identified with  
the D-instanton  measure, $Z_N$  (\ref{mudef}). 
The integral (\ref{zdef}) was evaluated for the   
case  $N=2$ in \cite{yi,sethi}.  The sub-integral over  configurations in which  
the elements of the Cartan subalgebra, $A_\mu^3, \psi^3$, are fixed  was  
evaluated in \cite{ggII}
and  corresponds to  two D-instantons at a fixed separation.
More recently, the integral has been explicitly  evaluated for general $N$
by deforming the integrand into a topological density 
\cite{moorenek} and numerical estimates have been made in \cite{nicolaia}.

The fact that the  integral (\ref{zdef}) is the same  as the
bulk integral which appears in the calculation of the Witten index
in the bound  state problem of D-particles 
is the statement of T-duality.  The radius $\beta$ of the euclidean circle
in (\ref{fulltrace}) shrinks to zero
size and  the insertion of $(-1)^F$ enforces supersymmetric
(periodic) boundary
conditions on the fermionic fields.
The conjectured form for the measure, $Z_N$ given by (\ref{mudef})   
therefore suggests that
the value of the bulk integral is given, for arbitrary $N$, by
\begin{equation}\label{result}{I^{(N)}_{bulk}= \sum_{\hat{m}|N}{1\over {\hat{m}}^2} = Z_N,}
\end{equation} %
where an overall constant has been fixed by noting that  
$I_{bulk}^{(1)} = 1$.
Note that for $N=2$ the value of  $I^{(2)}_{bulk}=5/4$ in (\ref{result})  
agrees with the explicit calculation of \cite{yi,sethi}. 

\subsection{The deficit term}

The appearance  of  a deficit
term is related to the presence of the  boundary terms  in (\ref{indexa}). For the $SU(2)$ case a heuristic
evaluation  of this 
term   was given in \cite{yi}  and a more rigorous argument in   \cite{sethi}.  
It was argued that since  the  boundary term   for the system of two  
D-particles arises from the region in which the particles are separated it  is obtained by 
treating the particles as identical {\it free} particles with
moduli
space $R^9/Z_2$. Since the Witten index vanishes for free particles the
deficit
term, $I^{(2)}_{def}$, must cancel the  bulk  term for the free system,
$I^{(2)}_{0\, bulk}$, where the subscript $0$ indicates the free theory.   But $I_{0\, def}^{(2)}$ is easily  
evaluated and is
equal to $1/4$.  In the following \cite{ggone} we
shall assume that this prescription generalizes to  $N>2$   
D-particles so that,
for {\it prime} values of $N$,
\begin{equation}\label{ndeff}{ I_{def}^{(N)} = -  I_{0\, bulk}^{(N)},}
\end{equation} %
where $I_{0\, bulk}^{(N)}$ is the bulk index for $N$ identical free  
particles
moving on $R^{9(N-1)}/S_N$.   For {\it non-prime} values, $N=\hat m n$,  the
generalization will take into account the regions of moduli space  
of $\hat m$
free charge-$n$ particles on $R^{9(\hat m-1)}/S_{\hat m}$.

In the $N=2$ case considered in \cite{yi,sethi} the only configuration that
contributed to the trace over the free two-particle states   was  
one with an odd permutation of the two particles. If the trace is expressed as a  
functional integral this includes only  configurations in which the two  
D-particles are described by a single  euclidean world-line that winds twice around  
the compact $\beta$ direction. In order to generalize this to arbitrary $N$  
we need to consider the action of $S_{N}$, which is the
Weyl group of $SU(N)$.  This can be parameterized
by matrices $M_{ab}$  acting  on the positions
of the $N$ particles  modded out by the overall translation
invariance, $X^i_a$ ($i=1, \cdots, 9$),  and the
fermions $\psi^\alpha_a$. The index $a=1,..,N-1$ labels the different
$U(1)$'s in the
Cartan subalgebra.
The action of  an element in $S_{N}$ is given by
\begin{equation}\label{weyl}{X_a^i\to M_a^b X_b^i, \qquad \psi_a^\alpha\to  M_a^b
\psi_b^\alpha,}
\end{equation} %
where the vector index runs over $i=1,\cdots, 9$  and the spinor
index $\alpha=1,\cdots 16$.

The fermion fields $\psi$ satisfy the following anticommutation
relations,
\begin{equation}\label{fermanti}{ \{ \psi_a^\alpha,\psi_b^\beta\}=
\delta_{ab}\delta^{\alpha\beta}.}
\end{equation} %
It is convenient to build up the fermionic Hilbert space by
defining fermionic
creation and anihilation operators
$\psih^\alpha_a, {\psih}_a^{\dagger\alpha}$
by
\begin{equation}\label{creat}{\psih^\alpha_a={1\over \sqrt
2}\left(\psi_a^{2\alpha-1}+i\psi_a^{2\alpha}\right),\quad
{\psih}_a^{\dagger\alpha} ={1\over \sqrt
2}\left(\psi_a^{2\alpha-1}-i\psi_a^{2\alpha}\right),\quad
\alpha=1,\cdots ,8,}
\end{equation} %
which satisfy $\{{\psih}_a^{\dagger\alpha},\psih^\beta_b\}=
\delta_{ab}\delta^{\alpha\beta}$.
A general wave function  $\mid \Psi\rangle$ can be expanded as
\begin{equation}\label{wavef}{ \mid \Psi\rangle = \left(\Psi^{(0)}(X_a)
+\Psi^{(1)}_{a_1}(X_a)\psih^\dagger_{a_1}+{1 \over2}
\Psi^{(2)}_{a_1a_2}(X_a)\psih^\dagger_{a_1}\psih^\dagger_{a_2}+\cdots
\right)\mid 0\rangle,}
\end{equation} %
where the vector and spinor indices are suppressed and the highest
term has
$8\times N$ fermionic creation operators acting on the Fock space vacuum.

The  expression for the bulk contribution to the Witten index for   
$N$ free D-particles (with $N$ a prime)  is given
by
\begin{equation}\label{deficitterm}{ I^{(N)}_{0\, bulk} =  \lim_{\beta\to 0} \tr(-1)^F  
e^{-\beta H}.}
\end{equation} %
The  trace  is taken over gauge invariant states in the
Hilbert space
and hence one has
to insert a
projector on states which are invariant under the Weyl  permutation  
group,
\begin{equation}\label{project}{{\cal P}={1\over N!}\sum_{\pi\in S_{N}} M_\pi.}
\end{equation} %
The matrix  $M_\pi$ representing $\pi \in S_N$  act as in (\ref{weyl}) on the
coordinates $X_a$ and the
fermions $\psih_a$ in the wave function $\Psi$ given in (\ref{wavef}).
The bulk index for the free theory  is therefore given by
\begin{equation}\label{traceb}{ I^{(N)}_{0\, bulk} =  \lim_{\beta\to 0}<\Psi| (-1)^F e^{-\beta H}
{\cal P}|\Psi>.}
\end{equation} %
The factor of  $(-1)^F$ counts bosons (even number of $\psih$) with
$+1$ and fermions (odd number of $\psih$) with $-1$. Because the fermions
transform under the symmetric group matrices $M$, the introduction of the
projector ${\cal P}$ can give a non-vanishing contribution.

The action of $M_\pi$ on   $\psih_a$ and   $X_a$ in (\ref{traceb})
factorizes into a trace over the Hilbert space built from the fermionic
creation
operators and a bosonic gaussian integral coming from the heat kernel
approximation for the free propagator in the limit $\beta\to 0$ \cite{yi}.
\begin{equation}\label{oneterm}{I^{(N)}_{0\, bulk}= \lim_{\beta\to 0}{1\over  
N!}\sum_{\pi\in S_N}\tr_\psi\left( (-1)^F M_\pi\right) \int   
\prod_{i=1}^9\prod_{a=1}^{N-1} dX^i_a
\; {e^{-(X-M_\pi X)^2/2\beta} \over (2\pi\beta)^{(N-1)9/2}}.}
\end{equation} %

For all values of $\alpha$  the fermionic trace is  given by
\begin{eqnarray}\label{fermtracenew}\tr_\psi \left((-1)^F M_\pi\right) & = &  
\langle 0\mid 0\rangle -
\sum_a \langle 0\mid \psih_a M_{\pi\,ab}\psih^\dagger_b\mid 0\rangle\nn \\
& + & {1\over 2}\sum_{a,b}\langle 0\mid  \psih_a   \psih_b
M_{\pi\, ac}M_{\pi\, bd}\psih^\dagger_c\psih^\dagger_d\mid  
0\rangle+\cdots\nn\\
& = & 1- \tr(M_\pi)+{1\over 2} \tr(M_\pi^2)-{1\over 2}  
\tr(M_\pi)^2+\cdots \\
& = & \det (1-M_\pi),\nn
\end{eqnarray} %
so that the trace over all eight fermion components gives a factor of
$\det(1-M_\pi)^8$.     The gaussian integration over the  
coordinates $X^i_a$
similarly gives a factor of $\det(1- M_\pi)^{-9}$

The determinant is easily evaluated by using  an explicit  
representation for
the matrices $M_\pi$.  Recall that  these represent the action of  
$S_N$ on the
elements of the Cartan subalgebra.   The roots of $SU(N)$ are the  
vectors in
$R^N$,
\begin{equation}\label{roots}{e_i-e_j,\quad\quad i\neq j;\quad i,j=1,\cdots,N,}
\end{equation} %
where $e_i$ the $i$th unit vector of $R^N$.
All the roots lie in a
($N-1$)-dimensional subspace $\cal{R}_\omega$  orthogonal to the
vector $\omega=e_1+e_2+\cdots+e_N=(1,1,\cdots,1)$ and an element
$\pi\in S_N$  acts as a permutation, $\pi: e_i\to e_{\pi(i)}$.

Two clases of permutations need to be distinguished:\hfill\break
\noindent (a){\it  Cyclic permutations}\hfill\break\noindent
These can be
represented   up to conjugation by the  $N\times N$ matrix
\begin{equation}\label{cycmat}{M_{c} = \pmatrix{0  & 1& 0 & 0 & \cdots & 0 \cr
                    0  & 0  &1 & 0 & \cdots & 0 \cr
                    0  & 0  &0 & 1 & \cdots & 0 \cr
                    \cdots & \cdots & \cdots & \cdots & \cdots &
\cdots \cr          1   & 0  &0  & 0 & \cdots & 0 \cr},}
\end{equation} %
which has eigenvalues
\begin{equation}\label{eigencyc}{\lambda_k=e^{2\pi i k\over N},\quad k=0,1,\cdots,N-1.}
\end{equation} %
The matrix  $M_{c}$ does not leave any of the roots  (\ref{roots})   
invariant and $\omega$ is
the unique eigenvector with eigenvalue  $\lambda_0=1$.   We are  
interested in evaluating $\det(1-M_\pi) = {\det}'(1-M_c)$,  where the prime indicates the
omission of the zero eigenvalue.  This is the determinant in the space
orthogonal to $\omega$ and is given by the product of non-zero  
eigenvalues,
\begin{equation}\label{detcyc}{{\det}'(1-M_c)=\prod_{k=1}^{N-1}(1-\lambda_k)
=2^{N-1}\prod_{k=1}^{N-1}\sin{\pi k\over N}= N. }
\end{equation} %

\noindent (b) {\it  Non-cyclic permutations}\hfill\break\noindent
These  can be represented by
\begin{equation}\label{ncycmat}{M_{nc}=\pmatrix{M_1 & 0 & 0 & \cdots& 0\cr
                      0      & M_2 & 0 & \cdots& 0 \cr
      \cdots&  \cdots & \cdots & \cdots& \cdots  \cr
0      & 0 & 0 & \cdots &M_p
 \cr}}
\end{equation} %
where $M_1$, $M_2$, $\cdots$, $M_p$,   represent cyclic
permutations of subsets
of elements and can be written in the form of (\ref{cycmat}).
The matrix $M_{nc}$ has $p>1$  unit eigenvalues  so  there are  $p-1$
eigenvectors $w_j \in R_\omega$ for which
$(1-M_{nc})w_j= 0$ ($j=1,\cdots,p-1$).  As a result,  some elements  
of the
Cartan subalgebra are left
invariant by $M_\pi$  and
\begin{equation}\label{nceq}{\det(1-M_\pi)= {\det}'(1-M_{nc}) = 0,}
\end{equation} %
where the prime again indicates the omission of the zero eigenvalue  
associated
with $\omega$.

To be more precise, a  zero eigenvalue of the bosonic determinant
$\det(1-M_{nc})$   should be interpreted as the inverse volume,   
$R^{-1}$  in
the limit $R\to \infty$, whereas the fermionic determinant vanishes
identically. This means that only  the cyclic permutations contribute to the  
index and the expression (\ref{oneterm}) reduces to
\begin{equation}\label{freeres}{I_{0\, bulk}^{(N)} = \sum_{c\in S_N} {1\over
N!}({\det}'(1-M_c))^{-1} .}
\end{equation} %
The non-vanishing determinant  ${\det}'(1-M_c)=N$ simply counts the winding
number
of a
fundamental  D-particle world-line  and the result  is  the obvious
generalization of the $N=2$ case.
The non-cyclic permutations do not contribute to the index but  have the
interpretation of $p$ disconnected  D-particle
world-lines winding around the euclidean circle.  This is a multi  
D-instanton
configuration and the vanishing of the determinant
is due  to the appearance of extra fermionic zero modes.
Since each of
the $(N-1)!$ cyclic elements of $S_N$ gives the same contribution  to
(\ref{freeres}) the result is
\begin{equation}\label{defresult}{I^{(N)}_{0\, bulk}= {1\over N^2} = - I_{def}^{(N)}.}
\end{equation} %
Combining this with  our earlier ansatz that $I^{(N)}_{bulk} =  Z_N =1+ 1/N^2$
leads to the conclusion that
\begin{equation}\label{resui}{I^{(N)}= I_{bulk}^{(N)} + I_{def}^{(N)} =1,}
\end{equation} %
for   prime values of $N$.  This is consistent with  the presence of at least
one bound state for every prime value of $N$.

Now consider  the  non-prime charge sector  in which $N =\hat m n$   
with $\hat m, n >1$.   It is
useful  to recall the interpretation of the D-instanton of charge $N$ as a  
single wrapped euclidean
D-particle world-line.   The contributions to the instanton measure  labelled
by $n$ were  associated with the world-line of a charge-$n$  
D-particle winding $\hat m = N/n$ times.   In the case when $N$ is prime
the only contribution to the deficit term comes  from $n=1$, $\hat m = N$,
which
corresponds to $\hat m$ windings  of a charge-$n=1$  D-particle  
world-line.   The
other possibility, $n=N$, $\hat m =1$, gives the net contribution of one to the  
Witten index of the threshold bound state.
It is therefore very compelling to {\it assume} that the contribution to the Witten index in the non-prime  
case, $N=\hat m n$, of a subset of $\hat m$ D-particles of  charge $n$ can be calculated in the same way as that of $N$ 
fundamental ($n=1$) D-particles.   In other words, we assume  a generalization of the   
property that was strongly motivated  in the $N=2$ case in \cite{yi,sethi} that the deficit term is  
determined by free supersymmetric particle dynamics with 16 supercharges.
This  means that  the space of free-particle states that enters in the
calculation of $I_{0\, bulk}$ should be enlarged to include the  
infinite
tower of charge-$n$ threshold bound states.  In that case a cyclic  
permutation
of $\hat m$ identical charge-$n$ particles  contributes  a term
\begin{equation}\label{nfree}{ I^{(\hat m)}_{0\, bulk}= {1\over {\hat m}^2},}
\end{equation} %
which  leads to a total deficit in the charge-$N$ sector of
\begin{equation}\label{nmtot}{ I^{(N)}_{def} = -\sum_{\hat{m}|N \atop \hat{m}>1}  
I^{(\hat{m})}_{0\,bulk } = -\sum_{\hat{m}|N\atop \hat{m}>1}{1\over
{\hat{m}}^2} ,}
\end{equation} %
consistent with the Witten index $I^{(N)}=1$ for all integer $N$.

The success of these assumptions points to the systematics that  
needs to emerge
from a more precise treatment of the integration over the boundary  
term when $N$
is not prime.  This suggests that there are regions of  moduli
space in which the non-abelian integration leads to  $\hat m$ charge-$n$
threshold bound states which behave as free particles. Related issues arise in the considerations of  
\cite{bacha,bachb}
concerning D-string instantons in $d=8$ type I string theories. 

The systematics of the ten-dimensional D-particle problem is consistent with the expected  features 
of D-particle quantum mechanics with less supersymmetry.  The clearest examples are in six 
and four dimensions.  The six-dimensional case is realized  in the compactification of IIA on $K3$ by 
wrapping  D2-branes around a two-cycle in $K3$ and scaling to the
limit in which  the two-cycle vanishes.  
Using duality with the heterotic string on $T^4$
identifies the resulting  D-particles as massless gauge particles 
which have no threshold bound states.  The four-dimensional case is realized by wrapping the 
type IIB D3-brane around a three-cycle in a 
Calabi--Yau threefold and again scaling to the limit of zero area as
in \cite{stromingerc}. The successful resolution of the conifold singularity in terms of 
light states again requires the absence of threshold bound states.  
The absence of threshold bound states in these cases implies  that 
the Witten index should vanish.  In both cases 
the  bulk term in the N-particle index can be evaluated explicitly 
(as in \cite{moorenek}) or extracted from the T-dual D-instanton problem (as in \cite{ggIII}) and has the form 
\begin{equation}\label{lowd}{I^{(N)\, D=4,6}_{bulk} = {1\over N^2}.} 
\end{equation} %
Furthermore, it is easy to see that in these cases  the deficit term is given by $I^{(N)\, D=4,6}_{def} = -1/N^2$ 
(assuming, as before,  that it is 
given by the free-particle index).  The extra factor of $N$ here 
compared to (\ref{defresult}) comes about because in these lower-dimensional cases the number of fermions is equal 
to the number of 
bosons.  This systematics suggests that 
\begin{equation}\label{dlowt}{I^{(N)\, D=4,6} = I^{(N)\, D=4,6}_{bulk} + I^{(N)\, D=4,6}_{def}=0,}
\end{equation} %
as expected.

\section{The AdS/CFT correspondence and higher derivative terms}

Up to now the discussion of the higher derivative terms has been in the context of   the low energy effective 
action of string theory or M theory compactified on a circle or two-torus.
We now turn to consider the effect of the higher derivative terms when the IIB string propagates in 
an \AdS5s5\ background (much of this section and  section 9 is based on \cite{bg,bgkr}).  
For our purposes it will be sufficient to consider the euclidean version of $AdS_5$, which has a 
boundary with topology $S^4$ (or $R^4$).  The metric will be parameterized by 
\begin{equation}\label{cfmetric}{
ds^2 = {L^2 \over \rho^2}(dx\cdot dx  + d\rho^2) + d{\omega_5}^2 =  {L^2 \over \rho^2} (dx\cdot dx + dy \cdot dy ),}
\end{equation} %
where 
$(x^\mu, y^i)$  ($\mu =0, \cdots, 3$, $i= 1,\cdots,6$) are ten-dimensional cartesian coordinates with
$\rho^2 =  y^2$ and $d\omega_5^2$ 
is the spherically-symmetric  constant curvature metric on $S^5$ with radius $L$. 
In  these coordinates   the $AdS$ boundary is located at $\rho =0$.  
The second equality in (\ref{cfmetric})    makes it obvious that the metric is conformally equivalent 
to the flat ten-dimensional metric.  This will prove to be important  later.  The only   
nonvanishing curvature components in this background are
\begin{equation}\label{adsxs}{ R_{_{MNPQ}} = - \frac{1}{L^2} (g_{_{MP}}g_{_{NQ}} - 
g_{_{MQ}}g_{_{NP}}) \qquad R_{mnpq} = + \frac{1}{L^2} (g_{mp}g_{nq} - g_{mq}g_{np})}
\end{equation} %
(upper case Latin indices, $M,N,\dots=0,1,2,3,5$, label  the
$AdS_5$ coordinates and lower case Latin indices, $m,n,\dots 
=1,2,3,4,5$ label the $S^5$ coordinates).  
The non-vanishing components of the Ricci tensor are 
\begin{equation}\label{ricci}{  
R_{MN} = - \frac{4}{L^2} g_{MN} \qquad 
R_{mn} = + \frac{4}{L^2} g_{mn} \, ,}
\end{equation} %
so that the 
scalar curvature vanishes.  The dilaton and $C^{(0)}$ are arbitrary constants and  the  only 
other field that is nonvanishing in the \AdS5s5\ background is $F_5$, which has the solution, 
\begin{equation}\label{ffives}{F_{_{MNPQR}} = \frac{1}{L} \varepsilon_{_{MNPQR}}, \qquad  F_{mnpqr} = \frac{1}{L} 
\varepsilon_{mnpqr},}
\end{equation} %
which corresponds to a configuration in which the  total \RR\ charge associated with $F_5$ is 
equal to ${\alpha'}^{-2}L^4 e^{-\phi}$, which must be an integer, $N$.
The isometry of this space is the product  $SO(4,2) \times SU(4)$.  These factors package 
together with fermionic symmetries into the supergroup $OSp(4,2|4)$ which  also acts as a 
superconformal symmetry group on the boundary.

This background is maximally supersymmetric (just like the
Minkowski vacuum) and there are 32 conserved 
supercharges that transform as a
complex chiral spinor of the tangent-space group, $SO(4,1) \times
SO(5)$.
In the basis where the ten dimensional $\Gamma_\Lambda$ matrices\footnote{In this section and the 
next the ten-dimensional gamma matrices will be denoted by upper case $\Gamma$ while lower case 
$\gamma$ will be reserved for four-dimensional gamma matrices.} a
re 
given by 
$\Gamma_M = \s_1\otimes\gamma_M\otimes I$ and
$\Gamma_m=\s_2\otimes I\otimes\gamma_m$, the supersymmetries are 
generated by  
the Killing spinors that satisfy  
\begin{equation}\label{kilspina}{  
D_\Lambda \epsilon - {1 \over 2L}(\s_1\otimes I\otimes I) 
\Gamma_\Lambda\epsilon = 0 ,}
\end{equation} %
which follows from the requirement that the  gravitino 
supersymmetry transformation should vanish. 
In this basis the complex chiral supersymmetry parameter reads
\begin{equation}\label{spindecomp}{
\epsilon=  \pmatrix{1\cr 0}
\otimes \zeta \otimes \kappa \, , }
\end{equation} %
where $\zeta$ is a
complex four-component $SO(4,1)$ spinor  and $\kappa$ is a complex
four-component $SO(5)$ spinor.  The Killing spinor equation 
(\ref{kilspina}) has components   
\begin{equation}\label{kilspon}{  
D_M \zeta_\pm - {1 \over 2L} \gamma_M \zeta_\pm = 0 ,}
\end{equation} %
\begin{equation}\label{kilspin}{
D_m \kappa_\pm  \mp i {1 \over 2L} \gamma_m \kappa_\pm =0 .} 
\end{equation} %
The  Killing spinors $\kappa_\pm$ on $S^5$ may
be  used to construct the Kaluza-Klein excitations of all the fields in the IIB 
gauged supergravity starting from the modes of the  massless  complex
singlet scalar. We will later make use of the Euclidean 
continuation of the Killing spinors on $AdS_5$, which requires strategically placed factors of $i$.

The metric (\ref{cfmetric}) can be obtained from the geometry of a stack of coincident D3-branes, 
where each D3-brane carries a unit charge associated with $F_5 = 
dC^{(4)}$.  The effective world-volume 
theory  of $N$ coincident D3-branes  is described  by $U(N)$ gauge
theory, where the integer $N$ is again the total \RR\ charge.  In terms of the low energy 
supergravity theory the classical D3-brane geometry interpolates from the   Minkowski region 
far away from the horizon to   the  near-horizon geometry, which is  
\AdS5s5. The transition region grows with $N$ and the  $N\to \infty$ limit considered 
in \cite{klebrev,maldacena}    is  one   in which  the horizon region fills  all space.  
This leads to the AdS/CFT conjecture  \cite{maldacena} which suggests that the bulk 
string theory is {\it equivalent} to  superconformal $\calN =4$ SU(N) Yang--Mills  
theory on the boundary\footnote{It has been argued in
 \cite{wittads,aharowitt}
 that the theory should be   $SU(N)$ and not $U(N)$.} (see also 
\cite{gkp,wittads}).  
The supergroup $OSp(4,2|4)$ is that associated with the superconformal $\calN =4$ Yang--Mills theory. 

According to this idea the effective action of type IIB supergravity, 
evaluated 
on a solution of the equations of motion   with prescribed boundary 
conditions, is equated with the generating  functional of connected  
gauge-invariant correlation functions in 
the  Yang--Mills   theory.   The  parameters of the $\calN=4$ Yang--Mills 
theory and the IIB  superstring on \AdS5s5\ are related by the dictionary,
\begin{equation}\label{dict}{  g = {g_{_{YM}}^2 \over {4\pi}} \, \qquad
2\pi\tilde  C^{(0)} =  \theta_{_{YM}}\, \qquad {L^2 \over 
\ap} = \sqrt{g^2_{_{YM}}N} \ , }
\end{equation}%
where $g=e^\phi$ and $\tilde C^{(0)}$ are  the arbitrary constant scalar string fields, $g_{_{YM}}$ is
the   Yang--Mills coupling and $\theta_{_{YM}}$ is the vacuum angle. 
The complex Yang--Mills coupling is therefore identified
with the constant boundary value of the complex scalar field 
of the IIB superstring,
\begin{equation}\label{idenscal}{
S =  {\theta_{_{YM}}\over 2\pi}  +  {4\pi i\over g_{_{YM}}^2} =
\tilde{C}^{(0)} +  {i\over g}  .}
\end{equation} %
This explicit connection between the bulk theory and the boundary  
theory can be  expressed symbolically in terms of the generating functions \cite{wittads,gkp,ffz},
\begin{equation}\label{main}{ 
\exp(-S_{_{IIB}} [\Phi_{m}(J)]) = \int DA \, \exp(-S_{_{YM}}[A] + 
{\cal O}_{\Delta} [A] J). }
\end{equation} %
The left-hand side is the generating function for the amplitudes in   IIB superstring theory or its low 
energy supergravity limit.  The effective action $S_{_{IIB}}$ is  
evaluated  in terms of the \lq 
massless' 
supergravity fields  and their Kaluza--Klein  
descendents, that we have generically indicated with 
$\Phi(z;\omega)$, where $\omega$ are the coordinates on 
$S^5$ and  
$z^M\equiv (x^\mu,\rho)$ ($M=0,1,2,3,5$ and $\mu=0,1,2,3$) are the 
$AdS_5$ 
coordinates.  The notation in (\ref{main}) 
indicates that this action depends on the boundary data,    $J(x)$, 
of the bulk fields.  The right-hand side of (\ref{main}) is the generating function of the correlation 
functions of the gauge-invariant  composite operators,  ${\cal O}(A)$,   to which $J$ 
couples  in the boundary  $\calN=4$ supersymmetric  Yang--Mills,       where the fluctuating 
Yang--Mills potential is denoted 
by $A$.

For general values of the dimensionless   ratio, $\alpha'/L^2$,   an expansion in powers of 
$\alpha'$ is not necessarily  a good approximation to the theory.  Only for $AdS_5$ scales 
that are large compared to the string scale, i.e. $\alpha' << L^2$, is t
he  curvature   small and the $\alpha'$ expansion can still be interpreted as a derivative 
expansion in the nontrivial background.  
It follows from (\ref{dict}) that in this regime  classical supergravity  is related to the 
Yang--Mills limit in which $\gy^2 N \to \infty$.  This requires $N\to \infty$ (since $\Sl2$ 
duality can always be used to map the large string coupling to a value that is not large).  
Since, according to 't Hooft \cite{thooft} the quantity $\gy^2 N$ is the 
coupling constant of large-$N$ Yang--Mills theory we see that the AdS/CFT conjecture 
relates the strong coupling limit of one theory to the weak coupling limit of the 
other.  
Several   tests of this conjecture have been made at the level of the  two-point and three-point correlations 
of currents 
\cite{twoandthree} 
based  on  the semiclassical approximation to the bulk supergravity 
($g <<  
1$) which is valid   in the limit $\alpha' /L^2 << 1$, which
corresponds to the large-$N$ limit  
with 
$g_{_{YM}}^2 
N$ fixed at a large value. 
 
The spectrum of IIB supergravity compactified on \AdS5s5\ is  described  in \cite{kiromvan}.  
Here we are most interested in the 
 identification of the lowest lying modes that form the 
supergravity supermultiplet.    While the massless dilaton is associated
with the constant   mode on  
$S^5$, \ie\ with the scalar spherical harmonic $Y_{\ell}$ with 
$\ell=0$, 
the other scalars in the supermultiplet are associated with 
excitations on the 5-sphere. 
In particular the real scalars $Q^{ij}$ ($i,j,k=1,2,\dots ,6$), with mass $m^2 = -4/L^2$ in 
the {\bf 20}$_{_{{R}}}$ of the 
$SO(6)$ isometry group of $S^5$ result from a combination of the trace
of the internal metric and the self-dual \RR\ five-form field,
$F^{(5)}$, with $\ell=2$ ($Q^{ij}$ are  quadrupole moments of $S^5$).  
Similarly the complex scalars $E^{AB}$ with mass $m^2 = -3/L^2$ and 
their 
conjugates (in the {\bf 10}) 
 are associated with the pure two-form fluctuations with
$\ell=1$ of the  
complexified antisymmetric tensor in the internal directions. 
The {\bf 15} massless vectors $V_M^{[ij]}$ that gauge the $SO(6)$
isometry group  
are in one-to-one correspondence with the Killing vectors of $S^5$ and
result  
from a linear combination with $\ell=1$ of the mixed components of the
metric  
and the internal three-form components of the \RR\ four-form
potential, 
$C^{(4)}$. The {\bf 6} complex  
antisymmetric tensors $B_{MN}^{[AB]}$ with $m^2=1/L^2$, which have  
first order 
equations of motion, result from scalar spherical harmonics with 
$\ell=1$. The analysis of the fermions is similar. 
The {\bf 4} dilatini\footnote{In this section and the next the dilatini will be indicated by the upper 
case $\Lambda$ in order to avoid confusion with the gaugino of the Yang--Mills theory which is 
denoted by lower case $\lambda$.}      $\Lambda^A$ with mass 
$m = - 3/(2L)$ are 
proportional to the internal Killing spinors $\kappa_+$. The {\bf 
20}$_{_{{C}}}$  
spinors $\chi_{BC}^A$ 
with mass $m=-1/(2L)$ correspond to internal components of the
gravitino  
with 
$\ell=1$. Finally the supergravity multiplet is completed by the 
massless {\bf 4}$^*$ gravitinos $\Psi_{MA}$ which  are 
proportional to the internal Killing spinors $\kappa_-$.  

The above fields  are those that act as sources for the superconformal currents in the boundary 
Yang--Mills theory.  The 
higher Kaluza--Klein  modes on $S^5$, which  have higher values of
$\ell$,  are bulk fields of  
dimension $\Delta =l$.  The boundary value of any of these fields
couples to a composite  gauge  
invariant operator in the Yang--Mills theory that can be expressed 
as a rank $l$ symmetric traceless tensor of $SO(6)$ (the isometry group of $S^5$) made from 
a product of $l$ superfields of the boundary theory.  These operators will not concern us in the 
following but are a vital part of the complete story.

Whereas the duality group for the type IIB theory compactified on $T^6$ is  $E_{6(+6), \IZ}$ 
the duality group for the \AdS5s5\ background is   $\Sl2$,  which is inherited from ten dimensions.
The  local symmetry group is  $U(4)$ where the $U(1)$ factor is a
remnant of the local symmetry of flat ten-dimensional classical
supergravity.  However, there is no 
 corresponding  $U(1)$ symmetry in the $\calN=4$  Yang--Mills boundary
theory where the R-symmetry is $SU(4)$. 
 This fits
  with the fact that in IIB string theory  the classical $SL(2,\IR)$
is replaced 
 by  (local) $\Sl2$ and the continuous  $U(1)$ symmetry is not  
present.

\subsection{$\calN =4$ Yang---Mills fields and supercurrents}

The   $\calN=4$  
four-dimensional Yang--Mills   theory  is    classically invariant under 
superconformal 
transformations as well as  under global $SU(4)$ transformations, 
which form the R-symmetry 
group of  automorphisms of the ${\cal N}=4$ supersymmetry algebra.  
The physical fields comprise six real scalars,  four Weyl 
spinors,   
and one gauge vector potential, all in the adjoint  representation 
of  the
gauge group, which may be arbitrary in general but should be $SU(N)$
for the most straightforward application of the AdS/CFT
correspondence. 
The real scalars,  $\varphi^i$ ($i=1, \dots,6$),    form
a {\bf 6} of the $SU(4)$ R symmetry group and can be written as
antisymmetric tensors in the defining representation of $SU(4)$,  
 $\varphi^{i}=\frac{1}{2} {\overline{t}^{i}}_{AB}\, 
\varphi^{AB}$ (where $\bar t^i_{AB}$ is a Clebsch--Gordan coefficient and $\varphi^{AB} = - 
\varphi^{BA}$ with 
${\overline 
\varphi}_{AB}=\frac{1}{2}\varepsilon_{ABCD}\varphi^{CD}$)\footnote{The superscripts $A,B=1,\dots,4$ 
label {\bf 
4}'s of $SU(4)$ while  subscripts label the {\bf 4}$^*$'s.}.
The spinors, $\lambda^{A},~{\overline \lambda}_{B}$ transform  as {\bf 4}  and 
{\bf 4}$^{*}$, respectively, and the vector $A_{\mu}$ is a  singlet 
of $SU(4)$.  The classical Minkowski signature action is given   by 
\begin{eqnarray}\label{action}S & = & \int d^{4}x \, \Tr \Big\{ 
    (D_{\mu}\varphi^{AB})(D^{\mu}{\overline \varphi}_{AB}) 
	-\frac{1}{2}i({\lambda^{\alpha}}^{A} 
	{\dd \gamma \cdot {D}}_{\alpha{\dot \alpha}} 
	{{\overline{\lambda}}^{{\dot \alpha}}}_{A}) - 
	\frac{1}{4}F_{\mu \nu}F^{\mu \nu}   \\
   &  -  & g_{_{YM}} {\lambda^{\alpha}}^{A}
 	[{\lambda_{\alpha}}^{B}, {\overline{\varphi}}_{AB}] 
	-g_{_{YM}} {\overline{\lambda}}_{{\dot \alpha}A}[{{\overline 
    {\lambda}} ^{{\dot \alpha}}}_{B}, \varphi^{AB}] + 2g^{2}_{_{YM}}  
	[\varphi^{AB},\varphi^{CD}][{\overline{\varphi}}_{AB}, 
	{\overline{\varphi}}_{CD}]\Big\},\nn
\end{eqnarray} %
where $\Tr$ denotes a trace over the $SU(N)$ colour 
indices and $D_{\mu}^{ab} = \delta^{ab} \partial_{\mu} + i g_{_{YM}} 
f^{abc} A_{\mu c}$ is the covariant derivative in the adjoint 
representation.  We are here using a standard chiral basis for the spinors with the $2 \times 2$ 
Pauli matrices $\sigma^{\mu}_{\alpha \dot \alpha}$ coupling the complex chiral spinors. 
 
The supersymmetry transformations on the
component  
fields  with parameters ${\eta_{\alpha}}^{A}$ and 
${\overline\eta^{\dot\alpha}}_{A}$ are given by,
\begin{eqnarray}
	\delta \varphi^{AB} & = & \frac{1}{2} ({\lambda^{\alpha}}^{A}  
    {\eta_{\alpha}}^{B}- 
	{\lambda^{\alpha}}^{B} {\eta_{\alpha}}^{A}) + \frac{1}{2}  
        \varepsilon^{ABCD} 
	{{\overline{\eta}}_{\dot \alpha}}_{C}  
	{\overline \lambda}^{\dot \alpha}_{D} \nn \\
	\delta {\lambda_{\alpha}}^{A}  & = & -\frac{1}{2} F^-_{\mu \nu}  
	{{\sigma^{\mu \nu}}_{\alpha}}^{\beta} {\eta_{\beta}}^{B} + 4i 
        (\gamma \cdot {D}_{\alpha {\dot \alpha}} \varphi^{AB}) 
        {\overline{\eta}}^{{\dot \alpha}}_{B} - 
	8g_{_{YM}}[{\overline{\varphi}}_{BC}, \varphi^{CA}] 
        {\eta_{\alpha}}^{B}\nn  \\
	\delta A_{\mu} & = & -i {\lambda^{\alpha}}^{A}  
	{\sigma^{\mu}}_{\alpha{\dot \alpha}} 
	{{\overline \eta}^{\dot \alpha}}_{A} 
	-i {\eta^{\alpha}}^{A} {\sigma^{\mu}}_{\alpha {\dot \alpha}} 
	{{\overline \lambda}^{\dot \alpha}}_{A} .\label{transs}
\end{eqnarray}%
These are ordinary global supersymmetries when the 16 components of $\eta_\alpha^{\ A}$ 
and $\bar\eta^{\dot \alpha}_{\ A}$ are constant but the symmetry extends to the thirty-two 
component superconformal symmetry when $\eta$ and $\bar \eta$ are linear fun
ctions of $x^\mu$,
\begin{equation}\label{etanew}{\eta_\beta^A \to  {1\over \sqrt{\rho_0}} \left(\rho_0 \eta^A_\beta + (x-x_0)_\mu 
\sigma^\mu_{\beta  \dot \beta} \bar \xi^{\dot \beta A}  \right)\equiv \left({1-\gamma^5\over 2}\right)
{1\over \sqrt{\rho_0}} \gamma_M z^M\zeta^{(0)},}
\end{equation} %
where $\bar \xi$ is the Grassmann spinor associated with the special  supersymmetries which is 
packaged together with $\eta$ into a four-component spinor in the second equality where
\begin{equation}\label{zetadef}{\zeta^{(0)}  =\pmatrix{\eta_\alpha^A \cr
                                    \bar \xi^{\dot \alpha\, A}\cr},}
\end{equation} %
$z^M \equiv (x^\mu - x_0^\mu,\rho)$ and  $\gamma^M \equiv (\gamma^\mu,\gamma^5)$.  Equation (\ref{zetadef}) 
is just a chiral projection on a spinor of $AdS_5$ (with tangent-space group $SO(4,1)$).  
The normalization factor of $\rho_0^{-1/2}$ in (\ref{etanew}) has been chosen for later convenience

The $\beta$ function for this theory vanishes which means that the quantum theory is expected to 
share the superconformal invariance of the classical theory.  The Noether  currents 
associated with these superconformal transformations, as well as   the
chiral $SU(4)$ transformations, form a supermultiplet with 256 component currents, including 
the energy-momentum tensor \cite{townsenda,dewit}. 
{}For example, the lowest member of the supermultiplet is the
scalar current that couples to $Q^{ij}$,
\begin{equation}\label{qcurr}{{\cal Q}^{ij}=\varphi^i\varphi^j-\delta^{ij} 
\varphi_k\varphi^k/6.}
\end{equation} %
Equation (\ref{qcurr}) only involves the bilinear terms in the current, but it  is important that the 
full nonabelian currents also contain higher-order contributions \cite{bgkr} which are crucial 
in verifying  the AdS/CFT correspondence \cite{freedmanb}.
The current that couples to the boundary value of the dilatino is 
\begin{equation}\label{lamcurr}{{{\hat \Lambda}_{\alpha}}^{A} =
	-{{\sigma^{\mu \nu}}_{\alpha}}^{\beta}  
	{F^{-}}_{\mu \nu} {\lambda_{\beta}}^{A}}
\end{equation} %
(again in the linearized approximation). 
The complete list of currents in the supermultiplet is given in \cite{dewit}.
These  superconformal
currents are in one to one correspondence with   the   fields of the supergravity multiplet.

{}From the AdS/CFT conjecture  the supersymmetrized Einstein--Hilbert action gives the leading 
large-$\gy^2 N$ contributions to the correlation functions of the superconformal currents.  
For example, a correlation function of $M$ stress tensors,  
\begin{equation}\label{mstress}{\langle T_{\mu_1\nu_1}(x_1) T_{\mu_2\nu_2}(x_2)\dots   T_{\mu_r\nu_r}(x_r)  
\dots T_{\mu_M\nu_M}(x_M)   ,}
\end{equation} %
corresponds to bulk superstring amplitudes in which the boundary values of the gravitons  
have polarizations, $h_{\mu_r\nu_r}$,  which are oriented
in the four-dimensional Minkowski directions, $\mu_r = M_r$, $\nu_r =
N_r$ where $M_r,N_r = 0,1,2,3$.  Since the coefficient of the Einstein--Hilbert term is 
$e^{-2\phi} {\alpha'}^{-4} = L^8 N^2$  the leading behaviour of the correlation functions 
obtained in this manner is of order $N^2$.   In general there is little  known 
about such  Yang--Mills correlation functions  at large 't Hooft coupling.  However, 
the free-field value  of the correlation functions of  two superconformal currents, 
such as $\langle T_{\mu_1\nu_1}(x_1^{\mu_1})  T_{\mu_2\nu_2}(x_2^{\mu_2})  \rangle$,
is known to be exact \cite{osborn,gk}  by virtue of  the relation of this correlation  
function to the $R$-symmetry anomaly, which is not renormalized due to the   
Adler--Bardeen theorem. An analogous Ward identity prevents the three-point correlation
function of superconformal currents (in the $\calN=4$ theory,  but not for $\calN<4$)  from receiving renormalizations  
beyond those of the free field theory \cite{freedmana}.   These free-field correlators of two and 
three superconformal currents are both proportional to  
\begin{equation}\label{sudef}{{\rm dim}\ (SU(N)) = N^2-1.}
\end{equation} %
The leading factor of $N^2$ corresponds to the coefficient of ${\alpha'}^{-4}g^{-2}$ 
in the Einstein--Hilbert action.  This would be the exact $N$-dependence if the gauge group were $U(N)$ but 
(\ref{sudef}) shows that there must be one other contribution to the Einstein--Hilbert action 
in the \AdS5s5\ background that arises at order $N^{-2}$ (order ${\alpha'}^4$) relative  
to  the leading term.  Such a term might come from a one 
string loop effect  which has not yet been elucidated.  Since the terms that concern 
us here are of order ${\alpha'}^3$ relative to the leading term they dominate over this effect.

\subsection{Generalities concerning higher derivatives terms}

We wish to consider the effects of the $o({\alpha'}^{-1})$ terms on  the AdS/CFT conjecture.  
Since these terms are suppressed by ${\alpha'}^3$ relative to the leading terms they will 
correspond to $O(N^{-3/2})$ corrections to the leading behaviour.  The 
interesting feature of these terms is that they manifest the $\Sl2$ duality in a very nontrivial fashion.

We saw from the definition of the $\calR^4$ term, (\ref{rfourdef}), that  
the only components of the curvature
which contribute   are those of the Weyl
tensor.   We can expand  $R_{\theta^4}$ (defined in (\ref{rthetdef})) in fluctuations around its classical value,  
$R^{(0)}_{\theta^{4}}$,
\begin{equation}\label{rexpa}{R_{\theta^4} =  R^{(0)}_{\theta^{4}} + \hat R_{\theta^4},} 
\end{equation} %
where $\hat R$ is the fluctuation\footnote{A more complete discussion, given in \cite{gkttwo}, shows that 
there is a   field definition  in which the off-shell fluctuations of the metric  can be restricted 
to those that only give fluctuations to the Weyl com
ponents of the curvature.}.  Since \AdS5s5\ is conformally flat the Weyl tensor vanishes and  it follows that 
\begin{equation}\label{rfourdefz}{R_{\theta^4}^{(0)}=0.}
\end{equation}%
This  leads immediately to several consequences:
\begin{itemize} 
\item[1.] $\calR^4 =0$ in the \AdS5s5\
background, which means that the dilaton equation of motion is  unchanged in this background.  
\item[2.] $\delta R^4 / \delta g_{\mu\nu} =0$ (since the differential is
proportional to $(R_{\theta^4}^{(0)})^3$), which means that the   
Einstein
equation is also unaffected.  It is also easy to see that none of  
the other
$o({\alpha'}^3)$ terms which are related to $\calR^4$ by supersymmetry  contribute
to the equations of motion. This means that the \AdS5s5\ background is
unaltered
by the presence of this term, so this background is a solution of  
the effective
equations of motion of string theory through $o({\alpha'}^3 )$,  
to all orders
in $g$ (a more general superspace argument  leads to the statement that this  background 
is exact to all orders in $\alpha'$ \cite{kallosha}).
\item[3.]  Equation (\ref{rfourdefz}) also implies that
\begin{equation}\label{twothree}{{\delta \over \delta g_{\mu\nu}} {\delta \over \delta
g_{\rho\sigma}} R^4 \big|_{AdS_5\times S^5} =0 =    {\delta \over \delta
g_{\mu\nu}} {\delta \over \delta g_{\rho\sigma}} {\delta \over \delta
g_{\tau\omega}} R^4 \big|_{AdS_5\times S^5}.}
\end{equation} %
This shows that there is no renormalization of the graviton two-point or
three-point functions in the \AdS5s5\ background at  
$o({\alpha'}^3)$. 
\end{itemize}

However, there is a non-zero four-graviton contribution from the $\calR^4$ term  which is
obtained by differentiating  four times with respect to the metric. This gives a contribution 
to the amplitude which adds to the
term which arises from the Einstein--Hilbert action, so that up to
$o({\alpha'}^3)$  the four-graviton amplitude is proportional to   
\begin{equation}\label{fourst}{{\alpha'}^{-4} e^{-2\phi} 
A_4^{String\,(1)} +k {\alpha'}^{-1} e^{-\phi/2} f^{(0,0)}(\tau,\bar \tau)  A_4^{String \,(2)},}
\end{equation}
where $A_4^{String\, (1)}$  is the classical
amplitude obtained from the Einstein--Hilbert action while
$A_4^{String\, (2)}$ is the contribution from the $\calR^4$ term
and $k$ is a well-defined constant.

{}For example, the contribution of the $\calR^4$ term to four-graviton   scattering is obtained by  contracting
the linearized $\calR^4$ vertex with four graviton propagators that propagate 
from the interaction point at  $(x_0^\mu,\rho_0)$  to the boundary 
at points $(x_{0^\mu\, r}, \rho_{0, r}=0)$, where $r$ labels the position of the $r$th boundary field.  
The  linearized curvature is
\begin{equation}\label{lincurv}{R_{\mu_1\omega_1\nu_1\tau_1} = D_{\omega_1} D_{\tau_1}
h_{\mu_1\nu_1},}
\end{equation} %
where  $h_{\mu\nu}$ is the linearized fluctuation of the metric around
its value in   \AdS5s5\ and $D$ is the \AdS5s5\ covariant derivative.  
The bulk-to-boundary graviton propagator in \AdS5s5\ is an obvious generalization \cite{tseytgrav} 
of the scalar and spin-1/2 propagators that will appear in later sections.  
The four-graviton interaction vertex can be expressed in the form (\ref{kinematic}) where  
the tensors $t_8$ in the kinematic factor $\tilde K$   have the form of  the product of four
inverse \AdS5s5\  metrics summed over various permutations of their indices.

Using the AdS/CFT conjecture  the $\calR^4$ term gives a  contribution to the correlation 
functions of  stress tensors in the boundary Yang--Mills field theory that are suppressed 
by a factor of $N^{-3/2}$ relative to the leading free-field contribution.
The vanishing of the graviton one-point function, translates into the  
statement that
the one-point function of the stress tensor $\langle T_{\mu\nu} \rangle$ vanishes, which follows from
conformal
invariance.  Similarly, the vanishing of the ${\alpha'}^{-1}R^4$ contribution to the
two-graviton and three-graviton  S-matrix elements in IIB supergravity translates into  
the statements
that the  correlation functions of two and of three stress tensors in $\calN = 4$  
Yang--Mills
theory  is not renormalized from the free-field value at  this order in $N^{-1}$.   
This is in accord with the exact results \cite{osborn,gk,freedmana} cited earlier.  

It is  
only when we
come to the four-graviton amplitude that the $\calR^4$ term contributes and
therefore the correlation  function of four stress tensors gets a new
contribution.
The supergravity calculations  give  the
following
expression for the momentum-space  correlation function of four  
stress tensors,
\begin{equation}\label{fourtens}{A_4^{YM}= \langle T_{\mu_1\nu_1}(1)T_{\mu_2\nu_2}(2) 
T_{\mu_3\nu_3}(3) T_{\mu_4\nu_4}(4) \rangle  = N^2 A_4^{(1)} + \tilde k N^{1/2} f^{(0,0)}(S,\bar S) A_4^{(2)} + \dots ,}
\end{equation} %
where  $A^{(1)}_4$ is the contribution  at leading order in  
$(\gy^2N)$ and $A_4^{(2)}$ is  the correction arising from the $\calR^4$ term (and an
irrelevant constant has been absorbed into $\tilde k$).  
The term $A_4^{(1)}$ comes from  the four-graviton amplitude
computed from the Einstein action in \AdS5s5, while $ A_4^{(2)}$ corresponds to the  
four-graviton vertex in the $\calR^4 $ term.  No further terms are necessary at 
this order in the ${\alpha'}$ expansion.      
The scalar field
$\tau$ is now  interpreted as the complex coupling constant, $S$.   

The second term in  (\ref{fourtens}) has a remarkable amount of information
concerning the Yang--Mills theory.  It is the first non-leading  
term in the
$1/N$ expansion but is an exactly known function of the (complex)  
coupling.  As we have seen, 
the  factor $f^{(0,0)}$    contains two  
terms that are perturbative in string theory and an infinite number of D-instanton terms. 
Therefore, in
the limit
$\gy\to 0$ with $\gy^2 N$ fixed and large, the expression  
(\ref{fourtens}) takes the
form 
\begin{eqnarray}\label{fourtenstwo}&& N^{-2} A_4^{YM}   = A_4^{(1)}  
+ \tilde k
A_4^{(2)} \left[2\zeta(3) \left({\gy^2 N\over 4\pi}\right)^{-3/2}
  +  {2\pi^2 \over 3 N^2} \left({\gy^2 N\over
4\pi}\right)^{1/2}\right. \\ 
&& \left. +  {(4\pi)^{3/2}\over N^{3/2}}  
\sum_{K=1}^\infty
Z_K  K^{1/2} \left(e^{- K \left(8\pi^2  \gy^{-2}  +  i \theta  
\right)} + e^{-K \left(8\pi^2  \gy^{-2}  - i \theta \right)}\right)(1 + o(\gy^2 /K ))\right],\nn
\end{eqnarray} %
which includes an  infinite series of instanton corrections.
We will return to a discussion of these corrections later.
An important property of this expression is its invariance under 
$\Sl2$ (which is interpreted as Montonen--Olive electromagnetic  duality). This transformation, which 
maps the small coupling regime to large coupling,
is manifest in (\ref{fourtens}) but  not  in (\ref{fourtenstwo}), which
is the 't Hooft expansion, and  is only valid when $\gy$ is very small.  The presence of 
noninteger powers of $\gy^2 N$ in this expansion is one indication of how far removed 
the strong coupling expansion is from weakly coupled perturbation theory.

\subsection{The scalar $AdS$ Green function and the instanton profile.}

In order to make use of (\ref{main}) to 
perform explicit detailed calculations of Yang--Mills correlation functions  we need to evaluate the  
bulk-to-boundary  
Green functions for the various fields of the bulk theory.  
These are defined as specific normalized limits of
bulk-to-bulk Green functions  \cite{gkp,wittads,freedmana} when one
point is taken to the $AdS$ boundary.  
The precise forms of these propagators 
depend on the spin and mass  
of the field. For example, the normalized bulk-to-boundary   
Green function for the $AdS$ Laplace operator for a dimension  $\Delta$ scalar field
is given by 
\begin{equation}\label{gdeff}{
G_{\Delta}(x,\rho,\omega; x',0,\omega') =  c_{_{\Delta}} K_{\Delta} 
(x^\mu, \rho; x'^{\mu},0),}
\end{equation} %
which is independent of $\omega$ and 
where  $c_{_\Delta} = \Gamma(\Delta)/(\pi^2 \Gamma(\Delta - 2))$ and 
\begin{equation}\label{greenfun}{
 K_{\Delta} (x^\mu, \rho; x'^{\mu},0) =  
{\rho^{\Delta} \over  (\rho^2 
+ (x-x')^2)^{\Delta}}. }
\end{equation} %
The expression (\ref{gdeff}) is appropriate for 
an `S-wave' process in which there are no excitations 
in the directions of the five-sphere, $S^5$.  
In terms of $K_{\Delta}$ the bulk field 
\begin{equation}\label{btob}{
\Phi_m (z;J) = c_{_{\Delta}}\int d^4x'  
K_{\Delta} (x,\rho; x',0) J_{\Delta}(x') }
\end{equation} %
satisfies the boundary condition as $\rho \to 0$,
\begin{equation}\label{boundcond}{
\Phi_m (x,\rho;J) \approx \rho^{4-\Delta} J_{\Delta}(x), }
\end{equation} %
since $\rho^{\Delta-4}K_{\Delta}$ reduces to a $\delta$-function on the 
boundary.  The conformal dimension 
of the operator is related to the $AdS$ mass of the corresponding bulk 
field by $(mL)^2 = \Delta(\Delta -4)$,  
so that $\Delta_{\pm} = 2 \pm \sqrt{4+(mL)^2}$ and only the positive 
branch,  
$\Delta = \Delta_+$, is relevant for the lowest-\lq mass' supergravity  
multiplet.  In the case of a massless scalar field ($\Delta_+ =4$) 
the propagator reduces to $\delta^{(4)}(x^\mu - x'^{\mu})$ in the 
limit $\rho \to 0$.  
In comparing the D-instanton contributions of the bulk theory with the Yang--Mills instanton 
contributions to the boundary theory  it is of significance, as we will see in section 9,   
that the Green function $K_4$ defined in    (\ref{greenfun}) has    the same
form (up to an additive constant)  as the   
profile of $e^\phi$ in the D-instanton solution  centered on the point $(x^{
\mu},\rho)$  in  
\AdS5s5\  and evaluated at the boundary point $(x^{\prime
\mu},0)$.   

We now turn to the Yang--Mills instanton.  Recall that 
that the instanton solution for the $SU(2)$ Yang--Mills potential gives a nontrivial 
nonabelian field strength of the form   
\begin{equation}\label{instprof}{F^{a-}_{(0)\mu\nu} (x^\mu; x_0^\mu, \rho_0)  =  - \frac{4}{g_{_{YM}}} 
{\eta_{\mu\nu}^a \rho_0^2 \over \left(\rho_0^2  
+ (x-x_{0})^2 \right)^2}  ,}
\end{equation} %
where $\eta_{\mu\nu}^a$ is the standard 't Hooft symbol, $\rho_0$ is the arbitrary 
scale and ${x_0}^\mu$ the position of the instanton on $S^4$.    Squaring $F^-$ and 
using properties of $\eta$ leads to the equality
\begin{equation}\label{insteq}{(F_{(0)}^{-})^2 = {4 \over \gy^2} K_4(x_0^\mu,\rho_0; x^\mu, 0),}
\end{equation} %
which is again equal to the scalar Green function for the $\Delta =4$ case  where 
the fifth  
coordinate $\rho$ is now  identified with the instanton scale. 
This is a key observation in  identifying   D-instanton  effects of 
the bulk theory with Yang--Mills instanton effects in  the boundary 
theory.   It is related to the fact that the moduli space of a
Yang--Mills instanton on $S^4$ has an $AdS_5$ factor.

\section{D-instantons and Yang--Mills instantons}

We now want to compare the leading order one-instanton contribution to the  supersymmetric 
Yang--Mills correlation functions  with the 
amplitudes of the IIB   superstring theory with appropriate boundary conditions.  
For illustrative purposes we will only consider the sixteen-$\Lambda$  amplitude  in 
the \AdS5s5\ background and the corresponding Yang--Mills sixteen-${\hat \Lambda}$ correlation
function \cite{bgkr}.  From either perspective 
the leading instanton contribution arises from the product of sixteen 
factors each carrying one single fermionic zero mode. Correlation functions of other 
superconformal currents can be calculated in analogous fashion.  For certain of these 
it is possible  to express the result very explicitly in terms of dilogarithms \cite{bgkr,brodiegut}.      

{}From the bulk type IIB point of view we are interested in the leading D-instanton 
contribution to amplitudes, which  may  either be extracted directly   from 
the exactly 
known $o({\alpha'}^{-1})$ terms in the effective action (method (a)  below) or 
deduced by integration over the fluctuations around the classical 
$AdS_5\times S^5$ D-instanton solution (method (c)).  From the Yang--Mills perspective 
we are interested in the instanton contributions to supercurrent correlators of the $\calN=4$ theory (method (b)).  

\vskip 0.3cm
{\it Method (a). Expansion of ${\alpha'}^{-1} f^{(12,12)}
\Lambda^{16}$ term in  bulk IIB supergravity} 
\vskip 0.1cm

The first method for obtaining this amplitude is to 
expand the function $f^{(12,-12)}$ in order   to extract the
one-instanton term.  The leading behaviour comes from the leading 
power of $g^{12}= \tau_2^{-12}$  in $\calF_{1,12}$ defined by
(\ref{expadef}) and (\ref{znpddef}).   We want to consider the situation in which
all sixteen fermions propagate to specific values
on the boundary.  
The Dirac operator acting on spin-1/2 fields in $AdS_{5}$ was given 
in \cite{henningsfet,kaviani} as
\begin{equation}\label{diracop}{
   \gamma\cdot D\Lambda  = {e_{{\hat L}}}^{M} \gamma^{\hat{L}}  
	\left( \partial_{M} +\frac{1}{4} 
	\omega^{\hat{M}\hat{N}}_{M} \gamma_{\hat{M}\hat{N}}
	\right)\Lambda
 =  	(\rho  \gamma^{\hat 5}  
	\partial_{5}  
	+ \rho \gamma^{\hat\mu} \partial_{\mu} - 2 \gamma^{\hat 5})
\Lambda \; , }
\end{equation} %
where ${e_{\hat{L}}}^{M}$ is the vielbein,
$\omega^{\hat{M}\hat{N}}_{M}$  the  
spin connection (hatted indices refer to the tangent
space) and $\gamma^{ \hat\mu}$ are the  four-dimensional Dirac 
matrices.  
Equation (\ref{diracop}) leads to the normalized 
bulk-to-boundary propagator of the fermionic field $\Lambda$ 
of mass $m=-3/2L$, associated with the composite operator  
${\hat \Lambda}$ of dimension $\Delta = \frac{7}{2}$, 
\begin{equation}\label{propsev}{ 
K^F_{7/2}(\rho_0,x_0;x) = K_{4}(\rho_0,x_{0};x) {1 \over 
\sqrt{\rho_0}}
\left( \rho_0 \gamma_{\hat5} +  {(x_{0}-x)}^{\mu}
	\gamma_{\hat\mu}	\right) ,}
\end{equation} %
which, suppressing all spinor indices, leads to  
\begin{equation}\label{boundarylambda}{
 \Lambda_{J}(x_0,\rho_0) = \int d^4 x K^F_{7/2}(\rho_0,x_{0};x) 
J_{\Lambda}(x) \, ,}
\end{equation} %
where $J_{\Lambda}(x)$ is a left-handed  boundary value of 
$\Lambda$ and acts as the 
source for  the composite operator  
${\hat \Lambda}$  in the boundary $\calN=4$ Yang--Mills theory satisfying $\hat \gamma^5 J_\Lambda = J_\Lambda$.
As a result, the classical action for the operator ${(\Lambda)}^{16}$
in the   
$AdS_{5}\times S^{5}$ supergravity action is
\begin{eqnarray}\label{sourcelsx} 
  S_{\Lambda} [J] & = & (const.)\, e^{-2\pi ({1\over g} -  
  i  C^{(0)})} g^{-12}  
  \, V_{S^5} \, \int \frac{d^{4}x_{0}d\rho_0}{\rho_0^{5}}\nn \\
  & &   \varepsilon_{16} \prod_{p=1}^{16}  
    \left[ K_{4}(\rho_0,x_{0};x_{p}) \frac{1}{\sqrt{\rho_0}}
	\left( \rho_0 \gamma^{\hat 5} +  
    {(x_{0}-x_{p})}^{\mu} \gamma_{\hat\mu} \right)  
    \, J_{\Lambda}(x_{p}) \right] \, ,
\end{eqnarray} %
where $J_{\Lambda}(x_p)$ is the wave-function of the dilatino
evaluated at the boundary  point $(x_p,0)$ and  we have set $e^\phi = g$ and $C^{(0)} = \tilde C^{(0)}$ 
(since the scalar fields are taken to be 
constant in the \AdS5s5\ background) and $V_{S^5}=\pi^3$ 
is the $S^5$ volume.  The spinor indices on the $J$'s are contracted
with the  rank-sixteen antisymmetric  tensor, $\varepsilon_{16}^{a_1a_2\dots a_{16}}$.
We have not kept track of  the overall constant normalization in this expression.  
The final integration over the moduli $x_0^\mu, \rho_0$ has not been carried out explicitly in (\ref{sourcelsx}).  
For certain of the other correlation functions the integration over the instanton moduli can be 
evaluated in terms of standard functions \cite{bgkr,brodiegut}.
According to (\ref{main})
the sixteen-$\hat \Lambda$ Yang--Mills correlation function is extracted from (\ref{sourcelsx})  
by differentiating with respect to all the $J_\Lambda$'s.

\vskip 0.3cm
{\it Method (b).  Sixteen-$\hat \Lambda$ correlator in the boundary Yang--Mills theory}
\vskip 0.1cm

We will now consider the correlation 
function of sixteen fermionic superconformal current operators in a
charge-one instanton background, 
\begin{equation}\label{lsx}{
G_{\hat \Lambda^{16}}(x_p) = 	\langle \prod_{p=1}^{16} g_{_{YM}}^2 
\hat \Lambda^{A_{p}}_{\alpha_{p}}(x_{p})  
	\rangle_{_{K=1}} \, , }
\end{equation} %
where 
$\hat \Lambda$ is defined in (\ref{lamcurr}).  The $SU(2)$ Yang--Mills instanton has sixteen 
supermoduli which correspond to the fermionic zero modes that are induced by the broken 
supersymmetries.   The special feature of the product of fields inside the   
correlator (\ref{lsx}), as well as the others  that  are  
related  by supersymmetry, is that they provide precisely the sixteen fermionic  
zero modes that are needed to give a nonzero result in the instanton 
background.     In particular, it is easy to see that $\hat \Lambda$ is linear in 
fermionic zero modes so a total of sixteen factors is needed for a non-zero result.  To leading order in $g_{_{YM}}$, 
$G_{\hat \Lambda^{16}}$ does not receive contribution from anti-instantons.

The fermionic zero modes can be deduced  by applying the broken  components of the 
supersymmetry transformations to the fields.  
The bilinear   
$\hat \Lambda_\alpha^{A}=- {\sigma^{\mu \nu}}_\alpha^\beta F^{-}_{\mu \nu} \lambda_\beta^{A}$ 
is proportional to the zero mode of 
${\lambda_\alpha}^A$ which will be denoted ${\lambda_{(0)\, \alpha}}^A$ and can 
be deduced from the second line of (\ref{transs}),
\begin{equation}\label{lamzero}{
\lambda_{(0) \alpha}^A \equiv \delta \lambda_\alpha^A  =  {1\over 2} 
F_{(0)\mu\nu}^- \sigma_\alpha^{\mu\nu\, \beta} 
{1 \over \sqrt{\rho_0}} \left( \rho_0\eta_\beta^A + 
(x-x_0)_\mu \sigma^\mu_{\beta\dot \beta }
\bar\xi^{\dot \beta A} \right) \, .}
\end{equation} %
The leading term in $\hat \Lambda$ in the one-instanton sector 
is simply obtained by using the 
instanton profile  ${F^{-}_{(0)\mu \nu}}$ (\ref{instprof}) and substituting each  
$\lambda_\alpha^{\ A}$ by  the corresponding 
zero mode, ${\lambda_{(0)\alpha}}^{A}$.   The resulting correlation function has the form 
\begin{eqnarray}\label{lsxsym}
G_{\hat \Lambda^{16}}(x_p) 
& = & (const.)\, 
g_{_{YM}}^{8} e^{-{8\pi^2 \over g_{_{YM}}^2} + i\theta_{_{YM}}} 
\int \frac{d^{4}x_{0} \, d\rho_0}{\rho_0^{5}}  
\int d^{8}\eta d^{8}{\overline \xi} \\ 
& &    \prod_{p=1}^{16} \left[  K_4(x_p^\mu,\rho_0; x_0^\mu,0) \,
\frac{1}{\sqrt{\rho_{0}}}
\left( \rho_0 \eta^{A_{p}}_{\alpha_{p}}+
{(x_{p}-x_{0})}_{\mu}  \sigma^{\mu}_{\alpha_{p}{\dot \alpha}_{p}}  
{\overline \xi}^{{\dot \alpha}_{p}A_{p}} \right)  \right] \ .\nn
\end{eqnarray} %
Integration over the fermion zero modes leads to the sixteen-index invariant  tensor 
$\varepsilon_{16}$.  Converting to four-component spinor notation  and using (\ref{main})  
this result coincides with  (\ref{sourcelsx}), up to the undetermined overall constant. 

One is entitled to ask why such a comparison of a D-instanton effect
with a $SU(2)$ Yang--Mills instanton effect is being made in the
first place.   In method (a) we used the expression for the
$o({\alpha'}^{-1})$ interaction in the \AdS5s5\ background which
should only be a good approximation to strongly coupled $SU(N)$  Yang--Mills 
in the limit $N\to \infty$. On the other hand  method (b) computed the effect
of an $SU(2)$ Yang--Mills instanton.  However, as will be described in
the final subsection, the $N$-dependence of these    correlation functions was shown 
in \cite{doreyone} to only affect the overall     coefficient of the correlation 
function and not its functional form. Furthermore, the coefficient has the expected large-$N$ limit, $N^{1/2}$.  
This will be reviewed in the last subsection along with the results in \cite{doreytwo,doreythree} 
which demonstrate that the large-$N$ limit of the
Yang--Mills calculation agrees perfectly with the leading D-instanton
contribution even in the multi-instanton sectors.  

\subsection{The classical D-instanton in flat space and in \AdS5s5}

Later in this subsection we will evaluate the single   
D-instanton terms in the sixteen-dilatino  amplitude in \AdS5s5\   
by semi-classical quantization in the background of a  
  D-instanton solution of type  
IIB supergravity (this is method (c)).  First we need to 
recall some properties of the classical D-instanton solution.

In flat ten-dimensional euclidean space the charge-$K$  
D-instanton solution is a finite-action euclidean 
supersymmetric (BPS--saturated) solution  in which
the  metric is trivial ($g_{\mu\nu} = \eta_{\mu\nu}$ 
in the Einstein frame) but the complex scalar $\tau = C^{(0)} + i 
e^{-\phi}$ has a nontrivial profile with a singularity at the 
position of the D-instanton.   The (euclidean) 
\RR\ scalar is related to the dilaton  by the BPS condition  
$\partial_\Sigma C^{(0)} = \pm i 
\partial_\Sigma e^{-\phi}$, while the dilaton solution is 
the harmonic function (correcting an error in \cite{ggp}) 
\begin{equation}\label{tendim}{
e^{\hat \phi^{(10)}}  = g + { 3 K {\ap}^4 \over \pi^4 |X-X_0|^8}. }
\end{equation} %
This is the classical solution  of the ten-dimensional 
Laplace equation,
$\partial^2e^\phi =0$,  outside  an
infinitesimal sphere centered on the point $X_0^\Lambda$ (where 
$X^\Lambda$ is the ten-dimensional coordinate 
and $X_0^\Lambda$ is the location of the D-instanton\footnote{Having run out of 
many conventional options we are here using upper case Greek letters to signify the ten-dimensional indices.}), 
$g$ is the asymptotic value of the string coupling and the 
normalization of the second term has a quantized value by virtue of a 
condition analogous to the Dirac--Nepomechie--Teitelboim condition 
that quantizes the charge of an electrically charged   
$p$-brane and of its magnetically charged $p'$-brane dual 
\cite{nepo,teit}.  
It is notable that the solution in (\ref{tendim}) is simply the Green 
function for a scalar field to propagate from $X_0$ to $X$ subject to 
the boundary condition that $e^\phi = g$ at $|X| \to \infty$ or 
$|X_0| \to \infty$.

We are now  interested in solving the equations of motion of the IIB 
theory in euclidean  \AdS5s5. 
The   BPS condition for a D-instanton in this background again  
requires $\partial_\Sigma  e^{-\phi}  = \pm i \partial_\Sigma C^{(0)}$ that 
leads to  
\begin{equation}\label{disteq}{ 
g^{\Lambda\Sigma} \nabla_\Lambda \nabla_\Sigma e^\phi = 0,}
\end{equation} %
and (in the Einstein frame) the Einstein equations are unaltered by
the presence of the D-instanton  so  that \AdS5s5\ remains a solution.
Equation (\ref{disteq}) is  
identical to  the equation for the Green function of a 
massless scalar propagating between the location of the D-instanton
($x_0^\mu, y_0^i$) and the point ($x^\mu,y^i$), which is the
bulk-to-bulk propagator (subject to the  boundary condition that it is
constant in the limits $\rho\to 0$ and  $\rho \to \infty$).
This is easy to solve
using the conformal flatness of \AdS5s5\ which implies that the
solution for the dilaton  
is of the form
\begin{equation}\label{conflat}{
e^{\hat \phi}  = g +  { \rho_0^4 \rho^4\over L^8} 
\left(e^{\hat \phi^{(10)}} - g \right),}
\end{equation} %
where  $\rho_0 = |y_0|$ and $e^{\hat \phi^{(10)}}$ 
is the harmonic function that appeared in the flat ten-dimensional 
case, (\ref{tendim}).
In evaluating D-instanton dominated  amplitudes 
we will only be interested in the case in which the point
$(x^\mu,y^i$) approaches the boundary ($\rho \equiv |y| 
\to 0$), in which case it is necessary to rescale the 
dilaton profile (just as it is necessary to rescale the scalar
bulk-to-bulk propagator, \cite{gkp,wittads}) so that the combination
\begin{equation}\label{classd}{
\rho^{-4} \left(e^{\hat \phi} - g \right) =   
{3 K(\alpha')^4 \over L^8 \pi^4}{  \rho_0^4\over  ((x-x_0)^2 + 
\rho_0^2)^4 }  \, , }
\end{equation} %
is  of relevance in the $\rho \to 0$ limit.     

As mentioned earlier,
the correspondence with the Yang--Mills instanton 
follows from the fact that $  \rho_0^4/ ((x-x_0)^2 + 
\rho_0^2)^4  = K_4$
is proportional to the instanton number  density, $(F_{(0)}^-)^2$, 
in the $\calN=4$ Yang--Mills theory.  Strikingly, the  scale size of the Yang--Mills
instanton is replaced by the distance $\rho_0$ of the D-instanton 
from the boundary. This is another indication of how the geometry of the
Yang--Mills theory is encoded in the IIB superstring. Note, in particular,  that 
as the D-instanton approaches the boundary $\rho_0 \to 0$, 
the expression for  $\rho^{-4} e^{\hat \phi}$ 
reduces to a $\delta$ function that corresponds to a zero-size Yang--Mills instanton.

The BPS condition implies that we can write the solution for the \RR\
scalar as
\begin{equation}\label{rrhat}{
\hat C^{(0)} = \tilde C^{(0)} + i f(x,y),}
\end{equation} %
where $\tilde C^{(0)}$ is the constant real part of the field (which
corresponds to $\theta_{_{YM}}/2\pi$) and 
\begin{equation}\label{fsoln}{
f =A -  {1\over g} + e^{-\hat \phi}.}
\end{equation} %
Since the action is independent of constant shifts of $C^{(0)}$ it does not 
depend on the arbitrary constant, $A$. 
In a manner that follows closely  the flat ten-dimensional case 
considered in the appendix of \cite{greengut}  the  action for a single 
D-instanton of charge $K$ can be written as 
\begin{equation}\label{actint}{
S_K = - {L^{10}\over {\alpha'}^4} \int  
{d\rho d^4x d^5 \omega \over \rho^5}  g^{\Lambda\Sigma} 
\nabla_\Lambda (e^{2\hat \phi} f \partial_\Sigma f),}
\end{equation} %
which reduces to an integral over the boundaries of \AdS5s5\ and the surface 
of an infinitesimal sphere centered on the D-instanton at $x=x_0$, $y=y_0$.  
With the choice $A=0$ in (\ref{fsoln}) the 
entire D-instanton action comes from the boundary of 
the infinitesimal sphere.  Substituting for $f$ from (\ref{fsoln} gives 
\begin{equation}\label{dinstact}{S_K = {2\pi |K|\over g},} 
\end{equation} %
which is the same answer as in the flat ten-dimensional case.  
On the other hand,  with the choice $A=1/g$  in (\ref{fsoln})  the  
expression (\ref{actint}) reduces to an integral over the  boundary 
at $\rho=0$  but the
total action remains the same as $S_K$ in (\ref{dinstact}).   
Remarkably,  in this case the boundary integrand  is {\it identical} 
to the action density  of the standard four-dimensional Yang--Mills instanton. 
It is crucial for this agreement that the integral (\ref{actint}) includes
an integral over the $S^5$ factor even though the classical moduli
space of the Yang--Mills instanton does not include an $S^5$.  
This is a concrete  manifestation of the holographic principle whereby
the physics of the bulk is encoded on the boundary \cite{thoofth,susskindh}.

Whereas the \AdS5s5\ metric remains unchanged by the presence of the
D-instanton in the Einstein frame it is radically altered in the
string frame where the instanton is manifested as a space-time
wormhole (as in the flat ten-dimensional case \cite{ggp}).  For finite
values of $K$  the 
dilaton becomes large in the Planck-scale neck and the classical solution 
is not reliable  in that region.   However, for very large
instanton number, the neck region becomes much larger than the Planck
scale so, by analogy with the D-brane examples studied in
\cite{maldacena}, it should be very interesting to study the
implications of the modified \AdS5s5\ geometry in the large-$K$ limit
of the large-$N$ theory.

\vskip 0.3cm
{\it Method (c).  Fluctuations around a classical D-instanton}
\vskip0.1cm

The D-instanton contribution to the amplitude with
sixteen external dilatini, $\Lambda_{\alpha}^A$, may now be 
obtained directly by
semi-classical quantization around the classical
D-instanton solution in \AdS5s5. 
  The leading instanton contribution can be determined by applying
supersymmetry transformations to the scalar field  which has an 
instanton profile given by  (\ref{classd}).   Since 
the D-instanton background breaks half the supersymmetries the
relevant transformations are those in which the supersymmetry
parameter corresponds to the  Killing  
spinors for the sixteen broken supersymmetries.  
These Killing spinors have  $U(1)$ charge 1/2  and are  defined by a 
modified version of (\ref{kilspina})
that includes   the non-trivial composite $U(1)$ connection, $Q_M$
\cite{schwarza},  that is made from  the IIB scalar field \cite{ggp}, 
\begin{equation}\label{modsixty}{
{\cal D}_M \zeta \equiv (D_M - {i\over 2} Q_M)\zeta  = {1 \over 2L} 
\gamma_M \zeta.}
\end{equation} %
Substituting the euclidean D-instanton solution into the expression
for the composite connection gives 
\begin{equation}\label{Qdef}{
Q_M =  {i\over 2} e^{-\hat \phi} \partial_M e^{ \hat \phi} }
\end{equation} %
with $\hat \phi$ defined
by (\ref{classd}).  The solution of (\ref{modsixty}) is
\begin{equation}\label{spinkil}{
\zeta_{\pm} = 
e^{-\hat\phi/4} {z_{M} \gamma^{\hat M} \over \sqrt{\rho_0}} 
\zeta^{(0)}_{\pm}, }
\end{equation} %
where $\zeta^{(0)}_{\pm}$ is a constant spinor satisfying 
$\gamma_5 \zeta^{(0)}_{\pm} = \pm \zeta^{(0)}_{\pm}$.  
 
The sixteen broken supersymmetry transformations associated
with $\zeta^{(0)}_{-}$  give rise to the  dilatino
zero-modes,
\begin{equation}\label{dilfd}{
\Lambda_{(0)} = \delta \Lambda = (\gamma^M \hat P_M) \zeta_{-}, }
\end{equation} %
where $\hat P_M$ is the expression for $P_M \equiv i\partial_M 
\tau^*/2\tau_2$  in the D-instanton background
\cite{greengut},
\begin{equation}\label{pmdef}{
 \hat P_M =   e^{-\hat \phi} \partial_M  e^{\hat \phi}.} 
\end{equation} %
Using the Killing spinor equation and the D-instanton equation
${\cal D}^M \hat P_M=0$  
it is easy to check (recalling that $P_M$ has $U(1)$ charge 2) that 
\begin{equation}\label{lambdeq}{
\gamma^M {\cal D}_M \Lambda_{(0)} = - {3 
\over 2L} \Lambda_{(0)},}
\end{equation} %
so that $\Lambda_{(0)}$ is a solution of the
appropriate massive Dirac equation.  
We will be interested in amplitudes with external states located on the  
boundary, in which case  we may use the fact that for $\rho \sim 0$,
\begin{equation}\label{farapp}{
\hat P_M \sim{1\over  g} \partial_M e^{\hat \phi}}
\end{equation} %
in (\ref{dilfd}),  which  leads to 
\begin{equation}\label{lamzer}{
\Lambda_{(0)} \sim  {4\over g} (e^{\hat \phi} -  g) \zeta_{-}\, .  }
\end{equation} %
This means that near $\rho=0$   the dilatino profile in the D-instanton 
background   is proportional to $\rho^{4} K_4(x_0,\rho_0;x, 0)$.  

As a result the leading contribution to the sixteen-dilatino  
amplitude again reproduces the corresponding sixteen-current 
correlator in ${\cal N}=4$  supersymmetric  
Yang--Mills theory. 
Explicitly,  the D-instanton approximation to the  amplitude with  
sixteen external dilatini, ${\Lambda_\alpha}^{A}$, at points on the 
$\rho=0$ 
boundary is (up to an overall constant factor)
\begin{eqnarray}
&&\langle \prod_{p=1}^{16} 
\Lambda^{A_{p}}_{\alpha_{p}}(x_{p},0) \rangle_J 
 =  (const.)\, g^{-12} e^{-2\pi K ({1\over g} +  i  C^{(0)}) } 
V_{S^5} \,  \int \frac{d^4x_0 d\rho_0}{\rho_0^{5}}  
\int d^{16}\zeta^{(0)}_{-}\nn  \\
& &    \prod_{p=1}^{16} \left[ K_{4} 
\left(x_0,\rho_{0};x_{p}\right) {1\over\sqrt{\rho_0}}   
\left(\rho_0 \eta^{A_{p}}_{\alpha_{p}}+{(x_{p}-x_{0})}_{\mu}  
\sigma^{\mu}_{\alpha_{p}{\dot \alpha}_{p}}  
{\overline \xi}^{{\dot \alpha}_{p}A_{p}} \right)
J_{\Lambda}(x_p) \right].\label{lsxdilfd}
\end{eqnarray} %
The overall factor of $(const.)\, g^{-12}$ in this expression should
be derived from the    normalization of the bosonic and
fermionic zero modes but it has been inserted by hand.  
Up to this  overall constant factor, the  amplitude (\ref{lsxdilfd})   
agrees with  (\ref{lsxsym}) and therefore with
(\ref{sourcelsx}).   
 
In similar manner the instanton profiles of all the fields in the
supergravity multiplet follow by applying the broken supersymmetries
to $P_M$ any number of times, just as they do in the flat
ten-dimensional case \cite{greengut}.  The single D-instanton
contributions to any correlation function can then be determined.

\subsection{Instantons in the limit of large $N$}

We now return briefly to the issue of comparing the D-instanton and
Yang--Mills instanton effects for gauge groups $SU(N)$ in the limit
$N\to \infty$.   The  string computation was valid only in the region
of large $g^2 N$ since we kept only the first term in the $\alpha'$
expansion that contributes to 
$\calR^4$.  
 In the classical 
$\calN=4$ Yang--Mills theory with a gauge group $SU(N)$ 
an instanton background
has $8N$ fermion zero modes, $2N$ for each of the $4$ adjoint Weyl
fermions.   It is a special feature of the $N=2$ case that the number
of fermionic zero 
modes coincides with the number of broken supersymmetries but for 
larger values of $N$ there are more fermionic zero modes in the
classical D-instanton background  than the number of broken
supersymmetries.  
However, most  of these modes receive perturbative corrections that 
gives them nonzero energy and the only zero
modes which are protected  are the sixteen which correspond to 
 broken supersymmetries.
In other words,  most of the classical zero modes are not protected by 
symmetries and do not cause the correlation functions  that we have been
discussing (such as the $\hat \Lambda^{16}$ correlator) to vanish.  

The measure for a  Yang--Mills instanton for the group $SU(N)$
was considered a long time ago in nonsupersymmetric theories.
The measure is defined as the jacobian for the transformation from 
bosonic
zero-modes  to the collective coordinates and for pure  $SU(N)$
Yang--Mills it  has the form \cite{bernard}
\begin{equation}\label{nnon}{ {(2\pi)^{4 N} \over 
(N-1)!(N-2)!}\left({\rho_0 \over g_{_{YM}}}\right)^{4N} 
{d\rho_0 d^4x_0 \over \rho_0^5}}
\end{equation} %
(where $\rho_0$ is an arbitrary scale),
which accounts for the number of ways of embedding the $SU(2)$
instanton solution in $SU(N)$
and  the coupling constant dependence is  due to the eight bosonic
zero modes \cite{thoofta}.
In supersymmetric theories this coefficient is  modified due to
the gaussian fluctuations of the other fields.  
In \cite{doreyone} the measure for a single  Yang--Mills instanton in
$SU(N)$ $\calN=4$ supersymmetric Yang--Mills was computed, including
the integration over  fermionic zero modes.
The result is that the factor 
(\ref{nnon}) is replaced by  (dropping irrelevant constants)
\begin{equation}\label{nonag}{\gy^8   2^{-2N} {(2N-2)! \over (N-1)! (N-2)!},}
\end{equation} %
so that the overall power of the coupling is the same as in the
$SU(2)$ case with $\calN=4$ supersymmetry that we considered earlier
but there is now an explicit $N$-dependent coefficient.
The  large-$N$ limit of this expression can be extracted by 
using Stirling's approximation for the
gamma functions with the result that the 
overall measure at large $N$ is proportional to
\begin{equation}\label{propn}{\gy^8 N^{1/2} \sim {\alpha'}^{-1} \tau_2^{-7/2},}
\end{equation} %
which agrees with the $K=1$ case of the  expression  
deduced  from the D-instanton measure in (\ref{coeform}).  

The large-$N$ limit of the contribution of  $K$
instantons to the protected  processes that we have been considering 
was derived in two  interesting recent papers 
\cite{doreytwo,doreythree}.  Making extensive use of the ADHM
construction this paper showed that the large-$N$ limit
is dominated by a saddle point which constrains the $K$ $SU(2)$ instantons to
a very special point in the multi-instanton moduli space.  
This is a point in the subspace of the full moduli space at  which
the instantons are in $K$ commuting SU(2) subgroups of SU(N).  
Clearly, this can only be precisely true for $N\to \infty$.  
For finite $N$ such configurations may dominate for $K<<N$ but when 
$K> N/2$ no such configurations exist.  Furthermore, it was shown in
\cite{doreytwo} that at  the dominant saddle point the instantons  
coincide in space  and have equal scale sizes --- they behave as a single charge-$K$ instanton.
Even more impressively, the evaluation of the large-$N$ limit of the
fluctuations of the fields around the coincident instantons
induces five new  effective moduli that parameterize the
five-sphere.  Finally, the overall coefficient is identified in
\cite{doreytwo,doreythree} with the partition
function of the zero-dimensional matrix model, $Z_K$.
The result is  that  the effective moduli space of the leading contribution of  $K$
Yang--Mills instantons to the protected processes is \AdS5s5.   The resulting 
measure coincides precisely with the leading term in the 
D-instanton measure in (\ref{coeform}) for general $K\ge 1$.
 
These results apply to the instanton corrections to the protected
processes at leading power of
the Yang--Mills coupling constant.  However, the full $\Sl2$-invariant
structure of the Yang--Mills correlation functions  suggested by the
AdS/CFT correspondence in \cite{bg} is  very much richer. We have
seen that in the limit $\gy^2 N\to \infty$ ($\alpha'<< L^2$) not only
must each instanton contribution to a correlation 
function  be accompanied by an infinite number of
perturbative corrections but there are vital non-instanton terms which
correspond to the tree-level and one-loop contributions in the string
theory.  These are the terms in (\ref{fourtenstwo}) that have fractional
powers of $\gy$ and  are not visible in weakly-coupled Yang--Mills
theory and therefore cannot be obtained by matching the semiclassical
approximations of the bulk and boundary theories.  Explaining the
nature of these terms from the Yang--Mills perspective would
appear to
be of some interest.

\vskip 0.4cm
{\it Acknowledgments}: 

I am very grateful to Pierre Vanhove and Constantin Bachas for useful
comments concerning the material in this article. 

\begin{appendix}

\section{Derivative expansion of the string tree amplitude}

The  tree amplitude for four-graviton scattering
in either of the type II superstring theories is given by
\begin{eqnarray}\label{treeamps}
A_4^{tree}   & = &{\alpha'}^{-4} e^{-2\phi}{\tilde K\over stu}
{\Gamma(1-\alpha' s)\Gamma(1-\alpha' t)\Gamma(1-\alpha' u) \over
\Gamma(1+ \alpha' s)\Gamma(1 + \alpha' t)\Gamma(1  + \alpha' u)}\\
& = &{\alpha'}^{-4} e^{-2\phi} {\tilde K  \over stu} \exp\left(\sum_{n=1}^\infty
{{\alpha'}^{2n+1} 2 \zeta(2n+1) \over 2n+1} (s^{2n+1} + t^{2n+1} +
u^{2n+1})\right),\nn
\end{eqnarray} %
where the second equality follows from elementary properties of
$\Gamma$ functions.  The overall kinematic factor $\tilde K$, defined
in (\ref{kinematic}), is the linearized expansion of $\calR^4$ and $s,t$, $u$
are the Mandelstam invariants  in string frame (and satisfy $s+t+u
=0$).   These are related to the Mandelstam invariants in  M-theory
coordinates (which we will denote $S$, $T$ and $U$)  by 
\begin{equation}\label{mandelstring}{
s =  {S\over  R_{11}}, \qquad t =  { T\over  R_{11}}, \qquad u = {U \over  R_{11}}.}
\end{equation} %
Every term in the exponent of (\ref{treeamps}) can be expressed as a
polynomial in $s$ and $t$ multiplied by $stu$, as can be seen from
the identity,
\begin{equation}\label{factout}{
 s^{2n+1} + t^{2n+1} + u^{2n+1} = stu\left[
\sum_{r=1}^n \sum_{q=0}^{2n-2r} \right.    {(2n+1)! \over r!
(2n+1-r)!}   (-1)^q s^{2n-1-r-q} t^{r+q-1}   \bigg],}
\end{equation} %
where $n\ge 1$.  This means that the massless poles only contribute to
the first term in the expansion of the exponential.

When expressed in terms of the Mandelstam invariants in the M-theory metric
the  expression (\ref{treeamps})  has the low-energy expansion,
\begin{eqnarray}\label{lowtree}
& & A_4^{tree}  \sim   \tilde K  \left( {1\over STU}  +
{2\zeta(3)\over R_{11}^{3}}   +{ 2 \zeta(5)\over R_{11}^{5}}(S^2+ST+T^2)
+ {2 \zeta(3)^2\over  R_{11}^{6}}STU    \right.\nn\\
& & \left. +
{2\zeta(7)\over R_{11}^{7}}(S^4  + 2S^3T +3 S^2T^2 +2 S T^3 + T^4)
  + { 2 \zeta(3)
\zeta(5)\over R_{11}^8} STU(S^2+ST + T^2) \right. \nn\\ 
& & \left. + {2\zeta(9)\over R_{11}^9}(S^6 + 4 S^5T + \cdots + T^6)   +
{4\over 3 R_{11}^9}\zeta(3)^3 S^2T^2U^2 + \cdots \right).
\end{eqnarray} %
The first term in this expansion combines with the kinematic factor,
$\tilde K$, to give the tree-level  amplitude that is
described by  the Einstein--Hilbert action of eleven-dimensional
supergravity. 
The term with coefficient $\zeta(3)$ can be identified (section 5 of these
notes) with a one-loop 
effect in four-graviton scattering in eleven-dimensional supergravity 
compactified on a circle  of radius $R_{11} =
e^{2\phi/3}$.  This explains the corresponding term in (\ref{treeamps})  when
$A_4^{tree}$   is  interpreted as  a type IIA  tree amplitude. 
When (\ref{treeamps}) 
is interpreted as a  IIB tree  amplitude the $\zeta(3)$ term  comes from the
zero-volume limit of M theory on $T^2$.  Similarly, 
there is some evidence \cite{greenmult} that the
term 
with coefficient $\zeta(5)$ in (\ref{lowtree}) can be identified with 
 a two-loop effect in eleven-dimensional supergravity.
The higher-order terms in  (\ref{lowtree}) give  terms in the effective
action with
derivatives acting on $\calR^4$ and other higher-derivative terms.

\section{Modular covariant derivatives.}

The various coefficient functions  in the effective action are
$(w,\hat w)$ forms, where $w$ refers to the holomorphic modular weight and
$\hat w$ to the anti-holomorphic  modular weight.  A nonholomorphic
modular form $F^{(w,\hat w)}$ transforms as,
\begin{equation}\label{ftranss}{F^{(w,\hat w)} \to F^{(w,\hat w)}\, (c\tau + d)^w
(c\bar \tau + d)^{\hat w},}
\end{equation} %
under the $\Sl2$ transformation (\ref{slact}).
Equation (\ref{ftranss})  describes a $U(1)$ transformation when $\hat w = -w$.

The modular covariant derivative,
\begin{equation}\label{derdef}{{\cal D}_w =i \left( {\partial \over \partial \tau} - i {w \over
2\tau_2}\right),}
\end{equation} %
maps  $F^{(w,\hat w)}$ into $F^{(w+2,\hat w)}$  while the
anti-holomorphic covariant derivative,
$\bar {\cal D}_{\hat w} =
{\cal D}^*_{\hat w}$,  maps  $F^{(w,\hat w)}$ into $F^{(w,\hat w+2)}$.
It is more convenient
for our purposes to define the covariant derivatives,

\begin{equation}\label{dertdef}{D_w = \tau_2 {\cal D} = i \left(\tau_2 {\partial \over
 \partial \tau}  - i {w \over
2} \right), \qquad \bar D_{\hat w} =
\tau_2 \bar {\cal D} = -i \left(\tau_2 {\partial \over
 \partial\bar  \tau}  + i {\hat w \over
2} \right) }
\end{equation} %
which change the $U(1)$ charge of $F$ by two units,
\begin{equation}\label{derivprop}{D_w F^{(w,\hat w)} = F^{(w+1,\hat w -1)}, \qquad
\bar D_{\hat w} F^{(w,\hat w)} = F^{(w-1,\hat w+1)}.}
\end{equation} %

The Laplace operator on the fundamental domain of $\Sl2$ is defined to be,
\begin{equation}\label{lapon}{\nabla^2_0 \equiv
\nabla^2 = 4 \tau_2^2 {\partial \over \partial \tau } {\partial \over
\partial \bar \tau} ,}
\end{equation} %
when acting on $(0,0)$ forms.  More generally, we shall be interested
in the Laplacian acting on $(w, - w)$ forms.   There are two such
Laplacians which are  defined by,
\begin{equation}\label{lapthree}{\nabla^2_{(-)w} = 4 D_{w-1} \bar
D_{-w} = 4 \tau_2^2{\partial \over \partial \tau } {\partial \over
\partial \bar \tau} - 2 iw \tau_2 \left( {\partial \over  \partial \tau } + {\partial
\over  \partial \bar \tau}\right) -  w(w-1),}
\end{equation} %
and
\begin{eqnarray}\label{lapagain}\nabla^2_{(+) \, w} & = &
4 \bar D_{-w-1} D_w =  4 \tau_2^2
{\partial \over \partial \tau } {\partial \over
\partial \bar \tau} - 2 iw\tau_2  \left( {\partial \over  \partial \tau } +
{\partial \over  \partial \bar \tau}\right) -  w(w + 1), \nn\\
& = & \nabla^2_{(-)w} - 2w .
\end{eqnarray} %

Now consider a $(w,-w)$ form that is an eigenfunction of the Laplace operator
$\nabla^2_{(-)\, w}$ with   eigenvalue $\sigma_w$,
\begin{equation}\label{eigenf}{\nabla^2_{(-)\, w}\, F^{(w,-w)}
= 4D_{w-1} \, \bar D_{-w} \, F^{(w,-w)} = \sigma_w \, F^{(w,-w)}.}
\end{equation} %
Applying $\bar D_{-w}$ to this equation gives,
\begin{equation}\label{neweq}{\nabla^2_{(+)w-1}\, F^{(w-1,-w+1)} = \sigma_w \,
F^{(w-1,-w+1)}.}
\end{equation} %
It is also easy to see that,
\begin{eqnarray}\label{eagers}\nabla^2_{(-)\, w-1}\, F^{(w-1,-w+1)} & = &
4D_{w-2} \, \bar D_{-w+1} \, F^{(w-1,-w+1)},\nn \\ 
& = & (\sigma_w + 2w-2)
\, F^{(w-1,-w+1)}
\end{eqnarray} %
where $F^{(w-1,-w+1)} = \bar D_{-w}\, F^{(w,-w)}$.  Repeating this for
$m$ steps gives
\begin{eqnarray}\label{eigenm}\nabla^2_{(-)\, w-m}\, F^{(w-m,-w+m)}
& = & 4D_{w-m-1} \, \bar D_{-w+m+1} \, F^{(w-m,-w+m)},\nn \\
& = & (\sigma_w + 2mw-m^2-m) \, F^{(w-m,-w+m)}.
\end{eqnarray} %
Similarly,
\begin{equation}\label{eigendu}{\nabla^2_{(+)\, w-m}\, F^{(w-m,-w+m)}
= (\sigma_w + 2mw -2w - m^2 +m) \, F^{(w-m,-w+m)}.}
\end{equation} %
This relation between eigenvalue equations will be useful in analyzing
the  equations that are
satisfied by the modular forms that enter in $S^{(3)}$.

In section 4 it is proved that $f^{(12,-12)}$ satisfies
\begin{equation}\label{laptwelve}
{\nabla^2_{(-)\, 12} f^{(12,-12)} = \left( -132 +
{3\over 4}\right)  f^{(12,-12)}.}
\end{equation} %
This is (\ref{eigenm}) with $w=12$, $m=0$ and $\sigma_{12} = -132+3/4$.
It is also argued in section 3 that 
$f^{(12,-12)}$ is related to $f^{(0,0)}$ by  
\begin{equation}\label{twelvedef}{f^{(12,-12)} =  
D^{12} f^{(0,0)}\equiv D_{11} \cdots D_1 D_0 f^{(0,0)}.   }
\end{equation} %
This equation together with (\ref{laptwelve}) imply that 
$f^{(0,0)}$ satisfies (\ref{lapzero}). 

More generally, let us denote a  
solution of the scalar Laplace
equation  with eigenvalue $\sigma = s(s-1) >1/4$ by
$E_s(\tau)$ \cite{terras},
\begin{equation}\label{eslap}{\nabla^2 E_s = s(s-1) \, E_s.}
\end{equation} %
We can express $E_s (\tau)$ in terms of the nonholomorphic Eisenstein series,
\begin{equation}\label{esdef}{E_s (\tau) = {1\over 2} \tau_2^s \, \sum_{(m,n)=1}|m\tau+n|^{-2s}, }
\end{equation} %
where $(m,n)=1$  indicates that $m$ and $n$ are coprime. 
The eigenfunctions $E_s (\tau)$ are singled out by their power law behavior 
near the boundary of the moduli space, which agrees with the known tree-level 
and perturbative contributions to the interactions that we are  considering.  
The solution to the Laplace equation (\ref{eslap}) is unique for a given $s$ assuming 
that $E_s$ is a modular function.  Therefore, if we assume that
$f^{(0,0)}$ is well-defined on the 
fundamental domain of $\Sl2$ the solution is unique and its behaviour near the 
$\tau_2\to \infty$ boundary agrees with the known tree-level and 
one-loop contributions.  
It follows from (\ref{twelvedef}) that 
$f^{(12,-12)}$ is also determined uniquely by its Laplace equation (\ref{laptwelve}).  
In this case there are no string calculations with which to compare the perturbative terms.qyn

\section{Some properties of Type IIB Supergravity}

\subsection{Spinors and gamma matrices }

The spinors that enter into the IIB theory are complex Weyl spinors.
The gravitino $\psi_\mu$  and dilatino $\lambda$ have opposite chiralities and the
supersymmetry parameter $\epsilon$
 has the same chirality as the gravitino.  The
complex conjugate of the product of a pair of spinors is defined by
\begin{equation}\label{compconj}{ (\lambda_a\, \rho_b)^* = - \lambda_a^*\, \rho_b^*.}
\end{equation} %
The conjugate of any spinor is defined by,
$\bar \lambda = \lambda^* \gamma^0$.
We will choose our metric to be space-like and the $\gamma$ matrices to be real
and satisfy the Clifford algebra,
\begin{equation}\label{cliffalg}{ \left\{ \gamma^\mu, \gamma^\nu \right\} = 2 \eta^{\mu\nu}.}
\end{equation} %
Noting that, 
\begin{equation}\label{gammzer}{\gamma^0 \gamma^\mu = - (\gamma^\mu)^T \gamma^0,}
\end{equation} %
it follows that 
two complex  chiral spinors of the same chirality, $\lambda_1$ and
$\lambda_2$,  satisfy the relations,
\begin{eqnarray}\label{lamprops}\bar\lambda_1\gamma^\mu \lambda_2
& = & - \bar\lambda_2^* \gamma^\mu
\lambda_1^*, \nn\\
\bar\lambda_1\gamma^{\mu\nu\rho} \lambda_2 & = &
\, \bar\lambda_2^* \gamma^{\mu\nu\rho} \lambda_1^*, \\
\bar\lambda_1\gamma^{\rho_1\dots \rho_5} \lambda_2 & = &
 - \bar\lambda_2^* \gamma^{\rho_1 \dots \rho_5} \lambda_1^*,\nn
\end{eqnarray} %
while two chiral spinors of opposite chiralities, $\lambda$ and $\epsilon$,
satisfy,
\begin{eqnarray}\label{newlams}
\bar\lambda \, \epsilon & = &\, \bar\epsilon^*\, \lambda^*,\nn\\
\bar\lambda \gamma^{\rho_1\rho_2} \epsilon  & = &  -\bar\epsilon^*
\gamma^{\rho_1\rho_2} \lambda^*,\\
\bar\lambda \gamma^{\rho_1\rho_2\rho_3\rho_4} \epsilon & = &
\bar\epsilon^*\gamma^{\rho_1\rho_2\rho_3\rho_4} \lambda^*.\nn
\end{eqnarray} %
The Fierz identity for ten-dimensional complex Weyl spinors can be
expressed as,
\begin{equation}\label{fierz}{\lambda_1^{a} \bar \lambda_2^{ b} = -{1\over 16} \bar \lambda_2
\gamma_\mu \lambda_1\, \gamma^\mu_{ab} + {1\over 96} \bar \lambda_2
\gamma_{\mu\nu\rho} \lambda_1\,  \gamma^{\mu\nu\rho}_{ab} - {1\over 3840}
\bar\lambda_2 \gamma_{\rho_1 \cdots\rho_5}\lambda_1\,  \gamma^{\rho_1\cdots
\rho_5}_{ab},}
\end{equation} %
where $\lambda_1$ and $\lambda_2$ are two chiral spinors of the same chirality.

An additional useful identity is,
\begin{equation}\label{lamidens}{\gamma^{\rho_1\dots\rho_5}\lambda_1\,
\bar\lambda_2\gamma_{\rho_1\dots \rho_5} \lambda_3 =0,}
\end{equation} %
where $\lambda_1$, $\lambda_2$ and $\lambda_3$ are three chiral spinors of the
same chirality.

Some gamma matrix identities that are useful in proving the various
relationships in the text are,
\begin{eqnarray}\label{appone}\tr(\gamma_{\mu\nu\rho} \gamma^{\rho_1\rho_2\rho_3}) & = & -16
\left(\delta^{\rho_1}_\mu \delta^{\rho_2}_\nu\delta^{\rho_3}_\rho
- \delta^{\rho_2}_\mu \delta^{\rho_1}_\nu\delta^{\rho_3}_\rho
+ \delta^{\rho_2}_\mu \delta^{\rho_3}_\nu\delta^{\rho_1}_\rho \right.\nn \\
& & \left. -\delta^{\rho_3}_\mu \delta^{\rho_2}_\nu\delta^{\rho_1}_\rho
+\delta^{\rho_3}_\mu \delta^{\rho_1}_\nu\delta^{\rho_2}_\rho
-\delta^{\rho_1}_\mu \delta^{\rho_3}_\nu\delta^{\rho_2}_\rho
\right).
\end{eqnarray} %
\begin{eqnarray}\label{appfive}\gamma^\mu\, \gamma_\sigma\, \gamma_\mu & = & -8
\gamma_\sigma,\nn\\
\gamma^\mu \,\gamma_{\sigma_1\sigma_2\sigma_3}\,
\gamma_\mu  & = &  -4 \gamma_{\sigma_1\sigma_2\sigma_3},\nn\\
\gamma^\mu
\gamma_{\sigma_1\dots\sigma_5}
  \gamma_\mu &  = & \, 0,\nn\\
\gamma^{\mu\nu\rho} \, \gamma_\sigma\, \gamma_{\mu\nu\rho}
& = &   - 288 \gamma_{\sigma},\\
\gamma^{\mu\nu\rho} \,
\gamma_{\sigma_1\sigma_2\sigma_3}\, \gamma_{\mu\nu\rho}
& = &  - 48 \gamma_{\sigma_1\sigma_2\sigma_3},\nn\\
\gamma^{\mu\nu\rho} \,
\gamma_{\sigma_1\dots\sigma_5}\, \gamma_{\mu\nu\rho}
& = & -14\gamma_{\sigma_1\dots\sigma_5} .\nn
\end{eqnarray} %

In section 4  the first two terms  in parentheses  on the
right-hand-side of  (\ref{nonpert}) are considered in detail.  The
precise notation used in those terms uses the following  definitions 
\begin{equation}\label{defnot}{(\lambda^r)_{a_{r+1} \cdots a_{16}} \equiv {1\over r!}
\epsilon_{a_1\cdots a_{16}} \lambda^{a_1} \dots \lambda^{a_r}, }
\end{equation} %
so that,
\begin{equation}\label{definv}{\lambda^{16} =
{1\over 16!}\epsilon_{a_1 \dots a_{16}}
\lambda^{a_1} \dots \lambda^{a_{16}},}
\end{equation} %
and
\begin{eqnarray}\label{ghatlam}\hat G \lambda^{14} & \equiv  & \hat  G_{\mu\nu\rho}
(\gamma^{\mu\nu\rho}\gamma^0)_{a_{15} a_{16}}  (\lambda^{14})_{a_{15} a_{16}},
\nn\\ & = &
{1\over 14!}\hat G_{\mu\nu\rho}(\gamma^{\mu\nu\rho}\gamma^0)_{a_{15}a_{16}}\,
\epsilon_{a_1 \dots a_{16}}   \lambda^{a_1} \dots
\lambda^{a_{14}}.
\end{eqnarray} %
The following   identities follow very simply from (\ref{defnot}),
\begin{eqnarray}\label{identb}(\lambda^{14})_{ab}\, \lambda^c & = &
(\lambda^{15})_b \, \delta_{a}^c -
(\lambda^{15})_a\, \delta_{b}^c, \nn\\
(\lambda^{15})_a\, \lambda^b & = & \delta_{a}^b \,\lambda^{16}, \\
 (\lambda)^{15}_a \, \lambda^a & = & 16 \lambda^{16},\nn
\end{eqnarray} %
and
\begin{equation}\label{identa}{ (\lambda^{14})_{ab}\, \lambda_c \lambda_d =
\lambda^{16}\, (\delta_{ac}\delta_{bd} - \delta_{ad} \delta_{bc}).}
\end{equation} %

\subsection{The fields and their supersymmetry transformations}

Here we will review various features of type IIB supergravity that
are useful in the body of the paper.  Most of this material can be found in
\cite{schwarzwest,schwarza} in a form that is adapted to the field definitions
in which the global symmetry is $SU(1,1)$ and the scalar fields
parameterize the coset space $SU(1,1)/U(1)$, which is the Poincar\'e
disk.  It is simple to
transform this to our parameterization in which the global symmetry is
$SL(2,\IR)$
and the scalars parameterize the coset space $SL(2,\IR)/U(1)$,
or the upper half plane.

The theory is then defined in terms of the following  fields:
the scalar fields can be parameterized by the frame field,
\begin{equation}\label{vdef}{V \equiv \pmatrix{V^1_- & V^1_+ \cr
                            V^2_- & V^2_+ \cr} = {1\over \sqrt{-2i\tau_2}}
\pmatrix{\bar \tau e^{-i\phi} & \tau e^{i\phi} \cr
                e^{-i\phi} & e^{i\phi} \cr},}
\end{equation} %
where $V^\alpha_\pm$ ($\alpha =1,2$) is a $\Sl2$ matrix that
transforms from the left by the global
$SL(2,\IR)$ and from the right by the local $U(1)$. Note that we are using a
complex basis for convenience.
A general transformation is then written as,
\begin{equation}\label{vtrans}{(V_+^\alpha,V_-^\alpha)  \to U^\alpha_{\ \beta}
\left( V^\beta_+ e^{i\Sigma},  V^\beta_- e^{-i\Sigma}\right),}
\end{equation} %
where $U$ is a  $SL(2,\IR)$ matrix and $\Sigma$ is the $U(1)$ phase.
An appropriate choice of $\Sigma$ fixes the gauge and
eliminates the scalar field $\phi$.  We will make the gauge  choice
$\phi=0$.   Since this gauge is not maintained by generic symmetry  transformations,
it is necessary to compensate a symmetry transformation with an appropriate
local $U(1)$ tranformation to maintain the gauge.   In particular, the
local supersymmetry transformations require compensating  local
$U(1)$ transformations.  The supersymmetry and $U(1)$
transformations of $V_-^\alpha$ are given by,
\begin{equation}\label{susyv}{\delta^{(0)} V_-^\alpha = i V_+^\alpha \bar \epsilon \lambda^* - i
\Sigma V_-^\alpha.}
\end{equation} %
This choice ensures that
the gauge $\phi=0$ is maintained if a local supersymmetry transformation is
accompanied by a $U(1)$ transformation with parameter,
\begin{equation}\label{sigtrans}{\Sigma = {1\over 2} (\bar \epsilon \lambda^* - \bar\epsilon^*
\lambda).}
\end{equation} %

The $SL(2,\IR)$ singlet expression,
\begin{equation}\label{qdeff}{Q_\mu =-i \epsilon_{\alpha\beta} \, V^\alpha_+
\partial_\mu V^\beta_- , }
\end{equation} %
is the composite $U(1)$ connection and transforms as $Q \to Q + \partial_\mu
\Sigma$ under infinitesimal local $U(1)$ transformations, while the
$SL(2,\IR)$
singlet expression
\begin{equation}\label{pdef}{P_\mu = - \epsilon_{\alpha\beta}V_+^\alpha \partial_\mu V^\beta_-,}
\end{equation} %
transforms with $U(1)$ charge $u_{P}= 2$.  In the gauge $\phi=0$, the
expression
for $P_\mu$ takes the simple form,
\begin{equation}\label{pfix}{P_\mu = {i\over 2} {\partial_\mu \tau \over \tau_2},}
\end{equation} %
while
\begin{equation}\label{qdef}{Q_\mu =  -i \epsilon_{ab} V_+^a\partial_\mu V_-^b= -{1\over 2}
{\partial_\mu\rho_1\over\rho_2}.}
\end{equation} %

The fermions comprise the complex chiral gravitino, $\psi^a_\mu$, which has
$U(1)$ charge $u_\psi=1/2$, and the
dilatino, $\lambda^a$, with $U(1)$ charge $u_\lambda = 3/2$. These two fields
have opposite chiralities.
The graviton is a $U(1)$ and $SL(2,\IR)$ singlet as is the antisymmetric
fourth-rank potential, $C^{(4)}$,
which has a field strength $F_5 = d C^{(4)}$.  As is well known, this field
strength has an equation of motion that is expressed by the self-duality
condition $F_5 = * F_5$, which cannot be obtained
from a globally well-defined Lagrangian.  For this reason, our
considerations are restricted to statements concerning the on-shell properties
of the theory where the
fields satisfy the equations of motion.

The two antisymmetric second-rank potentials,  $B_{\mu\nu}$ and
$C^{(2)}_{\mu\nu}$, have   field strengths $F^1$ ($NS\otimes NS$)  and $F^2$
($R\otimes R$)
that form an  $SL(2,\IR)$ doublet, $F^\alpha$.  It is very natural to package
them into the $SL(2,\IR)$ singlet fields,
\begin{equation}\label{gdef}{G = - \epsilon_{\alpha\beta} V_+^\alpha F^\beta ,
\qquad G^* =-\epsilon_{\alpha\beta} V_-^\alpha F^\beta,}
\end{equation} %
which carry $U(1)$ charges $u_G =+1$ and $u_{G^*}= -1$, respectively.

In a fixed $U(1)$  gauge,  a global $SL(2,\IR)$ transformation which acts on
$\tau$ by
\begin{equation}\label{slactt}{\tau \to {a\tau + b\over c\tau + d}, }
\end{equation} %
with $ad-bc=1$, induces
a $U(1)$ transformation on the fields that depends on their charge.
Thus, a field $\Phi$ with $U(1)$ charge $u_\Phi$ transforms as,
\begin{equation}\label{uonep}{\Phi \to \Phi \, \left({c\bar \tau + d \over c\tau +
d}\right)^{u_\Phi/2}.}
\end{equation} %
The higher derivative terms of interest to us only repect the
$\Sl2$ subgroup of $SL(2,\IR)$ for which $a,b,c,d$ are integers and
the continuous $U(1)$ symmetry is broken.

The supersymmetry of the action is naturally described in terms of combinations
of bosonic fields and fermion bilinears which are \lq supercovariant', 
which means that they 
do not contain derivatives of the supersymmetry parameter $\epsilon$ in
their transformations.  These combinations are,
\begin{eqnarray}\label{supcov}\hat G_{\mu\nu\rho} & = & G_{\mu\nu\rho} - 3 \bar\psi
_{[\mu}\gamma_{\nu\rho]}\lambda - 6i \bar\psi^*_{[\mu} \gamma_\nu \psi_{\rho]},\nn\\
\hat P_\mu & = & P_\mu - \bar \psi_\mu^* \lambda, \\
\hat F_{5\, \mu_1\dots,\mu_5} & =  & F_{5\, \mu_1\dots,\mu_5}
-5 \bar\psi_{[\mu_1} \gamma_{\mu_2\mu_3\mu_4} \psi_{\mu_5]}-
{1\over 16} \bar \lambda \gamma_{\mu_1 \dots \mu_5} \lambda .\nn
\end{eqnarray} %

We will now present the lowest-order supersymmetry transformations, suitably
adapted from those given in \cite{schwarza} to the $SL(2,\IR)$
parameterization. From (\ref{susyv}) and (\ref{sigtrans}), it follows that
\begin{equation}\label{tautrans}{\delta^{(0)} \tau = 2\tau_2\bar \epsilon^* \lambda,
\qquad \delta^{(0)}\bar \tau = -2\tau_2\bar \epsilon \lambda^*.}
\end{equation} %
It follows from the definition of $Q_\mu$ and the transformations of
$\tau$ and $\bar \tau$ that
\begin{equation}\label{qtrans}{\delta^{(0)} Q_\mu = - \bar\epsilon \lambda^* \, P_\mu \,+ \ c.c.}
\end{equation} %
Also, the supersymmetry transformation of the zehnbein is given by,
\begin{equation}\label{zehntrans}{\delta^{(0)} e^m_\mu = i(\bar \epsilon \gamma^m \psi_\mu +
\bar\epsilon^* \gamma^m \psi_\mu^*).}
\end{equation} %
The transformation of the dilatino is given, in the fixed $U(1)$
gauge, by
\begin{eqnarray}\label{lamtran}\delta^{(0)} \lambda & = & i\gamma^\mu
\epsilon^* \,\hat P_\mu - {1\over 24} i \gamma^{\mu\nu\rho}\epsilon
\hat G_{\mu\nu\rho} + \delta^{(0)}_\Sigma \lambda\nn\\
& = & i\gamma^\mu\epsilon^*\,\hat P_\mu +{i\over 8}  \gamma^{\mu\nu\tau}
\epsilon  \, \left(\bar\psi_{[\mu}\gamma_{\nu\tau]}
\lambda\right)  - i\gamma^\mu \epsilon^*\,
(\bar \psi_\mu^* \lambda) + \delta^{(0)}_\Sigma \lambda + \dots ,
\end{eqnarray} %
where we have only kept the terms that are needed in the body of this paper in
the second line. The $\delta_\Sigma$ arises from the compensating $U(1)$
gauge transformation,
\begin{equation}\label{compenlam}{\delta^{(0)}_\Sigma \lambda_a = {3\over 2}i \Sigma\, \lambda_a
={3\over 4} i\lambda_a
(\bar\epsilon \lambda^*)  - {3\over 4} i
\lambda_a (\bar\epsilon^* \lambda) .}
\end{equation} %
The gravitino transformation is given by,
\begin{eqnarray}\label{vartino}\delta^{(0)} \psi_\mu  & = & D_\mu \epsilon +
{1\over 480}i \gamma^{\rho_1\dots\rho_5}\gamma_\mu \epsilon \hat F_{\rho_1\dots
\rho_5}+ {1\over 96}\left(\gamma_\mu^{\ \nu\rho\lambda}\hat G_{\nu\rho\lambda}
-9 \gamma^{\rho\lambda} \hat G_{\mu\rho\lambda}\right)\epsilon^* \nn \\
& & -{7\over 16} \left(\gamma_\rho \lambda\, \bar \psi_\mu \gamma^\rho \epsilon^*
- {1\over 1680} \gamma_{\rho_1\dots\rho_5} \lambda \, \bar\psi_\mu
\gamma^{\rho_1\dots\rho_5} \epsilon^* \right)\nn\\
& & + {1\over 32} i \left[\left({9\over 4} \gamma_\mu\gamma^\rho +
3\gamma^\rho \gamma_\mu\right)\epsilon\, \bar \lambda
\gamma_\rho\lambda \right.\nn \\
& &\left.   -\left({1\over 24} \gamma_\mu \gamma^{\rho_1\rho_2\rho_3} +
{1\over 6} \gamma^{\rho_1\rho_2\rho_3} \gamma_\mu\right)
\epsilon\,\bar\lambda \gamma_{\rho_1\rho_2\rho_3}\lambda +
{1\over 960} \gamma_\mu\gamma^{\rho_1\dots\rho_5} \epsilon\, \bar
\lambda \gamma_{\rho_1\dots\rho_5} \lambda    \right] \nn\\
& & + \delta^{(0)}_\Sigma (\psi_\mu),
\end{eqnarray} %
where the compensating $U(1)$  transformation is given by
\begin{equation}\label{compenpsi}{\delta^{(0)}_\Sigma\psi_\mu
= {1\over 2} i \Sigma = {1\over 4} i\psi_\mu
(\bar\epsilon \lambda^*)  - {1\over 4} i
\psi_\mu (\bar\epsilon^* \lambda) .}
\end{equation} %

By using (\ref{fierz}) and (\ref{vartino}) extensively we may manipulate the variation of
$\gamma^\mu \psi_\mu^*$   into the form,
\begin{equation}\label{vargamps}{\delta^{(0)}(\gamma^\mu \psi^*_\mu)_a = -{3\over 4}i
\lambda^*_a (\bar \epsilon \lambda) +  {1\over 1920} i
(\gamma^{\rho_1 \dots \rho_5} \epsilon^*)_a  \left(\bar\lambda
\gamma_{\rho_1\dots \rho_5} \lambda \right) + \dots ,}
\end{equation} %
where we have only kept the terms bilinear in
$\lambda$, $\lambda^*$. This implies the relation,
\begin{equation}\label{compsi}{(\lambda)^{15}_a \, \delta^{(0)}(\gamma^\mu \psi^*_\mu)_a =   -
15 i \lambda^{16} (\bar\lambda \epsilon^*) + \dots,}
\end{equation} %
which we use in the body of the text.

\section{Fermions in  M theory on $T^2$}

The massless  fermions of the type II string theories on $S^1$ 
are identified with  particular projections of the eleven-dimensional
gravitino compactified on $T^2$.  In order to calculate the
$\lambda^{16}$ loop amplitude we need to identify the 
particular projection that isolates the spin-half  dilatino of
the type IIB theory.

Compactification of the IIB theory on a circle  $S^1$ of
circumference $r_B$
(in string frame)  in the
direction $x^9$
breaks the $SO(9,1)$ Lorentz symmetry to $SO(8,1)$.   The
complex chiral
spin-$\half$ fermion $\lambda$ simply becomes a  complex spinor of
$SO(8,1)$.  The gravitino
decomposes into a
nine-dimensional gravitino $\hat \psi_\alpha$ (where the $SO(8,1)$
vector index
$\alpha = 0, 1,\cdots, 8$) together with a second
complex  spin-$\half$ fermion,
\begin{equation}\label{newferm}{\chi^A = r_B\psi^A_9,}
\end{equation} %
where the factor of $r_B$ comes from the component $e_{99}$ of the
zehnbein.
The nine-dimensional gravitino is defined by shifting $\psi_a$,
\begin{equation}\label{gravidef}{
\hat \psi_a = \psi_a +  { 1\over 7} \Gamma_a \Gamma^9 \chi}
\end{equation} %
(where $a$ labels the nine-dimensional tangent space, $a = 0,1 \cdots,8$)
so that the kinetic term is diagonal.

We will now identify the components of the $T^2$ compactification of
eleven-dimensional gravitino,  $\Psi_{\hat{\mu}}$   
($\hat{\mu}=0,\cdots,9,11$), that correspond to the IIB fermions
on $S^1$.  Compactification  on $T^2$  breaks the local Lorentz symmetry
\ from $SO(10,1)$ to $SO(8,1)\times SO(2)$.   The
nine-dimensional fermions can then be organized into eigenstates of
$SO(2) \equiv U(1)$, which  is   related to  the
$U(1)$  in
the denominator of  the coset space of the IIB theory.
The  world indices  split into the
compact   directions $\sigma =9,11$ and the noncompact directions
$\alpha = 0,\cdots, 8$.    Making  a block diagonal ansatz, the
elfbein $e_{ \hat{\mu}}^{\ \hat{m}}$, is written as
\begin{equation}\label{block}{e_{\hat{\mu}}^{\ \hat{m}}=\pmatrix{
e^{\ s}_{\sigma}  &   0  \cr
0&  e^{\ a}_{\alpha} \cr}}
\end{equation} %
where  $a$ again labels the nine-dimensional tangent space  and $s$ the
two-dimensional tangent space ($s=1,2$).
The zweibein, $e^{\ s}_{ \sigma}$,  of  $T^2$ may be chosen, in a
special Lorentz frame, to be
\begin{equation}\label{zweibein}{
e^{\ s}_{\sigma} =\sqrt{\calV\over  \Omega_2}\pmatrix{
\Omega_2 & \Omega_1  \cr
0&1  \cr}}
\end{equation} %
Symplectic reparametrizations of the torus act as
$SL(2,\IR)$ matrices from the left and local  Lorentz
transformations
act
as $SO(2)$ transformations from the right. The  condition that
the zweibein remains  in the frame
(\ref{zweibein}) leads to the standard $\Sl2$ transformation of the complex
structure of the
torus (which is (\ref{slactt}) with $\tau$ replaced by $\Omega$)
and induces a specific $\Omega$-dependent $U(1)$ transformation on  the  fermions.

The  spin-$\half$ components of the compactified
gravitino are written in a
complex basis  $z=x_{11}+ ix_{9}$,  $\bar{z}=x_{11}
-ix_{9}$,   as   $\Psi_z = \Psi_{11} +  i \Psi_9$ and $\Psi_{\bar z} =
\Psi_{11} - i\Psi_9$.    To relate these 32-component spinors  to
$\lambda$ and
$\chi$ of the IIB theory we  shall first   organize them into
eigenstates of
the $U(1)$ rotations of  the compact $T^2$ generated by $i\Gamma^{11}
\Gamma^{9}/2
+ j_{vec}$, where $j_{vec}$ acts on the vector index so that
$j_{vec} \Psi_z
=\Psi_z$ and  $j_{vec} \Psi_{\bar z} = - \Psi_{\bar z}$.
Those components of the  spin connection   that have  tangent-space
indices in the  compact directions are given by
\begin{equation}\label{omegdef}{
\omega_{z, \alpha z} = {i\over 2} {\partial_\alpha{\Omega}\over\Omega_2},
\qquad
\omega_{ \bar{z}, \alpha\bar{z}} =  {i\over 2}  {
\partial_\alpha{\bar \Omega}\over \Omega_2} ,\qquad
\omega_{ z, \alpha\bar{z}}= \omega_{ \bar{z},\alpha z}= {1\over \calV}
\partial_\alpha\calV.}
\end{equation} %
The type IIB complex scalar field is identified with $\Omega$ so the  
quantities $P_\alpha = \omega_{z, \alpha z}$ and  $\bar
P_\alpha = \omega_{ \bar{z}, \alpha\bar{z}}$ correspond to the fields with the same 
symbol defined in appendix C.  The third real scalar field  in the nine-dimensional theory is 
$\omega_{ z,\alpha\bar{z}}=- 4\partial_\alpha \ln r_B^E/3$ .

We shall make use of projectors made out of  the eleven-dimensional $\Gamma$ matrices. 
The complex combinations,
$\Gamma_z={1\over 2}(\Gamma^{11}+i\Gamma^9)$ and
$\Gamma_{\bar{z}}={1\over 2}(\Gamma^{11}-i\Gamma^9)$, obey
\begin{equation}\label{gamproc}{
\Gamma_z^2=\Gamma_{\bar{z}}^2=0, \qquad
\{\Gamma_z,\Gamma_{\bar{z}}\}= 1.}
\end{equation} %
so that the products
\begin{equation}\label{combs}{
\calP_z= (\calP_{\bar z})^T =\Gamma_{\bar{z}}\Gamma_{{z}}={1\over
2}(1-i\Gamma^9\Gamma^{11}),\qquad
\calP_{\bar{z}}= (\calP_z)^T = \Gamma_{z}\Gamma_{\bar{z}}={1\over
2}(1+i\Gamma^9\Gamma^{11})}
\end{equation} %
(where the superscript $T$ indicates the transpose) are projectors,
satisfying
\begin{equation}\label{projdef}{
\calP_z^2=\calP_z, \qquad \calP_{\bar z}^2 = \calP_{\bar z}, \qquad\calP_z
\calP_{\bar z} =0.}
\end{equation} %

The action of these projectors on a 32-component real spinor  is to
replace it
with a complex sixteen component spinor  which is an eigenstate of
the $U(1)$
spin, $i \Gamma^{11}\Gamma^{9}/2$.   Therefore, we  may identify the four
spin-1/2 fermions of
the nine-dimensional compactified theory with the components,
\begin{equation}\label{fermcoms}{
\calP_z \Psi_z, \qquad \calP_{\bar z} \Psi_{\bar z},}
\end{equation} %
\begin{equation}\label{fermcomstwo}{ \calP_z
\Psi_{\bar z}, \qquad  \calP_{\bar z} \Psi_z.}
\end{equation} %
The components  (\ref{fermcoms}) have $U(1)$ charges $+\thh$ and
$-\thh$,   respectively,  while the components
(\ref{fermcomstwo}) are a mixture of the states with  $U(1)$ charges
$\pm \half$.
These fields are simply related to the spin-$\half$ fermions of the
IIB theory
by converting $\lambda$ and $\chi$ (and their complex conjugates)  from
eigenstates of  $\Gamma^{11}$ to eigenstates of $i\Gamma^{11} \Gamma^9$.
This leads to the identifications,
\begin{equation}\label{ferms}{\calP_{z}\Psi_z=
\Gamma_{\bar{z}}\lambda , \qquad \calP_{\bar{z}}\Psi_z= \Gamma_z \chi}
\end{equation} %
which gives  the eigenstates of $\Gamma^{11}$,
\begin{equation}\label{lamdeef}{
\lambda = (1 + \Gamma_z) \calP_z \Psi_z, \qquad \chi  = (1 +
\Gamma_{\bar z})\calP_{\bar z} \Psi_z.}
\end{equation} %

In terms of the real components
\begin{equation}\label{lamcoms}{
\lambda_1=\psi_{11}^+-\Gamma^9 \psi_9^-, \qquad
\lambda_2=\psi_9^+   +\Gamma^9\psi_{11}^-.}
\end{equation} %
where the superscripts ${}^\pm$ indicate the chirality (the value of
$\Gamma^{11}$).
Similarly, writing $\chi = \chi_1 +i \chi_2$ and multiplying
(\ref{ferms})  by
$(1+\Gamma^{11})$  leads to
\begin{equation}\label{chicoms}{
\chi_1 = \Gamma^9 \psi_9^+ - \psi_{11}^-,\qquad  \chi_2 = -\psi_9^-
- \Gamma^9 \psi_{11}^+.}
\end{equation} %

The remaining components,  $\Psi_a$ form the  nine-dimensional
gravitino
after a shift similar to  (\ref{gravidef}),
\begin{equation}\label{mgravi}{
\hat \Psi_a=  \Psi_a +  {1\over 7} \Gamma_a
\left(\Gamma_z\Psi_{\bar{z}}+\Gamma_{\bar{z}}\Psi_{z}\right)
 = \Psi_a +  {i\over 7} \Gamma_a \Gamma^9(\calP_z \chi - \calP_{\bar z}
\chi^*).}
\end{equation} %
The relations between the M-theory  fields and the IIB fields can
be confirmed by  comparing the way they behave under supersymmetry
transformations.

For completeness  the IIB fermions can be related to the IIA
fermions in a straightforward  manner.
The spin-1/2 fermions in the IIA theory follow by direct dimensional reduction
of eleven-dimensional supergravity \cite{campbellwest}. The
nine-dimensional fermions that arise from  the spin-$\half$ fields in ten dimensions are
\begin{equation}\label{lamad}{
\lambda_{A}=\psi_{11}+\Gamma^9\Gamma^{11}\psi_9,}
\end{equation} %
which decomposes into the chiral components,
\begin{equation}\label{lamchir}{
\lambda_{A}^+=\psi_{11}^+-\Gamma^9\psi_9^-, \qquad
\lambda_{A}^-=\psi_{11}^-+\Gamma^9\psi_9^+.}
\end{equation} %
Similarly,
\begin{equation}\label{chiad}{
\chi_{A}=\psi_9 +\Gamma^9\Gamma^{11}\psi_{11},}
\end{equation} 
which  has   chiral components
\begin{equation}\label{chichir}{
\chi_{A}^-=\psi_9^- +\Gamma^9\psi_{11}^+, \qquad
\chi_{A}^+=\psi_9^+ -\Gamma^9\psi_{11}^-.}
\end{equation} %
Comparing (\ref{lamcoms}) with (\ref{lamchir})  and  (\ref{chicoms}) with
(\ref{chichir}) 
gives the identification of fields of IIA and IIB in nine dimensions,
\begin{equation}\label{fermrels}{
\lambda_1 = \lambda_{A}^+, \qquad
\lambda_2 = \Gamma^9 \lambda_{A}^-,\qquad
\chi_2=-\chi_{A}^-,\qquad
\chi_1= \Gamma^9 \chi_{A}^+,}
\end{equation} %
which are in agreement with the world sheet T-duality
rules.

\end{appendix}

\end{document}